\documentclass[12pt]{elsart}
\newcommand{\ms}{$M_{\odot}$}
\newcommand{\msb}{$M_{\odot}$~}
\newcommand{\al}{$^{26}$Al}
\newcommand{\fe}{$^{60}$Fe}
\newcommand{\be}{$^{10}$Be}
\newcommand{\cl}{$^{36}$Cl}
\newcommand{\ca}{$^{41}$Ca}
\newcommand{\mn}{$^{53}$Mn}
\newcommand{\pd}{$^{107}$Pd}

\newcommand{\pu}{$^{244}$Pu}
\newcommand{\hf}{$^{182}$Hf}
\newcommand{\ct}{$^{13}$C}
\newcommand{\ctb}{$^{13}$C~}
\newcommand{\cm}{$^{247}$Cm}

\newcommand{\nb}{$^{92}$Nb}
\newcommand{\msun}{\ensuremath{M_\odot}}
\newcommand{\mdot}{\ensuremath{\dot{M}}}
\def\gtrsim{\:\lower 0.4ex\hbox{$\stackrel{\scriptstyle >}
{\scriptstyle\sim}$}\:}
\def\lesssim{\:\lower 0.4ex\hbox{$\stackrel{\scriptstyle <}
{\scriptstyle\sim}$}\:}
\journal{Nuclear Physics A}

\usepackage{epsfig,lscape,longtable}
\usepackage{rotating}

\begin{document}

\begin{frontmatter}

\title{Short-lived Nuclei in the Early Solar System: \\
Possible AGB Sources}

\author{G. J. Wasserburg}
\address{The Lunatic Asylum,
Division of Geological and Planetary Sciences, California Institute
of Technology, Pasadena, CA 91125} \ead{isotopes@gps.caltech.edu }

\author{M. Busso}
\address{Department of Physics, University of Perugia, via
Pascoli, Perugia, Italy, 06123} \ead{maurizio.busso@fisica.unipg.it}

\author{R. Gallino}
\address{Department of General Physics and Sezione INFN,
University of Torino, via P. Giuria 1, Torino, Italy, 10125; also
Centre for Stellar and Planetary Astrophysics, School of
Mathematical Sciences, Monash University, 3800 Victoria, Australia}
\ead{gallino@ph.unito.it}

\author{K. M. Nollett}
\address{Physics Division, Argonne National Laboratory,
Argonne, IL 60439-4843} \ead{nollett@anl.gov}


\newpage
\begin{abstract}
The abundances of short-lived radionuclides in the early solar system
(ESS) are reviewed, as well as the methodology used in determining
them. These results are compared with the inventory estimated for a
uniform galactic production model. It is shown that, to within a
factor of two, the observed abundances of $^{238}$U, $^{235}$U,
$^{232}$Th, $^{244}$Pu, $^{182}$Hf, $^{146}$Sm, and $^{53}$Mn are
roughly compatible with long-term galactic nucleosynthesis. $^{129}$I
is an exception, with an ESS inventory much lower than expected from
uniform production. The isotopes $^{107}$Pd, $^{60}$Fe, $^{41}$Ca,
$^{36}$Cl, $^{26}$Al, and $^{10}$Be require late addition to the
protosolar nebula.  $^{10}$Be is the product of energetic particle
irradiation of the solar system as most probably is $^{36}$Cl.  Both
of these nuclei appear to be present when \al\ is absent. A late
injection by a supernova (SN) cannot be responsible for most of the
short-lived nuclei without excessively producing $^{53}$Mn; it can
however be the source of $^{53}$Mn itself and possibly of
$^{60}$Fe. If a late SN injection is responsible for these two nuclei,
then there remains the problem of the origin of $^{107}$Pd and several
other isotopes. Emphasis is given to an AGB star as a source of many
of the nuclei, including $^{60}$Fe; this possibility is explored with
a new generation of stellar models.  It is shown that if the dilution
factor (i.e. the ratio of the contaminating mass to the solar parental
cloud mass) is $f_0 \sim$ 4$\times$10$^{-3}$, a reasonable
representation for many nuclei is obtained; this requires that
($^{60}$Fe/$^{56}$Fe)$_{ESS}$ $\sim$ 10$^{-7}$ to
2$\times$10$^{-6}$. The nuclei produced by an AGB source do not
include $^{53}$Mn, $^{10}$Be or $^{36}$Cl if it is very abundant.  The
role of irradiation is discussed with regard to $^{26}$Al, $^{36}$Cl
and $^{41}$Ca, and the estimates of bulk solar abundances of these
isotopes are commented on. The conflict between various scenarios is
emphasized as well as the current absence of an astrophysically
plausible global interpretation for all the existing data. Examination
of abundances for the actinides indicates that a quiescent interval of
$\sim$ 10$^8$ years is required for actinide group production.  This
is needed in order to explain the data on $^{244}$Pu and the new
bounds on $^{247}$Cm.  Because this quiescent interval is not
compatible with the $^{182}$Hf data, a separate type of $r$-process
event is needed for at least the actinides, distinct from the two
types that have previously been identified.  The apparent coincidence
of the $^{129}$I and trans-actinide time scales suggests that the last
heavy $r$ contribution was from an $r$-process that produced very
heavy nuclei but without fission recycling so that the yields at Ba
and below (including I) were governed by fission.

\end{abstract}


\end{frontmatter}

\newpage

\section{Introduction}
More than forty years ago John Reynolds \cite{rey60} at Berkeley
announced the discovery of an excess of $^{129}$Xe in a meteorite
($^{129}$Xe$^{\ast}$) and suggested that it could be ascribed to the
{\it in situ} radioactive decay of $^{129}$I ($\bar\tau = 23$
Myr). This was proven later, through a demonstration that excesses of
$^{129}$Xe$^{\ast}$ are directly correlated with stable $^{127}$I
\cite{jr61}. These isotopes of iodine are $r$-process products,
attributed to supernovae. It was immediately recognized \cite{wfh60}
that the observed abundance of $^{129}$I could be ascribed to the
long-term production of $r$-process nuclei in the Galaxy, provided the
solar system material had been isolated from the interstellar medium
for about $10^8$ years. We can thus notice how basic questions on the
presence of ``live'' radioactivities in the Early Solar System
(hereafter ESS) were associated with supernova sources since the
beginning of modern research efforts.  However, proper answers for the
sources of nuclei have then been looked for in many works and remain
an open issue today.

The search for radionuclides other than $^{129}$I was, broadly
speaking, not successful for a long period of time. There were many
negative or failed or erroneous efforts. Critical advances in
analytical techniques would govern the progress. These advances
involved great improvements in high-precision, high-sensitivity mass
spectrometry, analytical microchemistry, and sample
preparation. Crucial for all subsequent progress was the fall of the
Allende meteorite in 1969.  This fall was contemporaneous with major
efforts in some laboratories to prepare for lunar samples to be
returned by the Apollo missions. The Allende fall made it possible to
sample old materials from the solar nebula including early refractory
condensates, the Calcium and Aluminum Inclusions (CAIs) \cite{gros72}
\cite{gros80} \cite{mwd70}. Measurements of $^{129}$Xe$^\ast$ were
also done on these inclusions and gave the same results as found for
normal chondrites \cite{pl72}. These early-formed samples in Allende
then also yielded quantitative evidence of the existence of
radioactive nuclei with a much shorter meanlife. In particular,
$^{26}$Al ($\bar\tau$ = 1.03 Myr) had been sought earlier in
meteoritic material using precise methods \cite{stw70} but had not
been found in the available meteoritic or lunar samples. However, with
the fall of Allende, the important discovery by Clayton, Grossman and
Mayeda of large oxygen anomalies that represented substantial shifts
in the isotopic abundances of a major element \cite{cla73}, and the
demonstration of exceedingly primitive Sr -- very low
$^{87}$Sr/$^{86}$Sr ratios with no enhancements due to $^{87}$Rb decay
in some refractory inclusions -- led to the possibility that \al~
should be looked for in this material \cite{gpw73}.  Isotopically
anomalous Mg was found in CAIs \cite{gcom76}\cite{ldap74}.  There were
samples with both excesses and deficiencies in $^{26}$Mg (a few per
mil) in samples with 50 per mil enhancements in $^{16}$O. \al~ was
then found to have been present in CAIs through a clear correlation of
excess $^{26}$Mg $(^{26}$Mg$^{\ast})$ with $^{27}$Al in different
minerals with widely varying Mg/Al ratios \cite{lpw76}
\cite{lpw77}. Early suggestions by Urey \cite{ur55} and Urey \& Donn
\cite{urd56} indicated that the nuclide \al, which had just then been
discovered in a cyclotron target \cite{sim54}, was the only reasonable
source for the early heating and melting of planetesimals. The authors
noted that: ``If the problem of producing sufficient quantities of
$^{26}$Al can be solved, this nuclide should be of considerable value
in chemistry, metallurgy, and related fields'' \cite{sim54}. As
pointed out by Urey \cite{ur55}, the heating and melting of small
planetesimals required a short-lived radioactivity of a major element
as the heat source, since there would not be sufficient gravitational
energy to melt them.

A few years before the fall of the Allende meteorite, the presence of
a relatively long-lived species, possibly $^{244}$Pu ($\bar\tau = 115$
Myr), had been inferred from excesses of neutron-rich Xe isotopes in
planetary differentiates \cite{rk65}. Subsequent work showed that the
enrichment in unshielded Xe isotopes was directly associated with
excess fission tracks in meteoritic minerals rich in U, Th, and REE
\cite{can67} \cite{whb69a} \cite{whb69b} \cite{sp77} and required the
{\it in situ} fission of a transuranic nuclide (possibly \pu). The Xe
isotopic composition in these meteorite samples showed very large
excesses in $^{131,132,134,136}$Xe. The pattern was identical to that
resulting from \pu~ spontaneous fission subsequently established in
the laboratory by Alexander et al. \cite{alr71}. \pu-fission Xe was
also found in the CAIs from Allende \cite{pl72}. The hint of excesses
of $^{142}$Nd by Lugmair \& Marti \cite{lm77} pointed to the possible
presence of $^{146}$Sm ($\bar\tau = 148$ Myr), a $p$-process isotope
that $\alpha$-decays to $^{142}$Nd with cosmochronologic implications
\cite{aud72}. Excesses of $^{142}$Nd were found to be widespread, both
in ESS materials and in planetary differentiates (PD), and to be well
correlated with Sm in planetary differentiates \cite{lm77} \cite{jw84}
\cite{ppw89}; see also \cite{spw94}. Evidence for the $p$ nuclide
$^{92}$Nb ($\bar{\tau}=52$ Myr) was found by Harper \cite{harp96}. The
presence of $^{107}$Pd ($\bar\tau = 9.4$ Myr) was established by Kelly
\& Wasserburg \cite{kw78}; this nuclide, which is produced by both $r$
and $s$ processes, was found to have been present in a large number of
iron meteorites representing metal segregation in protoplanets or
planetesimals \cite{cw83} \cite{cw96} \cite{ch01}. The results on \pd~
showed that planetary cores formed very early in the history of the
solar system. The discovery of tungsten isotopic anomalies
(deficiencies) in $^{182}$W in iron meteorites and their correlation
with Hf in chondritic meteorites demonstrated the presence of \hf~
$(\bar\tau = 13$ Myr) in early planets at the time of core formation
\cite{hj96} \cite{hjsb96} \cite{lh95}\cite{lh96}.

Both the possible presence of \mn~ ($\bar\tau = 5.3$ Myr) in CAIs,
with a correlation of $^{53}$Cr$^\ast$ with Mn, suggested by Birck and
All\`{e}gre \cite{bir85} \cite{bir88}, and the subsequent clear
demonstration of abundant \mn~ in planetary differentiates
\cite{bir88}\cite{hkk98}\cite{ls98}, stimulated new attention to the
nucleosynthetic processes in supernovae (SNe II or SNIa) and in
spallation processes. The hints of \fe~ $(\bar\tau = 2.2$ Myr)
discovered by Shukolyukov \& Lugmair \cite{sluga} \cite{slugb} and the
subsequent recent demonstration of $^{60}$Ni excesses correlated with
Fe/Ni in chondrites, require \fe~ to be present at rather high
abundances in the early solar system \cite{th03} \cite{mos03}
\cite{mos03b}. This nucleus has connections to both supernovae and
asymptotic giant branch (AGB) sources, but not to spallation
reactions. There is the possibility of $^{205}$Tl excesses that would
indicate the presence of $^{205}$Pb ($\bar\tau = 22$ Myr, a shielded
nucleus), which, like $^{204}$Pb, is certainly from the $s$-process
\cite{cw87}.

The discovery of the very short-lived \ca~ $(\bar\tau = 0.15$ Myr) by
Srinivasan, Ulyanov \& Goswami \cite{sug94} and Srinivasan et
al. \cite{sri96} has very important ramifications. Although abundantly
produced in AGB stars, its short time scale may present a problem. At
the low abundance observed, it might also be produced by proton
bombardment, but it is correlated with $^{26}$Al. The important
discovery of \be~ $(\bar\tau = 2.3$ Myr) in the early solar system by
McKeegan, Chaussidon and Robert \cite{mcr00} has demonstrated the
presence of a nuclide that is not a product of stellar
nucleosynthesis, but requires proton bombardment of small solids.

In addition, Allende provided a whole host of small, but clearly
measurable, isotopic anomalies in many elements (a Pandora's box
of the nuclides), which demonstrated that incompletely mixed
material from different presolar sources was preserved in
macroscopic samples of solar system processed material in
meteorites. These ``isotopic anomalies'' and proposals of various
nucleosynthetic mechanisms caused lots of excitement. They also
resulted in some difficulty in unraveling the presence and
abundances of some short-lived nuclei in CAIs, as there was not
always a clear base line of initial isotopic composition in some
of these samples.

A prescient study by Black \cite{black72} led to the remarkable result
that almost pure $^{22}$Ne was present in chondrites. Black attributed
this to the preservation of presolar grains from Red Giant Branch
(RGB) stars. This led to a long and difficult chase to find such
grains. The result was in the major discovery of individual
unprocessed refractory presolar dust grains in chondritic meteorites
\cite{az93} \cite{az94}. These grains provide direct evidence on the
nature of potential contributors to the solar nebula and of aspects of
stellar nucleosynthetic processes that were previously not available
(see Section \ref{cdg}).  In particular, $s$-process abundance
patterns can now be compared with the observed abundances in
circumstellar dust grains. It is now known that dust grains from
diverse AGB stars are a substantial or major source of the condensed
matter that was incorporated into the early solar system. Reviews of
astronomical observations on dust are given by Draine \cite{draine1}
\cite{draine2} \cite{draine3}.

In this report we present a review of the short-lived radioactivities
($10^9 > \bar\tau > 10^5$ yr) that were present in the early solar
system and the potential stellar sources of these nuclides. This
panoply of short- to intermediate-lifetime nuclei provides the direct
connection to nuclear astrophysical processes and a host of exciting
and confusing possibilities in search of real explanations. For an
extensive review of the experimental data on the early solar system
abundances of short-lived nuclei, we refer the reader to McKeegan \&
Davis \cite{md04}. In particular, we will focus on the possible
addition of fresh stellar debris into the protosolar nebula from a
single source, with an emphasis on possible AGB contributions. Some
intermediate and longer lived nuclei will be shown to come most
plausibly from long-term galactic nucleosynthesis. We will also review
some of the characteristics of pre-solar circumstellar condensates
that are preserved in meteorites, relating to the production of \al~
in stellar models. A critical matter will be the nature of
nucleosynthetic yields of potential sources. A recent review of the
short-lived nuclei by Busso et al. \cite{bgw03} will be used as the
source for the present report. Substantial new results will be
presented concerning the characteristics of AGB stars of a range in
masses and with specific consideration of the results of cool bottom
processing (see Section \ref{cbproc}).

\section{Circumstellar Dust Grains}\label{cdg}

A wide variety of grains has been discovered in residues from
chondrites that showed gross variations of the isotopic ratios of
major elements not related to radioactive decay. These grains
demonstrated that presolar dust from a wide variety of stellar sources
was the net material from which the solar system formed (see reviews
by Anders \& Zinner on this discovery \cite{az93} \cite{az94}). A
recent review of ``Astrophysics of Stardust'' \cite{clan04} is a
useful guide here.

Some grains contained clear evidence of \al~ when they formed
\cite{zaal91}. Some small fraction of these grains were presumably the
carriers that made up the solar inventory of the short-lived
radioactive nuclides. However, most of the extrasolar grains were
sufficiently old that the radioactive nuclei in them had already
decayed before the solar system formed.  Until recently, only
refractory grains had been recovered. There is extensive recycling of
the grains within the interstellar medium (ISM) and in diverse stars;
only a limited sampling is available. Much earlier generations of
stellar debris are, of course, involved in making up the solar
inventory; this includes gas (atoms, molecules, and ions) and stellar
dust that has been cycled-recycled by a variety of processes in the
ISM (see Draine \cite{draine1} \cite{draine2}, \cite{draine3}). The
gas is depleted in ``non-volatile'' elements which are mostly resident
in the dust phase. The phases responsible for carrying the elemental
budget in the ESS involves all of the above components. Some ``live''
nuclei present in the ESS, such as $^{129}$I, may be present in the
gas phase and others (like \al~ or \hf) are certainly in some dust
particles.

\begin{figure}
\begin{center}
\end{center}
\caption{$^{14}$N/$^{15}$N versus $^{12}$C/$^{13}$C of
circumstellar SiC grains from the data compilation generously
provided by S. Amari. Only data for grains of types A and B,
mainstream, and Y are shown ($\sim 98\%$ of all data), because
only these are attributable to AGB stars.  The solar ratios are
shown by the dotted lines. The histogram at the bottom shows the
relative frequency of occurrence of $^{12}$C/$^{13}$C in the data,
on a linear scale. Note the well-defined peak labeled MS (gray
squares, Mainstream Grains). We have included the Y grains as part
of the MS grains, with $^{12}$C/$^{13}$C$ >
(^{12}\mathrm{C}/^{13}\mathrm{C})_\odot$ as they are readily
explained by AGB evolution for stars with $Z$ somewhat less than
$Z_\odot$. The rare A and B grains (circles) with low
$^{12}$C/$^{13}$C and an enormous range in $^{14}$N/$^{15}$N still
require explanation. Presentation of these results is in Refs.
\cite{ama01} and \cite{anll01} and references therein.  The region
accessible by AGB evolution at solar metallicity with C/O $> 1$ is
enclosed by the solid curve. The accessible regions for different
values of the metallicity $Z$ and of the stellar mass are as
indicated. Note that with decreasing $Z$ the envelope of
accessible C-N space extends to higher $^{12}$C/$^{13}$C.}
\label{nvc}
\centerline{\epsfig{file=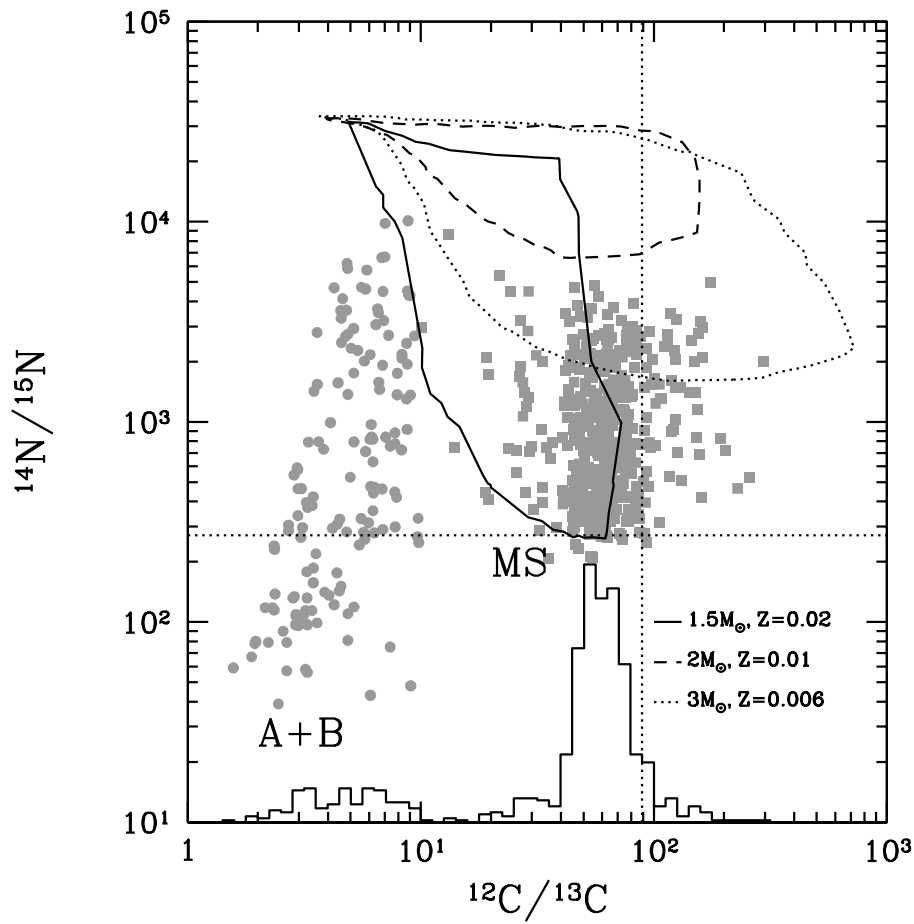,width=5in}}

\end{figure}

The $^{12}$C/$^{13}$C and $^{14}$N/$^{15}$N isotopic compositions of
circumstellar SiC grains recovered from meteorites are shown in
Fig. \ref{nvc}. A histogram showing the frequency of occurrence of a
given $^{12}$C/$^{13}$C ratio is shown at the base of the figure. It
is evident that the predominant population of grains lies in the range
of $40 < ^{12}$C/$^{13}$C $< 100$.  These are the so-called
``mainstream'' grains and represent the value expected for AGB
stars. There is a small population with very low $^{12}$C/$^{13}$C
that remains an unresolved problem (see \cite{nbw03}). The
$^{14}$N/$^{15}$N ratios require special attention (see Section
\ref{cbproc}). The $^{26}$Al/$^{27}$Al abundances in some of these
grains are shown in Fig. \ref{alc}; the observed range is $10^{-5} <
^{26}$Al/$^{27}$Al $< 2 \times 10^{-2}$.

Isotopic analyses of individual SiC grains for heavy elements
have recently become possible. This was primarily a result of
major instrumental development by M. Pellin and his colleagues at
Argonne National Laboratory (cf. \cite{nic98} \cite{npl96}
\cite{npc97}); similar measurements have now become feasible with
different techniques in other laboratories \cite{Marhas04}. The
results on a number of SiC grains from the ``mainstream''
population show clear and definitive enrichments in $s$-process
nuclei and deficiencies in $r$ and $p$ nuclei (see example in Fig.
\ref{moly}).  There is even evidence of $^{99}$Tc from SiC grains
\cite{sav04}.  The presence of these enrichments of $s$-process
nuclei greatly strengthens the assignment of this population of
carbide grains to AGB sources with significant neutron exposures.
These observations are in full accord with the direct astronomical
observations of enrichment of $s$-process ``elements'' in AGB
stars.

The discovery of circumstellar oxide grains in meteorites
\cite{hhws92} \cite{hfgw94} \cite{hhfw94} \cite{nag94} and the
important and extensive results by Nittler et al.
\cite{nag94}\cite{nag97}\cite{nawg98} revealed that most of these
grains also appear to come from AGB sources. Fig. \ref{oxy} shows
a compilation of oxygen isotopic data from individual refractory
oxide grains (corundum, hibonite, spinel). It can be seen that
the preponderance of the grains show $^{18}$O/$^{16}$O $<
(^{18}\mathrm{O}/^{16}\mathrm{O})_\odot$ and $^{17}$O/$^{16}$O $>
(^{17}\mathrm{O}/^{16}\mathrm{O})_\odot$. This is in general
accord with the effects expected from first dredge-up in Red Giant
stars (see Dearborn \cite{dear92}). However, observations on the
grains show overproduction of $^{17}$O and much more extensive
destruction of $^{18}$O that is far outside the range to be
expected for standard RGB or AGB models \cite{nag97} \cite{cwh99}.

A compilation of the available $^{26}$Al data is shown in Fig.
\ref{almg}. Again, as is the case for carbide grains, the data show an
enormous range in $^{26}$Al/$^{27}$Al, up to a few times
$10^{-2}$. Clear evidence of $^{41}$K$^\ast$ is found in some
circumstellar oxide grains from the decay of \ca. Measurement of
potassium isotopes in Ca-rich oxide grains (hibonites) show $^{41}$K
correlated with Ca/K. The inferred \ca/$^{40}$Ca for some grains is in
excellent agreement with the AGB model values, although other grains
show much lower values that still require explanation \cite{cwh99}
\cite{nit05}.

A most exciting discovery has been made of silicate grains from
circumstellar sources by Nguyen \& Zinner \cite{ngu04}, by Nguyen,
Zinner \& Stroud \cite{ngu05}, and by Hoppe, Mostefaoui \& Stephan
\cite{hoppe05}, all in the Acfer meteorite, and also by Mostefaoui,
Marhas \& Hoppe \cite{mostefaoui04} in Bishunpur.  Such presolar
grains were first found in interplanetary dust particles by Messenger
et al. \cite{messenger03}.  This new generation of measurements was
made possible by major technical developments by George Slodzian.
This permits analysis of sub-micron grains.  These circumstellar
silicate grains (long sought for) appear to be far higher in abundance
than other presolar grains.  In general, they appear to have their
origin in AGB stars.  This strongly emphasizes the points laid out
above about the importance of AGB contributions to the ESS.  These
observations open up an exciting area of research.

The results outlined above attest to the major contribution of
carbide, carbon, oxide, and now silicate grains from diverse AGB stars
to the initial solar system chemical and isotopic abundances. They do
not, however, identify any source that provided short-lived nuclei to
the ESS. In addition to the preponderance of grains attributable to
AGB stars, there are a relatively rare subset of some of the SiC and
graphite grains and a few oxide grains that are probably (sometimes
certainly) from SNe sources \cite{cwh99} \cite{nit05} \cite{scb05}
\cite{chw98}. These account for a fraction of about $10^{-2}$ of the
grains observed. If the available sampling is representative, then
this must limit the level of late contributions from a supernova. The
individual grains give us a view of possible contributing sources to
the short-lived nuclei, but they do not define these sources or the
amount contributed by them. Most meteoritic material has been
chemically and thermally processed in the solar system, so that the
individual components are typically homogenized.

\section{Review of Nucleosynthesis in AGB Stars}\label{revagb}

In considering the matter of short-lived nuclei, the basic
observational data are the abundances of the relevant radioactive
nuclei at some arbitrary ``initial'' time. As our main focus will be
possible AGB sources, we will first give a short review of AGB
nucleosynthesis considering the standard model. Then we will present a
more extensive review of cool bottom processing (which has caused
significant changes in the treatment of $^{26}$Al) and of AGB
contributions.

\subsection{Standard Models}

A thorough review of the status of stellar models is presented by
Straniero et al. \cite{sg05} in this volume. For low mass stars, after
exhaustion of the H in the core, energy is only produced by a
H-burning shell, leaving a He core. The star then develops a fully
convective envelope structure and ascends the H-R diagram to the red
giant branch (RGB), where surface convection first enters previously
radiative layers and mixes material that had previously experienced
there proton captures to the photosphere (first dredge-up).
Observations show that the model predictions for the first dredge-up
are in rough agreement with the spectroscopic abundances of C and O
isotopes in RGB stars of masses above $\simeq 2.5 M_\odot$ but not for
lower masses (see e.g. \cite{bgw99}). At the end of the RGB stage, He
is ignited and a C-O core evolves.  When the He at the center is
finally exhausted, energy is produced by H-shell burning, with regular
short interruptions from thermal instabilities at the top of the C-O
core.  In each interruption, the He shell burns for a short
time. Thus, the H shell and the He shell burn alternately over a very
narrow region of mass in the star. The He-shell ignition is followed
by the convective penetration of the envelope through the then
inactive H-burning shell (third dredge-up or TDU). This stage is
referred to as the thermally pulsing AGB phase (TP-AGB: see
\cite{ir83}). The third dredge-up mixes processed material from the
H-He zone interface region into the convective envelope and thus the
photosphere, where the changes in composition can be observed. This
involves addition of major reaction products made by proton reactions
(e.g., $^{13}$C, $^{14}$N) as well as by $\alpha$-capture reactions
and neutron captures (the $s$-process nuclei). TDU does not change the
O isotopic composition.

In the nineties, improvements in the input physics of stellar models
(equation of state, opacities) and the advent of cheap and fast
computers allowed the third dredge-up to be self-consistently modeled
down to low masses, thus yielding the first theoretical interpretation
of low-luminosity C stars \cite{str95} \cite{fl96}.  It is this class
of stellar models (see \cite{sg05}, this volume, and references
therein) that provides the temperature, density, electron density, and
effective chemical composition in each zone over the stellar
lifetimes. The input parameters for such models are: the initial
chemical composition of the star; its initial mass; and the rate of
loss of the envelope. In order to save computer time, detailed
nucleosynthesis calculations for neutron capture nuclei and other
minor species are, even now, often omitted in the complete stellar
models as they do not affect the energy budget and hence do not affect
the stellar structure. In these cases, the outputs of stellar models
are then used as a basis for post-process computations in which large
networks involving thousands of reactions can be used. However, the
major stellar codes (e.g. the Australian MSSSP, or the Italian FRANEC)
can now be run in individual control cases using the whole reaction
set \cite{kl03} \cite{str03}, thereby providing the nodes of a
grid. Stellar parameters for other cases can then be deduced by
suitable interpolations \cite{str03}.

In AGB stars two major neutron sources are at play: the
$^{22}$Ne$(\alpha,n)^{25}$Mg reaction \cite{agwc60} and the
$^{13}$C$(\alpha,n)^{16}$O reaction \cite{agwc56} \cite{agwc57}
\cite{grjl54}. Both reactions take place in the He shell, but only the
first is a direct and intrinsic consequence of the stellar
evolution. The neutron densities produced from $^{22}$Ne burning
reflect the temperature and thus the stellar mass. The reaction
products of neutron capture from this source do not depend on any
parameters other than the initial stellar abundances and the basic
stellar model. The second reaction producing neutrons is
$^{13}$C$(\alpha,n)^{16}$O and requires that matter rich in $^{12}$C
(from the He shell) must react with protons to produce $^{13}$C in a
layer of the He-H intershell called the {\it \ct~ pocket}.  This mixing
scenario requires some non-convective process to bring protons into
the intershell region where there is abundant $^{12}$C \cite{lug03b}.
A very good representation of the solar system ``main $s$-component''
can be obtained from a galactic chemical evolution model that uses the
outputs of AGB stars of different generations, including products from
both neutron sources \cite{tr99} \cite{ar99}, the dominant contributor
being the $^{13}$C$(\alpha,n)^{16}$O source. However, the strength of
this source must be treated as a free input parameter in the absence
of any self-consistent model for the required proton mixing.

The presentation here with regard to possible AGB sources to account
for some short-lived nuclei in the early solar system is an advanced
treatment of the earlier approach \cite{wbgr94}. In that report, we
used more approximate stellar models developed by other workers and
computed nucleosynthesis in AGB stars with a post-processing
calculation based on a schematic representation of the thermal
pulses. This involved two key parameters: 1) The neutron exposure
producing $s$-process nuclei; and 2) the mass of the star, which
controls the temperature of the H-burning shell for $^{26}$Al
production. These parameters then established a relationship between
the mass of the possible stellar source, the net neutron exposure, and
the amount of ejected AGB envelope that must mix with the ambient ISM
to provide the inventory of radioactive nuclei in the solar
nebula. The input parameters used were the observed ESS \pd~ abundance
as a monitor of the neutron exposure, and \al~ as a monitor of the
proton captures on $^{25}$Mg in the H-burning shell (and of its severe
destruction by neutron captures in the He shell, during thermal
pulses). Using these parameters, it was possible to obtain a
self-consistent abundance pattern for several other nuclides and to
predict the abundances of radioactive nuclei that had not been
observed. The \al~ produced in the standard model gave
$^{26}$Al/$^{27}$Al $\sim (1-3) \times 10^{-3}$. This model was
certainly not sufficient to provide the high $^{26}$Al/$^{27}$Al
observed in some circumstellar dust grains attributed to AGB stars
(see Figs.  \ref{alc} and \ref{almg}).

Recognition that additional proton processing of envelope material
just above the H-burning shell was necessary resulted from the
laboratory observations of oxygen isotopes in circumstellar dust
grains found in meteorites and from astronomical observations of
$^{12}$C/$^{13}$C in low-mass RGB stars.  This additional mixing and
reaction mechanism fundamentally alters the problem for AGB stars and
was discussed by Busso, Gallino \& Wasserburg \cite{bgw99} in their
review of short-lived nuclei. The result is that $^{26}$Al in AGB
stars may effectively be governed by reactions just above the H
burning shell due to some transport in a phenomenon called ``extra
mixing'' or Cool Bottom Processing (CBP). This process must also occur
in low mass stars on the Red Giant Branch \cite{char95} \cite{char04},
but would not produce much \al~ there due to the lower temperature of
the H shell.  The amount of this processing on the AGB is not a priori
known and is a free parameter. It is therefore not possible to use the
$^{26}$Al abundance as the basis for estimating the AGB input.  Thus
the effort to provide a self-consistent model is left open, since
there are no other radionuclides with lifetimes of $\sim 10^6$ yr that
are only produced by normal AGB evolution and whose early solar system
abundance has been established at a reasonably precise value. Because
of the importance of CBP in AGB stars, we provide an extensive summary
of the effects of this mechanism on both radioactive and stable
nuclei.

\section{Cool Bottom Processing}\label{cbproc}

Convection and mixing inside stars cannot be treated from first
principles in one-dimensional models.  Instead, these phenomena
are treated through approximations, applying stability criteria to
determine whether energy transport in a given layer is radiative
or convective.  Convective regions are described in the mixing
length formalism, and their compositions homogenize rapidly.
Radiative regions are assumed to be free of matter circulation.
However, there is clear observational evidence for partial mixing
in the radiative regions, so mechanisms must be available to drive
circulation there as well (e.g., Herwig  \cite{hw05} and
references therein). A prime example is weak activation of the CN
cycle attributed to material circulating below the convective
envelope of stars on the red giant branch, as mentioned above and
in \cite{char95}  \cite{hw05} \cite{gb91} \cite{bsij99}. Further
evidence may be found in the observation of an extremely metal
poor star with [Fe/H]~$= -2.72$ by Lucatello et al. \cite{luc}
(here $[{\rm Fe/H}] = \log({\rm Fe/H})_{star}-\log({\rm
Fe/H})_{\odot}$). These workers found that [Pb/Fe]~$=3.3$,
[C/Fe]~$= 2.6$ and ($^{12}$C/\ct)$ = 6$. This carbon isotopic
ratio could not be attained by standard AGB models as they would
yield ($^{12}$C/\ct)$ \gtrsim 4 \times 10^4$ and thus these
results are a further strong indication of CBP.

In models of cool bottom processing, it is assumed that slow
circulation begins at the bottom of the star's convective envelope
and carries envelope material down to layers dense and hot enough
for some nuclear processing to occur but not far enough to result in
significant energy generation.  A return flow maintains the stellar
structure in steady state, returning processed material to the fully
convective envelope.

Because the physical mechanism driving CBP is not known, its action
inside any given star is described in the models by two parameters, a
depth of mixing and a rate of mixing, specified in the following ways:
1) The maximum depth of mixing may be specified as maximum temperature
seen by the processed material ($T_P$) \cite{wbs95} \cite{nbw03}
\cite{mb00}, or as the fraction of the mass ($\delta M$) of the
radiative region above the hydrogen shell where the slow mixing takes
place \cite{dw96} \cite{ddnw98} \cite{wdc00}; 2) The rate of mixing is
specified either by the mass circulated per year ($\dot{M}$)
\cite{wbs95} \cite{nbw03} \cite{mb00} or by a parametric diffusion
coefficient ($D_\mathrm{mix}$) \cite{dw96} \cite{ddnw98}
\cite{wdc00}. The two sets of parameters (circulation rate and maximum
temperature versus diffusion coefficient and mass region) are
completely equivalent and indistinguishable as far as observable
effects are concerned, and relations between the two representations
are discussed in Nollett et al. \cite{nbw03}. Here, we use the
parameter set $(T_P/T_{\rm H}, \mdot)$ as in Messenger \cite{mb00},
where $T_{\rm H}$ is the temperature of the hydrogen-burning shell.

Computations of cool bottom processing may be performed by
post-processing, with the advantage that rapid exploration of the
CBP parameter space is then possible.  The mixing should move faster
than the rate of advance of the hydrogen shell, so as not to be
overtaken, and the rate of energy release that results from this
mixing should be very much less than the luminosity of the hydrogen
shell. For the stars of interest, this corresponds to $\dot{M}>
10^{-7}M_\odot$/yr and $\log (T_P/T_{\rm H}) \leq -0.1$   Moreover,
the rate of mass turnover in the cool bottom processing region
should be much lower than the rate of mass turnover in the envelope
above, corresponding to $\dot{M} \lesssim 10^{-4}M_\odot$/yr.  We
assume that CBP operates throughout the hydrogen shell-burning
intervals between thermal pulses.

Surface compositions produced by cool bottom processing result from
the interplay of three rules \cite{wbs95} \cite{nbw03}: 1) the
amount of nuclear processing increases with the depth of mixing
$(T_P)$; 2) the amount of nuclear processing in the circulation
stream decreases with $\dot{M}$ because a higher circulation rate
results in less time spent at high temperature by a fluid element;
3) the amount of processed material that ends up in the stellar
envelope increases with $\dot{M}$.

We previously explored the consequences of these rules, both
analytically and in numerical post-processing models, for the case of
the TP-AGB phase of a star with a 1.5\msun\ initial mass and initial
solar system composition \cite{nbw03}. In subsequent calculations
summarized here, we found that essentially the same consequences arise
in stars with initial masses of 2 and 3 \msun, and also at lower
metallicities. At masses beyond 3\msun, CBP is not possible, because
the hydrogen burning shell reaches temperatures of $T_{\rm H} > 10^8$
K and the bottom layers of the convective envelope are also hot enough
for proton captures to occur.  These conditions are called hot bottom
burning (HBB), and they are explored thoroughly by other authors
(cf. \cite{fclw98} \cite{lf99} \cite{lfc00}).

\subsection{Production of $^{26}$Al}

\begin{figure}
\begin{center}
\end{center}
\caption{$^{26}$Al/$^{27}$Al versus $^{12}$C/$^{13}$C for SiC grains
as compiled by S. Amari, restricted to the same AGB-derived grain
types as in Fig. \ref{nvc}, and again drawn from Refs.  \cite{ama01}
and \cite{anll01} and references therein. The range in
$^{26}$Al/$^{27}$Al is very great, from essentially no \al~ to
$^{26}$Al/$^{27}$Al $= 1.5 \times 10^{-2}$. Balloons outlining the
accessible Al-C space are shown for three cases for TDU+CBP with C/O
$\geq$ 1. The increase in the range of $^{12}$C/$^{13}$C with
decreasing $Z$ changes the boot of Italy into a loose slipper. The
insert shows the range of $^{26}$Al/$^{27}$Al produced by CBP at
steady state as a function of $T_P/T_H$. We have assumed a baseline of
$^{26}$Al/$^{27}$Al = 8 $\times 10^{-4}$ as a minimum from normal AGB
evolution without CBP (note that the total H shell production can be
up to 5-6 times higher than this value, as discussed in the text). The
cutoff in $^{26}$Al/$^{27}$Al results from assuming $\log T_P/T_{\rm
H} \leq -0.1$. The A+B grains are problematical as it is not possible
to produce the high $^{26}$Al/$^{27}$Al found in many B grains at the
low $T_P$ when $^{12}$C/$^{13}$C$\lesssim$10, while maintaining C/O
$\geq$ 1.}\label{alc}

\centerline{\epsfig{file=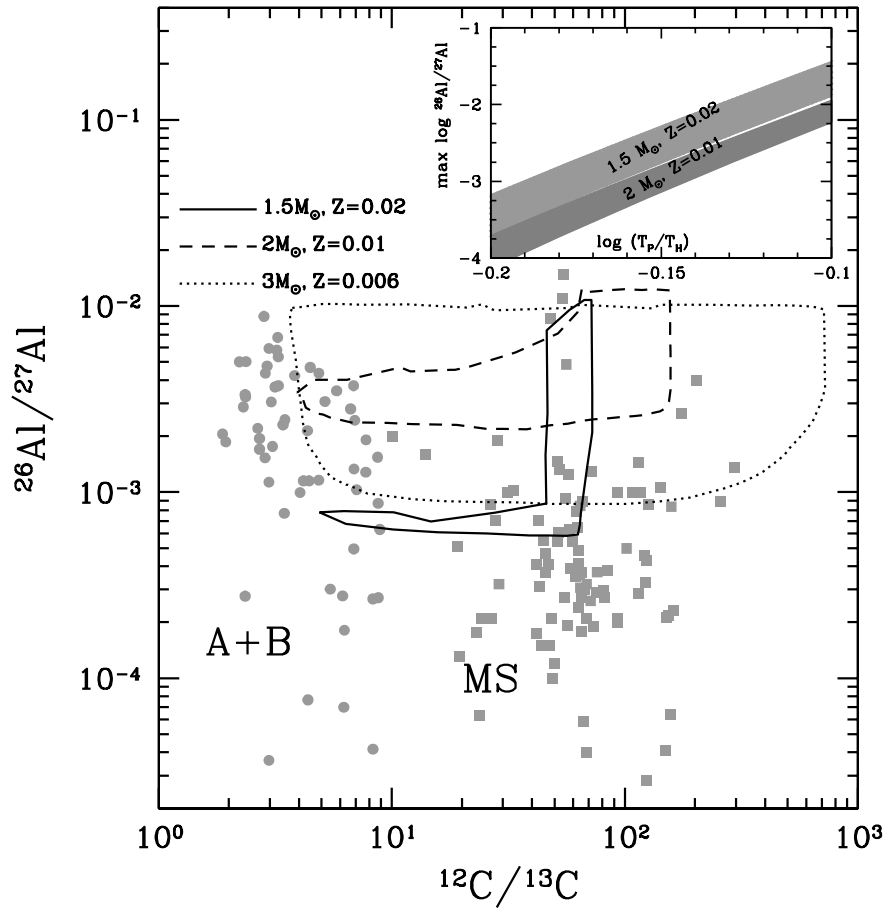,width=5in}}

\end{figure}

\begin{figure}
\begin{center}
\end{center}
\caption{The isotopic composition of Mo in a SiC grain. The values
represent the per mil deviations in the isotopic composition relative
to the solar values. The isotope $^{96}$Mo is a pure $s$ nuclide while
$^{98,100}$Mo are almost exclusively $r$ process and $^{92,94}$Mo are
pure $p$ process nuclei. The left panel shows the measured data
\cite{nic98} and the right panel shows the calculated values for an
$s$ process source \cite{lug03}. It is evident that this carbide grain
grew in the circumstellar environment of an AGB star with a very large
degree of neutron exposure. (Reprinted from Nicolussi et al.
\cite{nic98}, Copyright (1998), with permission from
Elsevier.)}\label{moly}

\centerline{\epsfig{file=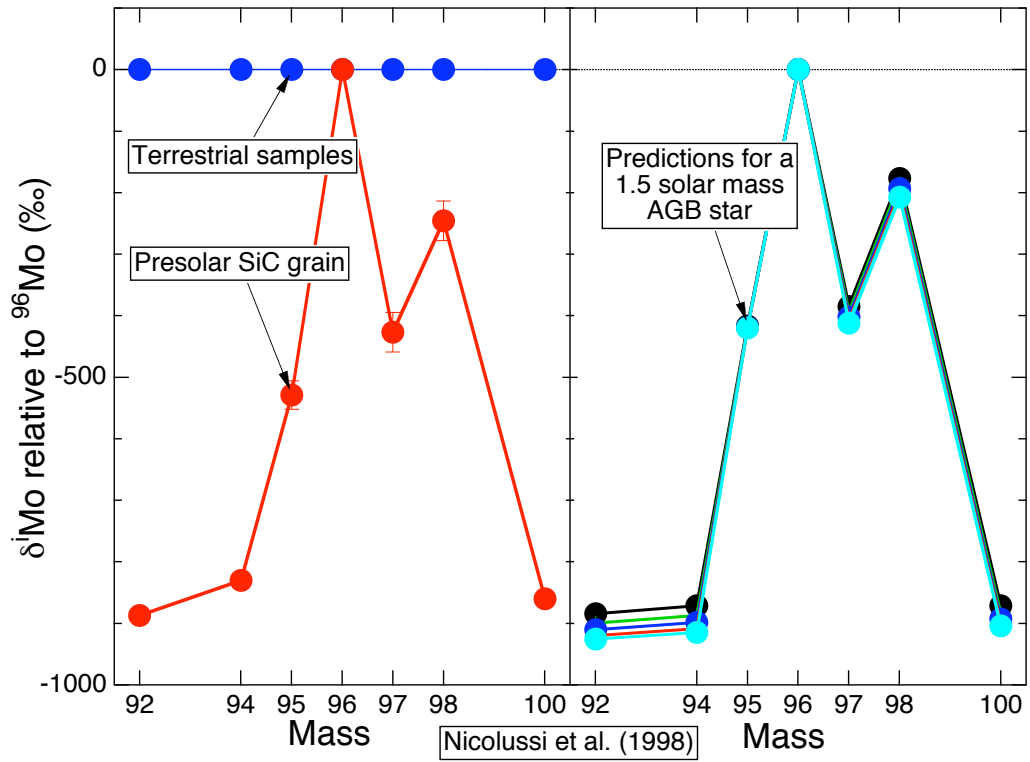,width=6in,angle=270}}

\end{figure}

\begin{figure}
\begin{center}
\end{center}
\caption{$^{18}$O/$^{16}$O versus $^{17}$O/$^{16}$O for circumstellar
oxide grains as compiled by Krestina et al.  \cite{kw02}. Most of the
results are from Nittler et al.  \cite{nag97} with major contributions
from Huss et al.  \cite{hfgw94}, Choi et al. \cite{chw98} and Krestina
et al.  \cite{kw02}. The oxide grains represent condensates formed
when C/O$ < 1$. The red dots represent data from Krestina et al.
\cite{kw02}. The solar ratios are as indicated. The trajectory
(dashes) emanating from the solar value (labeled $Z = Z_\odot = 0.02$)
represents the evolution of oxygen in AGB stars with TDU but no
CBP. Trajectories for different $Z$ values are also shown.  The
circles with black dots and associated numbers represent the stellar
masses. The region labelled HBB is where hot bottom burning occurs for
intermediate-mass stars.  The region to the left is accessible by CBP
for low-mass stars (after \cite{bsij99}). Note that there is a great
abundance of data lying far below the trajectories accessible by
normal AGB evolution, even for low $Z$. The symbols represent the
phases: corundum (circles ); hibonites (diamonds); and spinel
(squares).  Note the absence of grains in the HBB region as $^{18}$O
is destroyed and $^{17}$O is very greatly enhanced
\cite{bsij99}.}\label{oxy}

\centerline{\epsfig{file=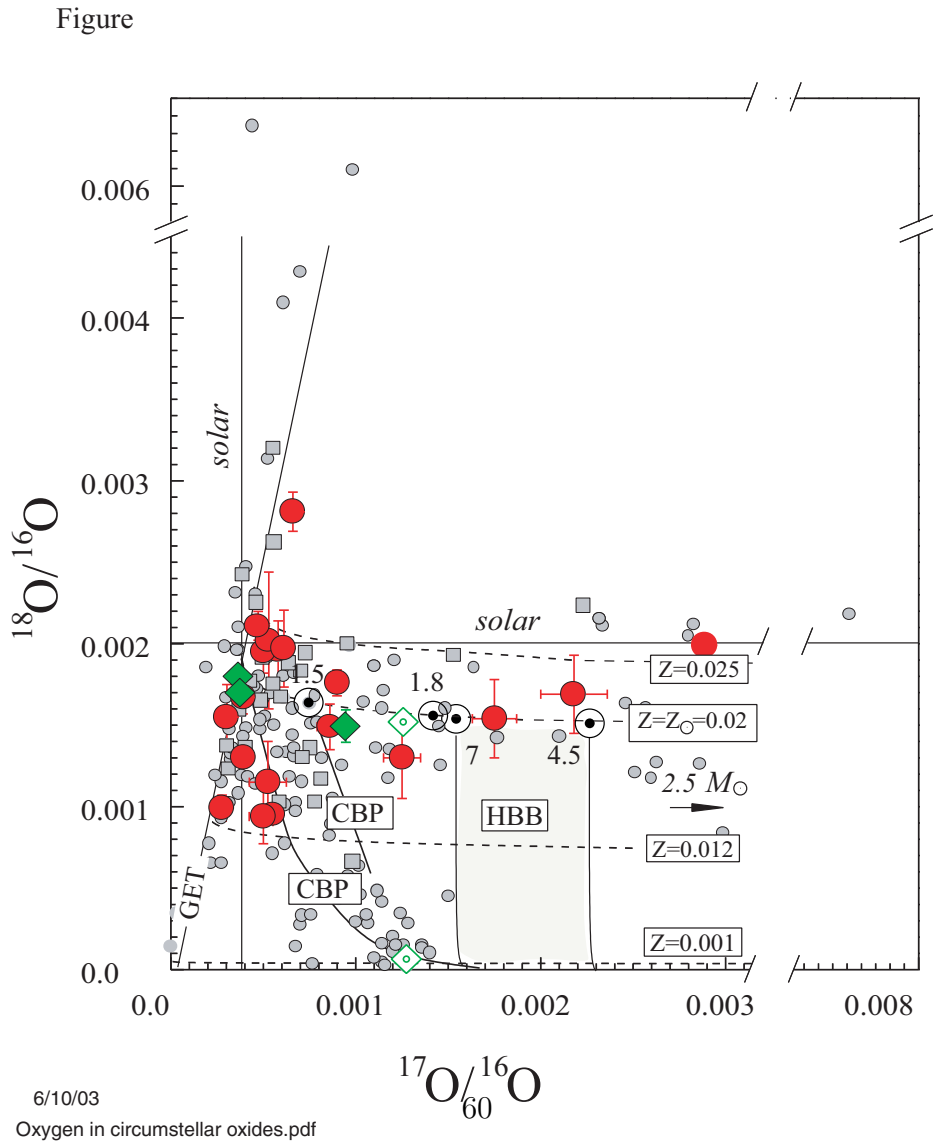,width=5.5in}}

\end{figure}

\begin{figure}
\begin{center}
\end{center}
\caption{Same diagram as previous figure but showing only those
circumstellar oxide grains with Al and Mg isotopic data available.
The value of $^{26}$Al/$^{27}$Al in each grain is color indexed as
indicated. The highest value observed is $^{26}$Al/$^{27}$Al$ = 2
\times 10^{-2}$. Note the high $^{26}$Al/$^{27}$Al near the
$^{17}$O/$^{16}$O equilibrium value with almost no $^{18}$O. Also note
grains (red and green) with high $^{17}$O/$^{16}$O and high
$^{26}$Al/$^{27}$Al but with $^{18}$O somewhat depleted but not
destroyed. The oxygen data are clear indications of AGB evolution
with CBP and the cases of high $^{26}$Al/$^{27}$Al require $\log
T_P/T_{\rm H} \sim -0.1$. Compilation after Krestina et al.
\cite{kw02}.} \label{almg}

\centerline{\epsfig{file=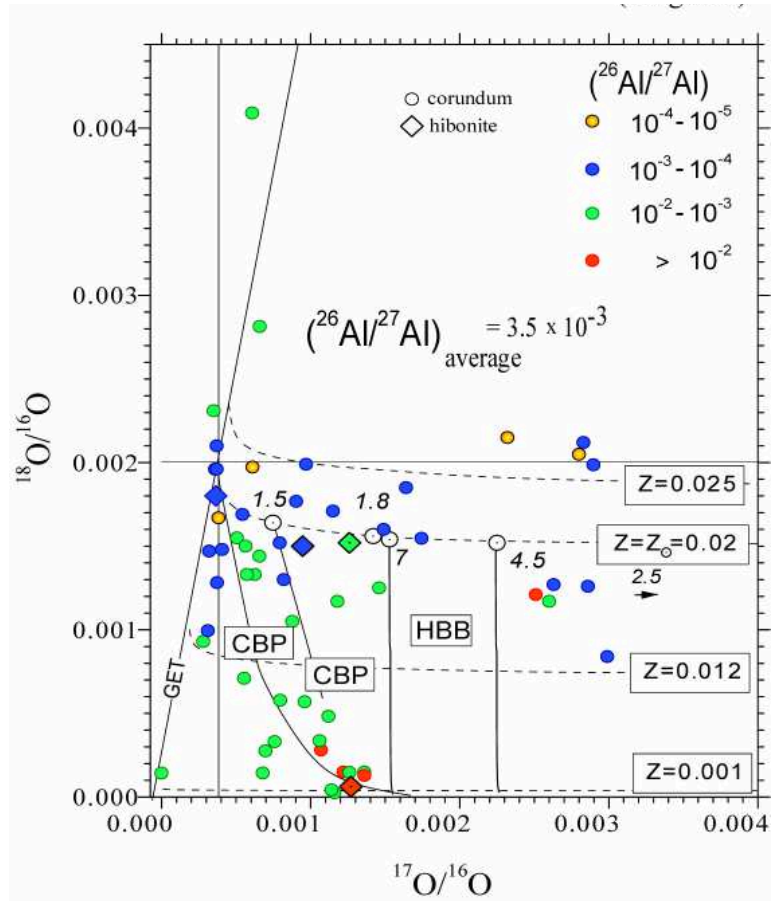,width=5in,angle=0}}

\end{figure}

The rate of $^{26}$Al enrichment of the stellar envelope does not
depend on $\dot{M}$, but depends only on $T_P$, through the rate for
the $^{25}\mathrm{Mg}(p,\gamma)^{26}\mathrm{Al}$ process.  The product
of the rate of $^{26}$Al transfer to the envelope with the TP-AGB
lifetime determines the maximum amount of $^{26}$Al that can be
made. For several different stellar models in the initial mass range
1.5--3.0 $M_\odot$, the envelope $^{26}$Al enrichment rate shows a
near-regularity when plotted against $\log(T_P/T_{\rm H})$ as shown in
the insert of Fig. \ref{alc}. This regularity holds rather generally,
but its normalization changes with time during the evolution of a
single star. It is a robust result that for a given stellar model, the
amount of \al~ that can be produced during the TP-AGB phase varies
from the level $^{26}$Al/$^{27}$Al$\simeq $ a few times $10^{-3}$
provided by third dredge-up to a few times $10^{-2}$.  Thus CBP can
adequately explain the population of $^{26}$Al/$^{27}$Al found in both
carbide and oxide grains (see Figs. \ref{alc} and \ref{almg}). Note
that without CBP, a large fraction of the \al~ that is produced is
subsequently destroyed in the He shell (cf. \cite{wbgr94}).  The
lowest values in the grains must then result either from stars without
CBP and with less efficient production in the H-burning shell, or from
more effective destruction of \al~ in the He shell (without CBP) than
predicted in the standard models.

\subsection{Destruction of $^{18}$O}

The case of $^{18}$O brings out the third rule listed above: the
amount of processed material brought to the envelope increases with
$\dot{M}$.  Even at $\log T_P/T_{\rm H} \simeq -0.2$, essentially all
of the $^{18}$O in the CBP stream is destroyed by the reaction
$^{18}\mathrm{O}(p,\alpha)^{15}\mathrm{N}$.  Thus the material
circulates down and returns to the envelope, and the $^{18}$O
remaining in the envelope is diluted with $^{18}$O-free material. The
envelope $^{18}$O/$^{16}$O declines exponentially with a time constant
$M_E/\dot{M}$, where $M_E$ is the mass of the stellar envelope.  For
the maximum \mdot\ considered ($10^{-4}$\msun/yr) and an envelope mass
of 0.7 \msun, this comes to $M_E/\dot{M}= 1.4\times 10^4$ yr,
considerably less than the total time spent in the TP-AGB
phase. Thorough depletion of $^{18}$O is therefore a sign of CBP, and
it sets in at lower $T_P$ than needed for significant $^{26}$Al
production.

It was suggested by Nittler at al. \cite{nag97} and Choi, Wasserburg,
\& Huss \cite{cwh99} that $^{26}$Al/$^{27}$Al and $^{18}$O/$^{16}$O
would be correlated. However, each of these ratios is sensitive to a
different CBP parameter.  With CBP, the $^{26}$Al production depends
almost entirely on $T_P$. The $^{18}$O depletion depends only on
\mdot~ with a relatively low temperature threshold $(\log T_P/T_{\rm
H} \geq -0.2)$. See Fig. 6 of \cite{nbw03}. It is thus possible to
obtain destruction of $^{18}$O with or without significant $^{26}$Al
production and substantial \al~ production with or without $^{18}$O
destruction; however, conditions on oxide versus carbide production
discussed below place further constraints on which compositions are
possible in a given type of grain.

\subsection{Equilibration of $^{17}$O/$^{16}$O}

Above a relatively low threshold in $T_P$, $^{17}$O/$^{16}$O in the
circulating material reaches an equilibrium by balancing rates for
reactions $^{16}\mathrm{O}(p,\gamma)^{17}\mathrm{F}(\beta^+
\nu)^{17}\mathrm{O}$ and $^{17}\mathrm{O}(p,\alpha)^{14}\mathrm{N}$.
Given the presently recommended reaction rates \cite{ang99}, the
equilibrium ratio has a broad minimum in exactly the temperature
range of interest for AGB stars, so that for all cases CBP should
produce $^{17}$O/$^{16}$O $\cong 0.0011$.  (This value may be
slightly changed because of a recent revision of the rates for
proton capture on $^{17}$O \cite{fox04}.)  Thus, when plotted on a
three-isotope plot for oxygen isotopes, abundances in the stellar
envelope after some CBP will lie on a mixing line connecting the
composition at the start of the AGB phase with a composition that
has the equilibrium value of $^{17}$O/$^{16}$O.  One then expects to
find grains heavily depleted in $^{18}$O that simultaneously have
$^{17}$O/$^{16}$O $= 0.0011\pm 0.0003$. Oxide grains fitting this
description have been found in the data of several groups of authors
(see Fig.  \ref{oxy}).

\subsection{Carbon and Nitrogen}

The C and N isotopes are also affected by CBP.  However, this is
more complicated because third dredge-up contributes significant
amounts of carbon to the envelope. The effects of CBP on
$^{12}$C/\ct\ and $^{14}$N/$^{15}$N mostly consist of processing the
large amounts of $^{12}$C added to the envelope by third dredge-up.
In the case of $^{15}$N the effect is simple: it has a very
effective destruction mechanism,
$^{15}\mathrm{N}(p,\alpha)^{12}\mathrm{C}$, that makes the effects
on $^{15}$N exactly analogous to those on $^{18}$O. The effects on
the carbon isotopes and on $^{14}$N depend in a somewhat complicated
way on $T_P$ and \mdot, following the three rules as outlined above.
In brief, CBP converts $^{12}$C to $^{13}$C, and then to $^{14}$N.

In the absence of CBP, the envelope abundance of $^{12}$C
increases with time, in a discrete jump following each thermal
pulse. Eventually, if enough $^{12}$C is brought to the stellar
envelope, the surface composition attains C/O$>1$ and the
consequent changes in chemistry follow.  At moderate $T_P$, CBP
lowers the ratio $^{12}$C/$^{13}$C in the stellar envelope by
converting some of the newly-formed $^{12}$C into $^{13}$C.  For
some values of \mdot, this can bring the envelope abundance ratio
down to $^{12}$C/$^{13}$C $\simeq 5$ while leaving
$(^{13}\mathrm{C}+^{12}\mathrm{C})/\mathrm{O}$ relatively
unchanged. If, however, $T_P$ is a little higher (or \mdot\ a
little lower), then the $^{13}$C is rapidly converted into
$^{14}$N.  At low values of \mdot, the ratio $^{12}$C/$^{13}$C
approaches $\sim 4$ from the CN-cycle equilibrium.  However, the
material returning to the envelope when $T_P$ is high is depleted
in carbon of both isotopes so that its main effect is to reduce
C/O without necessarily having a large effect on
$^{12}$C/$^{13}$C.  For high enough \mdot\ and $T_P$, a large
fraction of the $^{12}$C brought up by third dredge-up can be
converted into $^{14}$N, delaying or preventing the formation of a
carbon star. The results of CBP calculations show a well-defined
relationship between $^{12}$C/$^{13}$C and C/O as the CBP
parameters vary. Since $^{26}$Al is effectively a thermometer and
C/O a measure of $\dot{M}$ and $T_P$, the condition of a star is
rather well determined by \al~, $^{12}$C/$^{13}$C, and C/O.

Stars whose evolution into a carbon star has been prevented will
have large N/O abundance ratios, because of the conversion of
newly-produced carbon into nitrogen.  Such nitrogen-rich final
abundances (observed, perhaps, in planetary nebulae) can in
principle constitute evidence for CBP.  However, this is also a
fairly generic result of hot bottom burning, so that information
about the stellar mass is needed to make a clear assignment to CBP.
An apparent case of a planetary nebula with high enough N/O and low
enough progenitor mass for CBP to be important has been reported
\cite{pwz00}.

Without CBP, the production and dredge-up of C eventually make an
envelope with $\mathrm{C/O}>1$.  This should result in carbide grains
forming, while extreme destruction of C by CBP will give
$\mathrm{C/O}<1$ and should result in oxide grains forming \cite{sw95}
\cite{flod}.  We expect that if CBP were active in the stars that
formed presolar grains, abundances in the carbide grains should
reflect the parts of the CBP parameter space consistent with C/O$>1$,
while those in oxide grains should reflect CBP parameters that
preserve C/O $<1$.  This is broadly true, though there are some
discrepancies (see Fig. \ref{oxy}). In particular, it is difficult to
account for simultaneous high \al/$^{27}$Al and high $^{18}$O/$^{16}$O
in an oxide grain, because high \al/$^{27}$Al indicates CBP operating
for a long time, while high $^{18}$O/$^{16}$O implies low \mdot~ and
thus C/O $>1$ at late times. Thus the red point and some of the green
points in Figure 5 with $^{18}$O/$^{16}$O $>$ 0.001 are examples of
discrepancies (cf. Figs. 7b and 10 of \cite{nbw03}).

The nitrogen isotopic ratios observed in the circumstellar grains are
difficult to understand in terms of AGB evolution, even without
CBP. They should have high values of $^{14}$N/$^{15}$N because first
dredge-up brings a great deal of $^{15}$N-depleted and
$^{14}$N-enriched material to the stellar envelope.  Previous
calculations starting with solar-system initial composition have found
that the increase in $^{14}$N/$^{15}$N in the envelope at first
dredge-up is roughly a factor of five.  The mainstream grains show
$^{14}$N/$^{15}$N extending all the way down to the solar value (see
Fig. \ref{nvc}). This would require that the initial $^{14}$N/$^{15}$N
at stellar birth was roughly 50, while the lowest values observed
astronomically are around 100 (in the Large Magellanic Cloud)
\cite{chlm99}.  The astronomical data concerning the time evolution of
$^{14}$N/$^{15}$N in the Galaxy (summarized in \cite{wr94}) neither
support nor rule out such low $^{14}$N/$^{15}$N in stars that produced
the SiC grains.  The Galactic disc today shows generally higher
$^{14}$N/$^{15}$N than solar, with a wide scatter of 200--600, and the
Galactic center has a lower limit of 600 \cite{wr94}.  These facts
suggest that $^{14}$N/$^{15}$N increases with astration, whereas the
gradient of $^{14}$N/$^{15}$N with distance from the Galactic center
within the disc suggests the opposite trend.  We note that the recent
factor-of-two reduction in the
$^{14}\mathrm{N}(p,\gamma)^{15}\mathrm{O}$ rate based on measurements
at LUNA \cite{form04} and TUNL \cite{run02} \cite{run05} further
increases the amount of $^{14}$N in the dredged-up material and
exacerbates the $^{14}$N/$^{15}$N problem.

All of these considerations bear on the case without CBP.  The
addition of CBP during the AGB phase (and earlier, during the RGB
phase) further increases $^{14}$N/$^{15}$N so that almost none of
the observed grains are accessible with CBP if the initial
$^{14}$N/$^{15}$N $= (^{14}\mathrm{N}/^{15}\mathrm{N})_\odot$.
This was discussed by Huss et al. \cite{hhw97} who suggested that
the $^{18}\mathrm{O}(p,\alpha)^{15}\mathrm{N}$ cross section might
be grossly wrong.  This does not appear to be likely.  The most
direct solution to this severe problem would be that
$^{14}$N/$^{15}$N in the ISM, at the time the stars parent of the
circumstellar grains formed, was at least a factor of 5 - 10 lower
than in the sun, so that the assumed initial isotopic composition
of N in all models is incorrect. More direct observations of
$^{14}$N/$^{15}$N are clearly required in the proper metallicity
domain. In general, the assumption of solar abundances for the
precursor stellar sources cannot be valid.

\section{Intermediate Mass Stars (IMS)}\label{IMS}

In AGB stars the formal H-burning shell (where maximum energy
generation from hydrogen burning occurs) is always a radiative
zone. However, for IMS models, convection from the envelope extends
into the top of the broad layer where physical conditions are suitable
for proton captures. The temperature of the H shell is very high in
these cases (above $10^8$ K) and the bottom of the convective envelope
can reach temperatures of several $10^7$ K.  Here burning occurs in
fully convective conditions, so that the consumed H-fuel is
continuously replenished and the efficiency of nucleosynthesis becomes
quite high. It occurs primarily through the CNO cycle, but also
through reactions involved in the Ne-Na and Mg-Al cycles, so that
these layers are a suitable site for production of several nuclei,
from $^{7}$Li, \ct, $^{14}$N, up to $^{23}$Na. Further, the isotopic
mix of Al and Mg can be heavily affected, and \al/$^{27}$Al ratios
close to unity may result \cite{kl03b}. This is contrary to what
occurs in LMS experiencing CBP where there is no fuel replenishment or
extra H burning.

This process is called Hot Bottom Burning (HBB), following a
prediction by Renzini and Voli \cite{rv81} made well before any
stellar model could confirm their hypothesis. The exact mass at which
HBB is found in evolutionary codes is dependent on the modeling. There
is a consensus that stars of $Z = Z_\odot$ experience HBB above $\sim
(5-5.5)~ M_\odot$, and that this mass limit decreases with
metallicity, so that for population II stars, the phenomenon may be
found at masses as low as 3 to 4 \ms.  Concerning the nucleosynthesis
processes of relevance here for short-lived nuclei, the results by
Karakas and Lattanzio \cite{kl03b} and Karakas \cite{kar03} confirm
the expectation that HBB produces \al~ at higher efficiency than
CBP. Due to the high $^{12}$C consumption in the CNO processing,
however, HBB inhibits the formation of a C star (except in very
special cases), so that these high \al/$^{27}$Al ratios must be
associated only with O-rich compositions. A circumstellar grain of
MgAl$_2$O$_4$ with shifted Mg isotope composition that is consistent
with HBB has been found \cite{nittler}.

\section{Estimates of the ISM inventory of short-lived nuclei}

In the simplest possible model of galactic nucleosynthesis for a
system evolving for a time duration $T$, the inventory of
radioactive nuclide $R$ relative to a stable nuclide $I$ uniformly
produced in the same process is:
\begin{equation}
[N^R(T)/N^I(T)]_{UP} \simeq \frac{P^R p(T)\bar\tau_R}{P^I<p>T}
\label{nrs}
\end{equation}
Here UP refers to uniform production (in the galaxy), $P^I <p>$ is the
average stellar production rate of $I$ over time $T$ and $P^R p(T)$ is
the production rate of $R$ near the time when production ceased
\cite{sw70}.  If we assume that $p(T)$ is constant, then we may
calculate relative abundances in such an ISM just using the values of
$\bar\tau_R$, the duration ($T \sim 10^{10}$ yr) and estimates of
$P^R/P^I$. All UP calculations were done using the explicit equation:
\begin{equation}
N^R(T) = P^R\bar\tau_R(1 - e^{-T/\bar\tau_R}) \label{Pi}
\label{nit}
\end{equation}
This should serve as a guide. If the nuclei selected (for each
pair $R,I$) are typically produced at the same site in a source,
we may then obtain rather well-defined estimates of the resulting
abundance ratios depending on the assumed nuclear physics. If the
ISM is isolated from further production of nuclei at some time
prior to onset of formation of the solar nebula, then the time
interval between injection and the formation $(\Delta_1)$ results
in further decay by the factor $e^{-\Delta_1/\bar\tau_R}$.

This approach gives some insights; however, it is a bit deceiving as
the events are discrete ones, not continuous. Let us consider the
events of type ``a'', producing $R_a$ and $I_a$ nuclei, with a fixed
recurrence interval of $\delta_a$, and where the yields of nuclei in a
single stellar event are $\tilde{p}^{R_a}$ and $\tilde{p}^{I_a}$. The
appropriate scaling factor is the ratio of the galactic time scale to
the recurrence time, $T/\delta_a$, which is the number of events
(typically $10^2 - 10^3$).

Then we obtain:
\begin{equation}
N^{R_a}/N^{I_a} = \tilde{p}^{R_a}/\tilde{p}^{I_a}
\left(\frac{\theta\delta_a}{T} + \frac{\bar\tau_R}{T}\right)
\end{equation}
for $\delta_a/\bar\tau_R << 1$, and where $\theta = 0$ or 1
depending on whether the sampling time is before or just after the
last contributing event. In this case the first term is negligible
whether $\theta = 0$ or 1.

For many nuclei with $\bar\tau_R \leq 10^7$ yr,
$\delta_a/\bar\tau_R \gtrsim 3$, and hence we obtain the following
expression:
\begin{equation}
N^{R_a}/N^{I_a} = \tilde{p}^{R_a}/\tilde{p}^{I_a}
\frac{\delta_a}{T}(\theta + e^{-\delta_a/\bar\tau_R})
\end{equation}
If $\theta = 1$, then the second term with the exponential becomes
negligible. It is thus the granularity of the production that governs
the inventory of the shorter-lived nuclei \cite{wp82}.  For many of
these short-lived nuclei, the inventory is controlled by the last
event contributing to the ISM.  In this case, if the ISM is sampled at
a time $\Delta_1$ after the last event:
\begin{equation}
N^{R_a}/N^{I_a} \sim \tilde{p}^{R_a}/\tilde{p}^{I_a}
\frac{\delta_a}{T} e^{-\Delta_1/\bar\tau_R}
\end{equation}
Thus, these nuclei must in some way be considered  as resulting from
late injection.

For the abundances in the ISM using the UP model and eqn.  \ref{nit}
we used estimates of the production rates of both the radioactive
$(R)$ and stable $(I)$ nuclei.  All pertinent ratios for UP are given
in Table 1 for the cases $\Delta_1 = 0$, 5 Myr, 10 Myr, and 70 Myr,
where $\Delta_1$ is the time after the termination of uniform
production and $e^{-\Delta_1/\bar\tau}$ corrects for the subsequent
decay of isotope $R$ without any additional injection into the ISM.
For the actinides we counted progenitors following \cite{b2fh} and
\cite{fh60}, and obtained
$(^{244}\mathrm{Pu}/^{232}\mathrm{Th})^{UP}_{ISM} = 6 \times 10^{-3}$
and $^{247}$Cm/$^{232}$Th $= 1.1 \times 10^{-3}$ (see \cite{wbg96} and
section 11 in this paper). For Ca, Pd, I and Cs we adopted the choices
discussed in \cite{bgw99}. If we consider $^{26}$Al, then
$(^{26}\mathrm{Al}/^{27}\mathrm{Al})^{UP}_{ISM} \sim
p^{^{26}\mathrm{Al}}/p^{^{27}\mathrm{Al}} \times 10^{-4}$. As
$^{26}$Al is produced at a very low level
$(p^{^{26}\mathrm{Al}}/p^{^{27}\mathrm{Al}} \sim 10^{-3} - 10^{-2})$
in SNe II and other sources, the ISM value will be extremely low. The
inventory of \al~ in the Galaxy today (\al~$\sim 2 - 3 M_\odot$) is
obtained from $\gamma$-ray measurements (see a review in
\cite{dchk04}).  This gives an observed ratio of
$(^{26}\mathrm{Al}/^{27}\mathrm{Al})_{\mathrm{ISM}} \approx 6 \times
10^{-6}$; we can notice that the continuous production computed using
present SN models cannot match this, but is low by at least an order
of magnitude. For the cases of $^{60}$Fe and $^{53}$Mn, we may use the
calculations from SNe II. Yields for these two nuclei have been
calculated by both Woosley \& Weaver \cite{ww95} and Rauscher et
al. \cite{rau02}, and are in rather good agreement, with about the
same values being obtained by these workers for both 15 $M_\odot$ and
25 $M_\odot$ stars. For $^{135}$Cs, $^{129}$I, and $^{107}$Pd, the
$p^R/p^I$ ratios are close to unity \cite{bgw99} \cite{wbg96}. For
$^{41}$Ca we used a nominal value of
$p^{^{41}\mathrm{Ca}}/p^{^{40}\mathrm{Ca}} \sim 10^{-3}$. For
$^{36}$Cl we used $p^{^{36}\mathrm{Cl}}/p^{^{35}\mathrm{Cl}} \approx
10^{-2}$ from \cite{rau02}. Note that each nuclide in an isotopic pair
may represent different astrophysical sites and processes. In many
cases stellar sources are not well established.

We list nominal values for the observed abundances in the ESS in
the fifth column of Table 1. These are essentially the same as
given in \cite{bgw99}. Comparison of these determinations with the
calculated UP values may be used as a guide as to which nuclei
might be available in the general ISM and which require special
sources. We will discuss critical aspects of the measured values
in meteorites in a later section.

If we now compare the values for UP with the ESS measurements, it
is evident that, between the actinides and \mn,  the estimated ISM
inventory is approximately sufficient to provide the observed
abundances for many short-lived  nuclei. For \al, \ca, and \be,
the model of uniform galactic nucleosynthesis is insufficient,
particularly if the time scale prior to solar system formation is
$\sim 10^6$ years. It follows that, if (\al/$^{27}$Al)$_{ESS}$ =
5$\times$10$^{-5}$, then the material of the solar system did not
form from the average ISM, but must have sampled a hot spot.

As \be~ can not be produced by stellar nucleosynthesis, it must
come from some  irradiation process. It is therefore these three
nuclei that require some very late addition.

Evidence for $^{129}$I is found in a wide variety of ESS materials and
even in terrestrial gas samples. There is a range of
$^{129}$I/$^{127}$I found in meteorites (cf. review by Ott
\cite{ott00} and report by Whitby et al. \cite{whit0}). We immediately
see that $^{129}$I$_{UP}$ is grossly overproduced (by a factor of 23
to 50) as compared to the observed $^{129}$I/$^{127}$I
ratios. $^{129}$I is a pure $r$-process product \cite{agwc93}
\cite{ctc93} and cannot be produced in the $s$ process \cite{kbw89}
\cite{wbgr94} \cite{wbg96}. This has long been recognized and is the
basis of the argument that the last ``$r$ process'' source enriching
the local ISM was $\Delta_1(^{129}$I) $\sim (0.7\ \mathrm{to}\ 1)
\times 10^8$ yr before the solar system formed (see Table
\ref{mlt}). The more recent discovery of \hf~ and its abundance in the
ESS (cf. \cite{hj96} \cite{hjsb96} \cite{lh95} \cite{lh96}) and
important new results by Yin et al. \cite{yin02} and Kleine et
al. \cite{kle02} led to a further discrepancy with $^{129}$I. The \hf~
nuclide cannot be significantly produced in AGB stars and must be
completely dominated by an $r$-process source \cite{wbgr94}
\cite{wbg96}.  There is no time interval $\Delta_1$ that will allow
the two $r$-process nuclei, \hf~ and $^{129}$I, to be in
agreement. This then required that the $r$-process was not a single
process producing both $^{129}$I and \hf~ but now had to be considered
as two or more processes, one of which produces low mass $r$-nuclei
(at or below Ba) while the other source produces heavy $r$-nuclei (Ba
and above) as proposed by Wasserburg, Busso \& Gallino
\cite{wbg96}. It then developed that the common and long-time practice
of attributing the $r$-process site to any standard SNe II has major
flaws. Observations on low metallicity halo stars show clearly that
heavy $r$-process nuclei are not associated with any production of all
the nuclei between oxygen and germanium.  This includes the ``iron''
peak, which was usually assumed to be co-produced with the
$r$-nuclei. This result was implied by the early observations by
McWilliam et al. \cite{mpss95} and \cite{mcw98}.  Much more extensive
work on low-metallicity halo stars (cf.  \cite{scl03} \cite{hill02}
\cite{sne00}) showed that Fe-group and other nuclei with $A<130$ were
fully decoupled from heavy $r$-process nucleosynthesis (including U
and Th) \cite{qw02} \cite{qw03}. This then requires that those SNe II
that may be the site of a heavy $r$-process must have masses $8
M_{\odot}\leq M \leq 11 M_{\odot}$ (with small envelopes).  Another
possibility is that the heavy $r$-nuclei might be the result of {\it
AIC} -- accretion-induced core collapse -- (to make a neutron star
with concurrent production of large neutrino fluxes), or even possibly
be related to explosive nucleosynthesis on white dwarfs by some other
mechanism (see \cite{qw03} and references therein). The source of some
of the ``light'' r-process nuclei is also a matter of current study
\cite{mey02}. Basic issues and aspects of the ``$r$-process'' are
discussed by \cite{tbm01}\cite{astq04}\cite{hond04}\cite{kral}. For
the contributions of Type Ia SNe, see \cite{nitr}. The $r$-processes,
and their sites and sources, are thus still not fully understood or
identified (see review by Qian \cite{qianrev}). We further wish to
emphasize the point that the ratio Th/Eu is not constant in the
different $r$-processes, but there is a distinct change in yields
between Eu and Th.  This then invalidates the often-used Th/Eu
chronometer, as pointed out by Qian \cite{qian02}. Those inferences
about the yield pattern involving very heavy $r$-nuclei and those of
intermediate mass are confirmed in this work (see Section
\ref{majprob}).

Although \pd~ is found to be abundant in the UP model for $\Delta_1 =
5 \times 10^6$ yr, it is grossly underabundant if the time interval
$\Delta_1^{^{107}{\rm Pd}} \sim \Delta_1^{^{129}{\rm I}} \sim 7 \times
10^7$ yr as inferred from $^{129}$I. If \pd~ is from the $r$-process,
it is most plausibly produced with the lower mass $r$-process nuclei
along with $^{129}$I, possibly also with the Fe group. \pd~ should
then be absent as a residue of long-term galactic nucleosynthesis. It
was a version of this conflict that led to the joint effort of the
Torino and Caltech groups to seek a solution considering a late AGB
injection, since \pd~ is readily produced in these stars. This led to
connections and predictions for the much shorter-lived nuclei produced
by AGB stars. For these AGB contributions there will also be \fe~ and
other nuclides produced that will add on to the possible ISM
inventory. From the above arguments it follows that radioactive iron
group nuclei (e.g. \mn~ and \fe) are not to be associated with the
heavy $r$-process nuclei. If they are associated with the light
$r$-nuclei then they should also be extinct if $\Delta_1 \sim 7 \times
10^7$ yr. This leaves a problem with \mn~ as it cannot be produced in
AGB stars (see Section \ref{SNeII}).

The lifetime of $^{244}$Pu is so long that small changes in the
history $(\Delta_1)$ play no role if $\Delta_1 << 10^8$ yr. We first
note that using the standard estimated yield for the actinides, the
calculated value of $(^{244}\mathrm{Pu}/^{232}\mathrm{Th})^{UP}_{ISM}$
is within a factor of two greater than that found in the ESS. This is
fully consistent with the transactinides being part of the general ISM
inventory from which the solar system formed.  For $^{247}$Cm there is
only an upper bound on its abundance. The experimental approach used
is to find variations in $^{235}$U/$^{238}$U caused by variations in
the Cm/U ratios in different phases in meteorites and different
meteorite samples. The strict interpretation of the results depends on
estimates of the chemical fractionation of Cm from U. There is, of
course, some basic uncertainty in the estimated nucleosynthetic yield
of these nuclei. The study by Chen \& Wasserburg \cite{cw81a}
\cite{cw81b} gave an upper limit of $(^{247}$Cm/$^{235}$U)$_{ESS} < 2
\times 10^{-3}$. A new report by Stirling et al. using more precise
techniques on bulk meteorites but with very limited U/Nd elemental
fractionation for chondrites, gives a bound of $1.0 \times 10^{-4}$
\cite{stir05}. The most direct interpretation of the results is that
$^{247}$Cm is in low or very low abundance compared to the UP
model. As mentioned, the measured abundance of $^{244}$Pu is low by a
factor of two compared to the UP value. This would then appear to
require a significant time interval between the termination of heavy
$r$-nucleosynthesis and formation of the solar nebula if the counting
of precursors is a reasonably reliable estimate of the relative yields
for actinides. We will return to this matter in the discussion.

For the actinides $^{238}$U, $^{235}$U, $^{232}$Th, and $^{244}$Pu,
the observed abundances are certainly within a factor of two of those
anticipated from the UP model (cf. \cite{wbg96}).  Further, $^{146}$Sm
is in accord with the observations, although its stellar origin still
remains somewhat uncertain. As both $^{146}$Sm and $^{144}$Sm are
$p$-process nuclei, the estimated production ratio should not be in
gross error. The $p$ nuclide $^{92}$Nb $(\bar\tau \sim 52$ Myr) was
discovered in the ESS by Harper \cite{harp96}, with
$^{92}$Nb/$^{93}$Nb $\sim 0.7 \times 10^{-5}$. We do not have any
basis for inferring precise yields for this nuclide, as estimates of
$^{92}$Nb/$^{93}$Nb are widely diverse, and further the index isotope,
$^{93}$Nb, is almost pure $s$-process.

The $^{182}$Hf abundance has been revised \cite{yin02}
\cite{kle02} to the value
$(^{182}\mathrm{Hf}/^{180}\mathrm{Hf})_{ESS} = 1.0 \times
10^{-4}$. These workers have established the abundance of
$^{182}$Hf in the ESS and the initial reference ratio for bulk
solar W isotopes. The UP value is a factor of 4 greater than the
ESS value and also appears to require some time interval for decay
$(\Delta_1 \sim 18$ Myr). In general, all of the heavy $r$ nuclei
appear roughly consistent with long term galactic nucleosynthesis
with no requirement of significant very late addition from a
special source.  The general inferences given above are in accord
with those in \cite{wbg96}. Some significant time interval
$\Delta_1$ appears to be required in order to obtain a plausible
match to the observations of the shorter-lived heavy $r$-nuclei.
This is an important matter as it generates further conflicts (see
Section \ref{majprob}).

\section{Injection from a single stellar source}\label{inject}

We now consider a model of injection of freshly synthesized
nuclear material from a single stellar source into the protosolar
molecular cloud. It is assumed that the pre-existing material and
the injected material are reasonably well-mixed. Then the
equations governing the mixture for short-lived nuclei (taken
originally to be absent) are as follows.

Let $N^R_{SC}$ be the number of short-lived radioactive nuclei $R$ in
the solar cloud (SC); $N^I_{SC}$ be the number of stable $I$ nuclei
(of the same element as $R$) in the protosolar cloud. Note that in
some cases we use another radioactive nuclide as an index isotope for
the ratio (e.g. $^{235}$U/$^{238}$U), and this changes the formalism
because decay must be considered. $N^{R}_{ENV}$, $N^{I}_{ENV}$ are the
numbers of nuclei ($R$ and $I$) in matter in the stellar envelope that
are injected into the solar cloud. Then $N^I_{SC}$ is the sum of the
previous ISM inventory ($N^I_0$) and of the late stellar addition
($N^I_{ENV}$), while for short-lived nuclei $N^R_{SC}$ contains only
the stellar contribution ($N^R_{ENV}$). One has:
\begin{eqnarray}
\frac{N^R_{SC}}{N^I_{SC}} \equiv \alpha^{R,I}_0& = &
\frac{N^R_{ENV}}{N^I_0 + N^I_{ENV}} \nonumber\\
  & = & \frac{\left( N^R/N^I\right)_{ENV}N^{I}_{ENV}}{N^I_0 +
  N^I_{ENV}} \nonumber\\
  & = & \frac{\left( N^R/N^I\right)_{ENV}q^I_{ENV}M_{ENV}}{q^I_0 M_0
  + q^I_{ENV}M_{ENV}}
  \end{eqnarray}
  We define $q^I_{ENV}$ and $q^I_0$ to be the number of stable $I$
  nuclei per gm of matter in the envelope and in the polluted ISM
  cloud. Note that the $q$ values depend on metallicity.  For the case
  where the ejected stellar envelope is not well-mixed, the problem is
  more complex: in that case the terms
  $\left(N^R/N^I\right)_{ENV}q^I_{ENV}$ must be summed over all the
  contributing subunits of the parent star. This can be especially
  important for a SNII. Thus,
  \begin{equation}
  \frac{\left( N^R/N^I\right)_{ENV}}{\alpha^{R,I}_0}
  \frac{q^I_{ENV}}{q^I_0} \approx
  \frac{M_0}{M_{ENV}} \equiv 1/f_0
  \label{ENV}
  \end{equation}
This is the condition obtained at the time of injection and instant
mixing with the local ISM. It relates the ratio of the radioactive
nuclide $R$ to the stable nuclide $I$ in the envelope to that in the
unpolluted cloud. The term $M_0/M_{ENV}$ is then the ratio of the mass
of the cloud to the mass of injected stellar envelope. Note that the
term $f_0$ is the dilution factor and will be used later. This follows
the treatment in \cite{wbgr94} with some modification as they used the
ratios in the He shell, not in the envelope.

For a self-consistent solution for a variety of isotopic pairs
($R,I$), it follows that the right hand side of equation (\ref{ENV})
should be the same for all $\left(N^R/N^I\right)$.  This is a rather
strict constraint and connects the relative abundances in the
hypothesized stellar envelope to the initial values in the polluted
protosolar cloud. Note that all the terms on the left hand side are
only dependent on the model of stellar nucleosynthesis. A model which
must assume a separate stellar source for each radioactive isotope
``$R$'' is certainly not attractive. Such proposals are commonly found
in the literature.  The approach used here will be to search for a
self-consistent solution for several radioactive nuclei. However, it
will become evident that a single source can not provide the observed
results and that a blend is required.

For a model to be reasonable, the dilution factor $f_0$ must be the
same for all species and the times $\Delta_1^R$ for each radioactive
nucleus ``$R$'' should be self-consistent (no negative values), and
physically compatible with the values of $\alpha^{R,I}_{obs}$. These
rules apply to any model of injection by a single source with
subsequent mixing. The degree of homogenization need not be complete,
and disagreements between the model and observed abundances by factors
of say $\sim 2$ may be acceptable considering the uncertainties.
Factors as large as 10 or more are not compatible with this approach
nor with the observations.

\section{Meteoritic observations on short-lived nuclei}

Samples of meteorites are the principal source for determining the
presence and the abundances of short-lived nuclei in the early solar
system. Meteorites represent objects formed early in solar system
history when low mass objects (protoplanets) accumulated from
condensed material with some volatile elements. They may be aggregates
of material processed chemically and thermally within the solar system
and mixed together with a small amount of residual presolar grains
that survived the processing. Material aggregated from a heated
portion of the solar nebula would thus contain such a mixture. A small
planet aggregated from original presolar grains and which was then
heated and partly melted would also produce highly heterogeneous
materials in different zones depending on the wide range of degrees of
chemical processing and preservation of the original material. Even
the most ``primitive'' meteorites show evidence of extensive chemical
and thermal processing. They represent ``averages'' of bulk solar
material with only a small amount of preserved presolar grains (cf.
\cite{hlew95}).  The new results on presolar silicate grains
\cite{ngu04} indicate that the net abundance of presolar grains is
much higher in unmetamorphosed meteorites.  As yet no macroscopic
object has been found that is simply a mechanical aggregate of
unprocessed presolar material.

The small objects in chondrites, which were melted and crystallized at
some early time, would have locally chemically fractionated phases
that crystallized together. These would show correlated isotopic
enrichments of the daughter isotope due to the radioactive decay of
the parent in different parts of the same small object. It is this
chemical fractionation between mother and daughter element, occurring
while some short-lived nuclei are still abundant, that gives direct
evidence of both the presence and abundances of these nuclei at that
time and place. The time and place is, in general, not known. The bulk
chemical composition and morphology of some objects (CAIs, chondrules)
defines the type of chemical-physical processing. It does not define a
time or place. The most common assumption is that these objects formed
in the solar nebula, necessarily at very high dust-to-gas ratios
\cite{el00} \cite{bpjw04}. They may, in fact, be produced from a
protoplanet. Attributing objects to direct formation in the solar
nebula may not clarify the issues at hand. We note that the accretion
disks found around stars in the early stages of formation are, in
general, cool. The only zones that show evidence of being ``hot'' are
those very close to the star. In no cases are temperatures of $10^3$ K
observed in any part of the disk. As a result, it is not evident that
a hot solar nebula can be assumed for formation of the high
temperature materials observed in chondrules (including, and in
particular, CAIs).  However, there are more developed views of the
thermal structure of the solar nebula (see \cite{cassen01}).  The
X-wind scenario of Shu, Shang, \& Lee \cite{ssl96} would certainly
provide a hot processing zone.  It was proposed by Wood \cite{wood84}
that shock heating could play a major role in both CAI and chondrule
formation.  The significance of shock wave heating and the dynamics of
formation of chondrules in the solar nebula has become a focus of
serious study
\cite{cuzzi04}\cite{connolly98}\cite{cuzzi03}\cite{boss05}.  It is
possible that the nanodiamonds (which have bulk solar
$^{13}\mathrm{C}/^{12}\mathrm{C}$ but have been considered as presolar
grains by many workers) may be products of shock in the solar nebula.
NanoSIMS analysis may aid in clarifying this issue if evidence of
large $^{12}\mathrm{C}/^{13}\mathrm{C}$ variations can be found in the
somewhat reduced sampling volumes available by this technique.  The
protoplanets heated by decay of \al~ and \fe~ would certainly provide
a high temperature planetary regime. The formation of CAIs and
chondrules remains a fundamental problem of meteoritics. Assuming them
to be products of nebular processes may in some cases lead us
astray. The ``primitive'' objects (e.g. chondrites) were parts of
asteroids which are small planetary bodies. They will, to varying
degrees, show some effects of planetary metamorphism. If they were
from very small bodies or near the surface of small planetary bodies,
they would have undergone less thermal and chemical processing.

Alternatively, the meteorite may represent extensive melting and
chemical segregation of protoplanets (cores, mantles, crust). They
would be essentially homogenized isotopically and only show the
effects of any remaining short-lived nuclei by isotopic enrichments of
the daughter isotope after there is large chemical fractionation on a
planetary scale. The material in these protoplanets will also undergo
long term heating after the planet formed. The two time scales (to
form ``primitive'' objects and to form planetary differentiates, etc.)
may not in general be identifiable or resolvable. The major
protoplanetary differentiation events are plausibly presumed to occur
later, after aggregation of the dust.

The relative abundance of a radioactive nuclide $R$ relative to a
stable isotope $I$ as \underline{observed} in some meteoritic
material is defined as $\alpha^{R,I}_{obs}$. Different objects may
show different values of $\alpha^{R,I}_{obs}$. As time passes in
going from some assumed initial state of the protosolar nebula
(with the original values $\alpha^{R,I}_0
e^{-\Delta_1/\bar\tau_R}$) to the state when the particular
meteoritic material formed, the relationship between these two
values is given by
\begin{equation}
\alpha^{R,I}_{obs} = \alpha^{R,I}_0 e^{-\Delta_2/\bar\tau_R
-\Delta_1/\bar\tau_R}
\end{equation}
where $\Delta_2$ is the time interval from the initial state
($\Delta_2 = 0$) to the time of formation of the differentiated
object. There are no direct measurements of $\Delta_1$ or
$\Delta_2$.

It is possible to measure the age $(T)$ of formation (isotopic
equilibration) of an object relative to the present time. More ancient
objects should better record short-lived nuclei. Thus there should be
a correlation of age with the abundance $\alpha^{R,I}_{obs}$ of a
radioactive nuclide $R$. Even in the cases where there appears to be a
correlation, this does not fix the time $T_0$ when $\Delta_2 = 0$
(i.e., the initial state of the solar nebula). Note that for a
long-lived nuclide ($\bar\tau \gtrsim 10^9$ yr), an accuracy of $\sim
10^6$ years out of 4.56 Gyr requires that the lifetime be known to
better than $\sim 0.1 \%$. There are no absolute lifetimes known to
that level of precision. The best that can be done is to use
self-consistent ages by a particular long-lived parent-daughter
system. The most useful of these is the $^{207}$Pb - $^{206}$Pb method
based on an assumed fixed $^{235}$U/$^{238}$U ratio and assuming
closed system behavior. This method uses rather precisely measured Pb
isotopic ratios, not elemental abundances.  However, it is not
possible to demonstrate closed system behavior from the $^{207}$Pb
-$^{235}$U and $^{206}$Pb - $^{238}$U systems at the level of
precision required, even ignoring uncertainties in decay constants,
analytical data, and model assumptions. It is, for the most part, the
presence of the short-lived species that indicates that the object
formed at a value of $T$ that is close to the initial state. However,
major efforts at determining a chronology based on the
$^{207}$Pb-$^{206}$Pb system have led to important advances and appear
to provide valuable information \cite{cw81b} \cite{akh02} \cite{gop}
\cite{zg02}. In seeking to obtain precise ages of formation, the
problems of properly recognizing later metamorphism, element
redistribution, and the accretion of matter that had previously been
metamorphosed, still remain and must be addressed (cf. \cite{agr05}).
The complexities of obtaining precise $^{207}$Pb-$^{206}$Pb meteorite
ages is extensively discussed by Tera and Carlson \cite{tera99} and
Tera, Carlson, and Boctor \cite{tera97}.

For the short-lived nuclei, we are therefore left with a set of
values, $\alpha^{R,I}_{obs}(\Delta_i^R)$ ($i=1,2$), corresponding to
different values of $\Delta_1$ and $\Delta_2$ which are not known.
If, from independent arguments, we can assume that $(\Delta_1^{R^\ast}
+ \Delta_2^{R^\ast}) << \bar\tau_{R^\ast}$ for some nuclide $R^\ast$,
then the value of $\alpha^{R^\ast,I}_{obs} \approx
\alpha^{R^\ast,I}_0$.  Then the relationship between the dilution
factor and the abundances in the stellar envelope for $R^\ast$ and $I$
for a given stellar model is known from equation \ref{ENV}. This then
fixes the dilution factor for all the other nuclei $R$ for that
stellar model.  Given a proposed stellar model for the contaminating
source, $\alpha^{R^\ast,I}_0$ is determined by the dilution factor.
All shorter-lived species should be explained by the model with the
same dilution factor for all species with positive $\Delta_i(R)$.

\section{What do SNe II Produce?}\label{SNeII}

\subsection{A SNe II source for $^{26}$Al, $^{41}$Ca, $^{53}$Mn, and $^{60}$Fe}
\label{SNradio}

The issues to be discussed here are the questions of which short-lived
nuclei in the ESS might have been supplied by a `local' triggering
supernova and whether there are diagnostic characteristics that would
identify SN contributions. We will take the diverse, and generally
unmixed ejected material from a type II SN source as representing the
bulk composition of the dispersed ejecta after the explosion.  We
denote this composition with subscript ``ENV'' in analogy to our
notation for lower mass stars. We will not discuss the $r$-process
contributions because of the need for diverse $r$-sources and also
because of the questions relating to SNe II as sources of heavy
$r$-nuclei as discussed earlier. Our focus will be on the lighter
nuclides produced by SNe II following available models. There are
direct observational data showing that SNe II produce the Fe group
nuclei (including, and in particular, $^{56}$Ni).  In an earlier
report \cite{wgb98}, following the work of Timmes et al. \cite{tim95},
we further investigated the problem of whether \al, \ca, \fe~ and \mn~
could be adequately provided by a SN II event, using the models of
Woosley \& Weaver \cite{ww95} as a basis. As recognized by
\cite{wgb98} and \cite{tim95}, the average number ratio of
(\al/\fe)$_{ENV}$ in the SN ejecta of these models is about 8.6, from
a wide range of stellar models, the only clear exception being for $M
\simeq 13 M_\odot$. It was proposed that the (\al/\fe) ratio be used
as a test of the SN II model. It was found that to match the
$(^{26}\mathrm{Al}/^{27}\mathrm{Al})_{\Delta_1}$ ratio of $\sim 5
\times 10^{-5}$, and the
$(^{41}\mathrm{Ca}/^{40}\mathrm{Ca})_{\Delta_1}$ ratio of $\sim 1.5
\times 10^{-8}$, a mixing ratio of $\sim 3 \times 10^{-4}$ was
required and that this source yielded
$(^{60}\mathrm{Fe}/^{56}\mathrm{Fe})_{\Delta_1} \sim 10^{-6}$. In
addition, the inferred relative abundance of \mn~ implies that
$(^{53}\mathrm{Mn}/^{55}\mathrm{Mn})_{\Delta_1}$ is about
$10^{-3}$. This is far above the observed value shown in Table 1.  A
SN II model would also require that several percent of all $^{16}$O in
the protosolar cloud, as well as large contributions of other major
isotopes, was from a single SN II event.

The yields of nuclear species produced in models of SNe II have been
extensively studied by many workers (cf. \cite{ww95} \cite{nom}
\cite{rau02} \cite{clim03}) and define a rather self-consistent set of
results for major elements. In particular, a new generation of SN II
nucleosynthesis models was computed by Rauscher et al. \cite{rau02}
for a wide range of masses. An up-to-date set of reaction rates (both
experimental and theoretical) was used, in conjunction with upgrades
in the evolutionary code. These calculations did not include a
parametric $r$-process model but did treat neutron capture reactions
during the normally-occurring intermediate and advanced evolutionary
stages.  Explosive nucleosynthesis from a parameterized final collapse
and bounce process was also included. As a result, several neutron
capture nuclei are produced. In the Rauscher et al. \cite{rau02}
paper, Tables 8 and 9 (published electronically) give the yields of
both radioactive and stable nuclei. Using these results and the
published ejected masses, we have calculated the corresponding
$q^I_{ENV}/q^I_0$ and $(N^R/N^I)_{ENV}$ values in the ejected bulk
envelope, or ``wind'', of a 15 \msb SN for selected isotopes as shown
in Table \ref{sln}.  From these $q^I_{ENV}$ values we have calculated
the dilution factor ($f_0$) required to give the
$(^{26}\mathrm{Al}/^{27}\mathrm{Al})_{ESS}$ and the
$(^{41}\mathrm{Ca}/^{40}\mathrm{Ca})_{ESS}$ values. The time
scale for \ca~ effectively defines a value for $\Delta_1$.  The
resulting values of $(N^R/N^I)_{\Delta_1}$ are shown for this couple
of $f_0$ and $\Delta_1$ values. It can be seen that most short-lived
nuclei are produced abundantly, with the only exception being \pd,
which is somewhat low.  The principal conclusion is that if \al~ and
\ca~ are from a SN II source, then the ratio
$(^{60}\mathrm{Fe}/^{56}\mathrm{Fe})_{\Delta_1}$ must be very high
$(\sim 5 \times 10^{-5})$ and the
$(^{53}\mathrm{Mn}/^{55}\mathrm{Mn})_{\Delta_1}$ must be extremely
high $(\sim 3 \times 10^{-3})$, which is implausible. Essentially, the
same results are found using the 25 \msb model of Rauscher et
al. \cite{rau02} as shown by \cite{bgw03}. These conclusions are
essentially the same as would be obtained with the yields by Woosley
\& Weaver \cite{ww95}; see also \cite{bgw99} and \cite{wgb98}.

There is one discrepancy of note that pertains to the \al~ and \fe~
budget. Rauscher et al. \cite{rau02} give an average \al/\fe~ ratio of
approximately 1.5. This is lower by a factor of about 5.7 compared to
what was previously found by Woosley \& Weaver \cite{ww95}. The ratio
\al/\fe\ obtained by Rauscher et al.  \cite{rau02} is now in direct
conflict with the limits set by $\gamma$-ray observations
\cite{smith03} \cite{har04} \cite{pran05}, which give a steady-state
value for the Galaxy in accord with \cite{ww95}. Recently, new models
have been presented by Limongi \& Chieffi \cite{clim03} in which the
\al/\fe~ ratios are more similar to those of Woosley \& Weaver
\cite{ww95}.

\subsection{The oxygen conundrum}
\label{oxygen}

The addition of major elements (e.g. O, Si, Mg, Fe, Ti, etc.) by an SN
II can be seen by considering the fraction $\delta N^I/N^I$ of the
total solar inventory from such a postulated event. This gives $\delta
N^I/N^I = (q^I_{ENV}/q^I_0) f_0$.  For major elements like oxygen and
iron, $q^I_{ENV}/q^I_0 \sim 10^2$, so that for $f_0 = 3 \times
10^{-4}, \delta N^I/N^I = 3 \times 10^{-2}$, thus requiring a very
substantial fraction of the major elements from a single late
source. As indicated in section 2, the overwhelming abundance of
presolar grains (both oxide and carbide) are from AGB stars and the
hallmarks of SNe II in the grain population are quite uncommon.  The
only evidence of major isotopic shifts relative to the average
``solar'' values is in oxygen. The discovery by Clayton et. al.
\cite{cla73} was critical to stimulating work in isotopic studies of
meteorites.  However, there are no large (few percent level) effects
observed in Mg, Si, Ca, Fe, Ni, Sr, Ba, etc., that have been found in
related materials.  The $^{16}$O excess has most frequently been
interpreted as the result of injection by a supernova.  However, there
is no correlation of other isotopic shifts with the $^{16}$O
enrichment. Maintaining a separate gas reservoir enriched in $^{16}$O
from a supernova source without any evidence of effects in other
nuclei is not readily understandable.  It follows that the $^{16}$O
enrichment is not plausibly associated with nucleosynthetic processes;
there must exist chemical-physical processes that cause the isotopic
shifts. The discovery of $^{16}$O-enriched oxygen without shifts in
$^{17}$O/$^{18}$O in the laboratory by Thiemens and Heidenreich
\cite{thiemens}, and in the stratosphere and laboratory by
Mauersberger et al. \cite{mau99} \cite{mau} testifies to such
processes. These effects, which occur in oxygen-rich environments and
which were not explainable by classical isotopic fractionation
mechanisms, have now been explained on sound theoretical grounds by
Gao, Chen and Marcus \cite{marc02}. However, the explanation of the
oxygen isotopic effects in the high-temperature condensed phases found
in meteorites and in bulk meteorite samples remains a serious
problem. A complex theoretical model has been proposed by Marcus
\cite{marc04}, but this remains to be tested. A classical mechanism of
self shielding in the gas phase was early suggested by Navon and
Wasserburg \cite{nv85} as a possible source of the $^{16}$O
anomaly. These workers showed that the self shielding did not apply to
the ozone cases cited above. They also argued that the solar nebula
would most plausibly provide a CO-rich environment and that self
shielding might provide the isotopic effects observed.

Recognition of the difficulties with a nucleosynthetic source for the
$^{16}$O anomaly has grown over the years (cf. \cite{cl02}).  As a
result, the self shielding mechanism in a CO nebula environment has
recently attracted considerable interest by several groups \cite{cl02}
\cite{yuri} \cite{cl05} \cite{chak} \cite{lyons}. Because of the
importance of self shielding in molecular clouds, this problem is the
subject of ongoing study, although there is no clear support for the
required oxygen effects \cite{and04}.  The $^{16}$O problem is
important and remains unsolved.  A key question is: what is the bulk
isotopic composition of the sun?  It has long been known that the
solar isotopic abundance of oxygen is not known precisely enough from
astronomical observations.  If the bulk solar oxygen is like CAIs,
then the remaining planetary material is depleted in $^{16}$O relative
to the bulk solar system.  If it is like average planetary oxygen,
then the CAIs require an enrichment in $^{16}$O.  All models must, of
necessity, assume an initial oxygen isotopic composition.  A recent
effort to determine the oxygen isotopic composition of the solar wind
by Hashizume and Chaussidon \cite{hashizume05} points toward a
$^{16}$O-enriched composition for oxygen implanted by the solar wind
on grains of metal from the lunar soil.

In any self shielding or photolysis mechanism, the basic problems
are: finding the specific molecular reaction paths necessary to
produce the isotopic effects in the gas phase  appropriate to the
bulk gas composition and molecular speciation; considering the
problem of isotopic exchange in the gas phase; finding the means
of sequestering fractionated materials; preserving the isotopic
effects in condensed phases; and then heating the sequestered
condensed material so that the effects are present in the final
material -- but with back reactions for some phases with
``normal'' oxygen. This is a complex problem in chemical physics,
cosmochemistry and meteoritics. A sequestration by precipitation
and removal of icy objects has been suggested \cite{krot}. Self
shielding definitely merits further study, but as yet it is not a
``slam dunk'' solution to the problem.

\subsection{A SNe II source for only $^{53}$Mn and $^{60}$Fe}
\label{MnFe}

In summary, for a SN II source, if one matches the ESS values of
\al/$^{27}$Al and \ca/$^{40}$Ca, and uses the Rauscher et al.
\cite{rau02} models (see Table 2), one finds the following results:

(i)  The mixing ratio of $f_0 = 3 \times 10^{-4}$ corresponds to
$3 \times 10^4$ \msb of ISM for $\sim 10$ \msb of ejecta. This is
roughly compatible with the amount of mass necessary to slow down
 the SNe II ejecta to local velocities \cite{fb96}. The time scale $\Delta_1$
becomes $\sim 10^6$ yr.

(ii)  The \mn/$^{55}$Mn is very far above any of the observed values
for the ESS.

(iii)  The  \fe/$^{56}$Fe is also far above the existing ESS data.

(iv) The addition of substantial amounts ($\sim 3$ \%) of the solar
system inventories of many of the lighter elements (stable nuclei) by
a SN II would produce large isotopic shifts that are not observed. The
sole observed shift is in $^{16}$O and is not correlated with any
effects in other elements (Mg, Ca, Fe, Si, etc.).

We conclude that a SN trigger to the formation of the solar system
with injection of short-lived nuclei, particularly including \al~ or
\ca, is not an acceptable scenario. If we had used the Woosley \&
Weaver \cite{ww95} models the results would be essentially unchanged.

We note that the \mn~ overproduction has led Meyer \& Clayton
\cite{mc00} to suggest that the source was a peculiar massive star
that ejected only the external envelope rich in \al~ and \fe~ and
not the innermost region rich in \mn. This {\it ad hoc} model is
difficult to evaluate at the present time without further
predictions based on the same idea. However, as \fe~ is produced
in layers external to the innermost regions producing \mn,
increasing the mass cut (to reduce \mn) would leave, at least, the
\fe~ problem unresolved.

However, if we instead consider that a supernova source is responsible
for the \mn~ in the protosolar system, then the matter is greatly
changed.  Adopting this assumption, we have calculated the dilution
factor to obtain \mn/$^{55}$Mn $\sim 1 \times 10^{-4}$. This is in the
range of the highest estimated values for the ESS \cite{bir85}. In
this case, $f_0 = 7 \times 10^{-6}$. The abundances of the other
nuclei were calculated for this dilution factor (see last column, Table
\ref{sln}), keeping $\Delta_1 \sim 1$ Myr as before. It is clear that
this model is sufficient to also provide \fe~ at the highest level
currently considered in the ESS (\fe/$^{56}$Fe $\geq$ 10$^{-6}$,
\cite{mos03b}). Then all of the other nuclei (\al, \ca, \pd) are
essentially absent. If the \mn~ in the ESS were decreased to
(\mn/$^{55}$Mn) $\sim 10^{-5}$ (which is compatible with many
observations, see Section \ref{mn}), then $f_0 \sim 7 \times 10^{-7}$
and the \fe~ would also approximately agree with an ESS value as
currently proposed by Tachibana \& Huss \cite{th03} and Tachibana et
al. \cite{tac05}, if one takes $\Delta_2 \le$ 1 Myr. As there is no
other stellar source for \mn, we conclude that a component of the
solar inventory of short-lived nuclei must be from SN II; this only
supplies the \mn~ and some of the \fe, but no other short-lived
nuclei. In considering both \mn~ and \fe~ we note that \fe~ requires a
stellar source and such a source can only be a supernova or an AGB
star. A SNIa origin is possible for Fe group nuclei \cite{nom1}
\cite{nitr}: this might include \mn, but probably not \fe, which
derives from neutron captures.

We conclude that SNe II are excluded from further consideration for
providing \al, \ca, \pd, etc. Instead, providing \mn~ and \fe~ from a
SN II source is appealing, as the mass fraction that needs to be
contributed to the ESS to account for them in a late addition is very
small (with only small contributions to the other nuclei, e.g. 0.1\%
of the oxygen). If the SN II contribution \mn~ and \fe~ is from an
earlier event (say $\Delta_1 \sim 10^7$ yr), then the dilution factor
would become significantly larger and there would be much less \fe.
The general inventory of $^{53}$Mn in the ISM should be quite
sufficient to provide the plausible ESS values if the refreshment time
by SNe II is less than $\sim$15 Myr.

\section{What do AGB Stars Produce?}\label{agb}

A new generation of AGB yields for short-lived nuclei has been
computed for this report. These calculations cover a range of stellar
masses and metallicities. These new results are a substantial
improvement over those discussed in Wasserburg et al.  \cite{wbgr94}
and in Busso et al. \cite{bgw99}. The first results of these
calculations were outlined by Busso et al. \cite{bgw03} and by Gallino
et al. \cite{gal04}. The new results are an improvement over the
earlier calculations, as they include use of: i) cross sections from
Bao et al. \cite{bao00} plus several subsequent updates; ii) an
improved technique for estimating the rate of bound-state electron
captures on \ca~ in the stellar envelope, obtained by averaging its
lifetime over the distribution of temperature and density of the whole
convective zone; and iii) the use of a finer grid for the basic
stellar evolution models than used earlier. As discussed previously
(Sec. \ref{revagb}) it is both impractical and unnecessary to perform
all the network computations directly within complex and
time-consuming stellar evolution codes. This is particularly important
when one wants to build a very large database of stellar yields, for
several different values of the stellar mass and metallicity. As far
as rare species and neutron-capture products are concerned,
post-process calculations remain a suitable tool, provided their input
parameters can be estimated safely. Recently, Straniero et
al. \cite{str03} computed a grid of stellar models with the explicit
goal of generating reliable interpolation tools for deriving the basic
detailed stellar models necessary to calculate nucleosynthesis in AGB
stars.These rules apply to masses in the range 1 to 3 \msb and
metallicities from 1/6 $Z_\odot$ to $Z_\odot$.

Based on the rules recommended by Straniero \cite{str03}, we
prepared a  finer network of models for calculating
nucleosynthesis in AGB stars. All  results discussed here are
taken from these revised models. The complete  stellar reference
models were all calculated with the FRANEC evolutionary code. When
this last and the MSSSP code \cite{fl96} \cite{kl03} \cite{lug01}
are used with the same treatment for mass loss and convective
mixing, very similar results are obtained. This confirms the
reliability of the evolutionary scenario. For details of the
models and a complete review of stellar evolution preceding the
late TP-AGB see Straniero et al. \cite{sg05} in this volume. Mass
loss was always simulated with the Reimers' parameterization,
setting the values of the  $\eta$ parameter to 0.3 (1.5 \msb
cases), 0.5 (2.0 \ms), 1.5 (3 \ms) and 5 (5 \ms).

In all of the calculations presented here in Tables \ref{qenv} to
\ref{ejecta} for low and intermediate masses, the \al~ contribution
only takes into account its production in the H shell, its engulfment
into the He-layers and burning through ($n,p$) and ($n,\alpha$)
captures in the convective pulses, and subsequent free decay in the
envelope (no CBP). The balance between these processes typically gives
rise to envelope ratios \al/$^{27}$Al of $\sim 5 \times 10^{-3}$ for
low mass stars. When considering the net \al~ inventory, this ``base
line'' production should be added to the larger yields that may be
derived from either CBP (low mass stars) or HBB (see Section
\ref{IMS}). Note that our 5 \msb models do not include hot bottom
burning in the envelope. For recent extensive calculations of HBB
yields, especially for \al~ and the ensuing effects on Mg isotopes,
the reader is referred to Karakas and Lattanzio \cite{kl03b}.

Concerning short-lived species produced by neutron captures, their
yields depend on the $^{22}$Ne($\alpha,n)^{25}$Mg reaction and the
\ct($\alpha,n)^{16}$O reaction (see Section \ref{revagb}). As this
second process comes from a mixing mechanism of unknown nature that is
treated parametrically, we present two sets of results: one
considering only neutron captures in the convective pulses from
$^{22}$Ne burning, and one that includes both sources assuming a \ct~
pocket as adopted in previous works (the 'standard' choice of
\cite{gabl98} \cite{bgw99} \cite{bus01}).  This choice corresponds to
$4 \times 10^{-6}$ \msb of \ct~ in LMS and to $4 \times 10^{-7}$\msb
of \ct~ in IMS (see Section \ref{revagb}). We note that the existence
of a wide intrinsic spread of \ct~ concentrations, from a few
$10^{-6}$ \msb down to the complete absence of the \ct~ pocket in the
intershell zone, is demonstrated by observations of a wide range of
$s$-process enhancements at a given metallicity in AGB stars
\cite{bus01} \cite{abia01} \cite{abia02}. This is also shown by the
isotopic patterns in presolar grains (see e.g.\cite{ama01}) and by
$s$-process distributions in extremely metal poor stars
\cite{trav}. The results for each stellar mass thus depend only on the
amount of \ct~ burned and on the metallicity.

We show here results for stars of 1.5, 2, 3 and 5 \ms, for both
solar and 1/3 solar metallicity. \al~ is always produced in
significant amounts, however less efficiently (by typically a
factor of 3) than in CBP or in HBB; its final yield then depends
on the contributions from these processes.

As can be seen from equation \ref{ENV}, the parameters that determine
the abundances in the protosolar molecular cloud are the relative
abundances, $(q_{ENV}^I/q_0^I)$, of each stable nuclide $I$ in the
envelope relative to the ISM, and the ratio $(N^R/N^I)_{ENV}$ in the
envelope.  Note that in our earlier work \cite{wbgr94} we gave
$q^I_{\mathrm{He}}/q_0^I$ and $(N^R/N^I)_{\mathrm{He}}$ in the
production shells.  Here we explicitly calculated the values of
$N^R/N^I$ and $q^I_{ENV}/q^I_0$ in the envelope over the evolution of
the star as done in \cite{bgw03} \cite{gal04}. The values of
$q^I_{ENV}/q^I_0$ are presented in Table \ref{qenv} and are shown in
Fig. \ref{qagb}. As can be seen, for no \ct~ pocket, for a given
stellar mass and a given $Z$, the $q^I_{ENV}/q^I_0$ for all of the
stable nuclei are close to $Z/Z_\odot$. However, with a \ct~ pocket
there is a substantial to great increase in $q^I_{ENV}/q^I_0$ for
those elements of higher atomic number. This rule applies to all the
cases shown, though they may differ from one another in other respects
according to their \ct/Fe ratios (i.e., metallicity) and to their
residual envelope masses (i.e., to the achieved dilution of C-rich and
$s$-process-rich matter with unprocessed material).

The values of $(N^R/N^I)_{ENV}$ are given in Table \ref{nrenv} for
the same models as above. It can be seen that, in general, for
light nuclei $(N^R/N^I)_{ENV}$ is grossly the same with or without
a \ct~ pocket, while the $(N^R/N^I)_{ENV}$ of species heavier than
iron increase for increasing neutron dose per seed nucleus.
As a result, the presence of a \ct~ pocket tends to give widely
varying values of $(M_0/M_{ENV})\alpha^{R,I}$ from one pair of
nuclides to another. This makes it very difficult to reproduce
isotopic ratios in heavy and light elements simultaneously, as it
overproduces the heavier nuclei from neutron capture on the Fe
seed. As a result, in seeking to explain a wide range of short-lived
nuclei as well as \pd, we focus on AGB models {\bf with no \ct~
pocket}. Such stars can represent a substantial fraction of all AGB
stars. The consequences of a standard $^{13}$C pocket are discussed in
Section 12: it can only provide some heavy nuclei.

\subsection{Determination of the dilution factor from \pd}

The value of (\pd/$^{108}$Pd)$_{ESS}$ is well known since the
discovery by Kelly \& Wasserburg \cite{kw78}. Extensive work by Chen
and Wasserburg \cite{cw83}\cite{cw96} established internal isochrons
for several iron meteorites of different classes with \pd/$^{108}$Pd$=
2 \times 10^{-5}$ with a typically small range.  The recent
high-precision results by Carlson \& Hauri \cite{ch01} confirm this
abundance and we may consider the value as well-established. The very
high value of \pd/$^{108}$Pd reported by Carlson \& Hauri \cite{ch01}
for a small isotopic shift in $^{107}$Ag/$^{109}$Ag on a bulk Allende
sample was not confirmed by Woodland et al. \cite{wrl03}. As
$\bar\tau_{^{107}\mathrm{Pd}} = 9.4 \times 10^6$ years, there will not
be a major difference between the ESS value and the initial value at
the time of injection into the ISM if $\Delta_1$ and $\Delta_2 \leq 5
\times 10^6$ years. This short time scale for formation of some
differentiated protoplanets is made clear by the discovery of \al~ in
basaltic achondrites (planetary crusts) as found by Srinivasan et
al. \cite{sgb99}. Thus (\pd/$^{108}$Pd)$_{ESS} \approx
$(\pd/$^{108}$Pd)$_{SC}$ and this value may be used to determine the
dilution factor for a stellar model (see Sec. \ref{inject}).

\begin{figure}
\begin{center}
\end{center}
\caption{The enhancement factors for stable nuclei with respect to
solar concentrations for stable isotopes of interest here, as
reached in the envelope of AGB star models with either low mass
(upper panel) or intermediate mass (lower panel), as discussed in
the text. Abundances refer to the last pulse computed. Cases are
shown with a standard \ctb pocket (ST) and with no \ct~pocket.}
\label{qagb}

\centerline{\epsfig{file=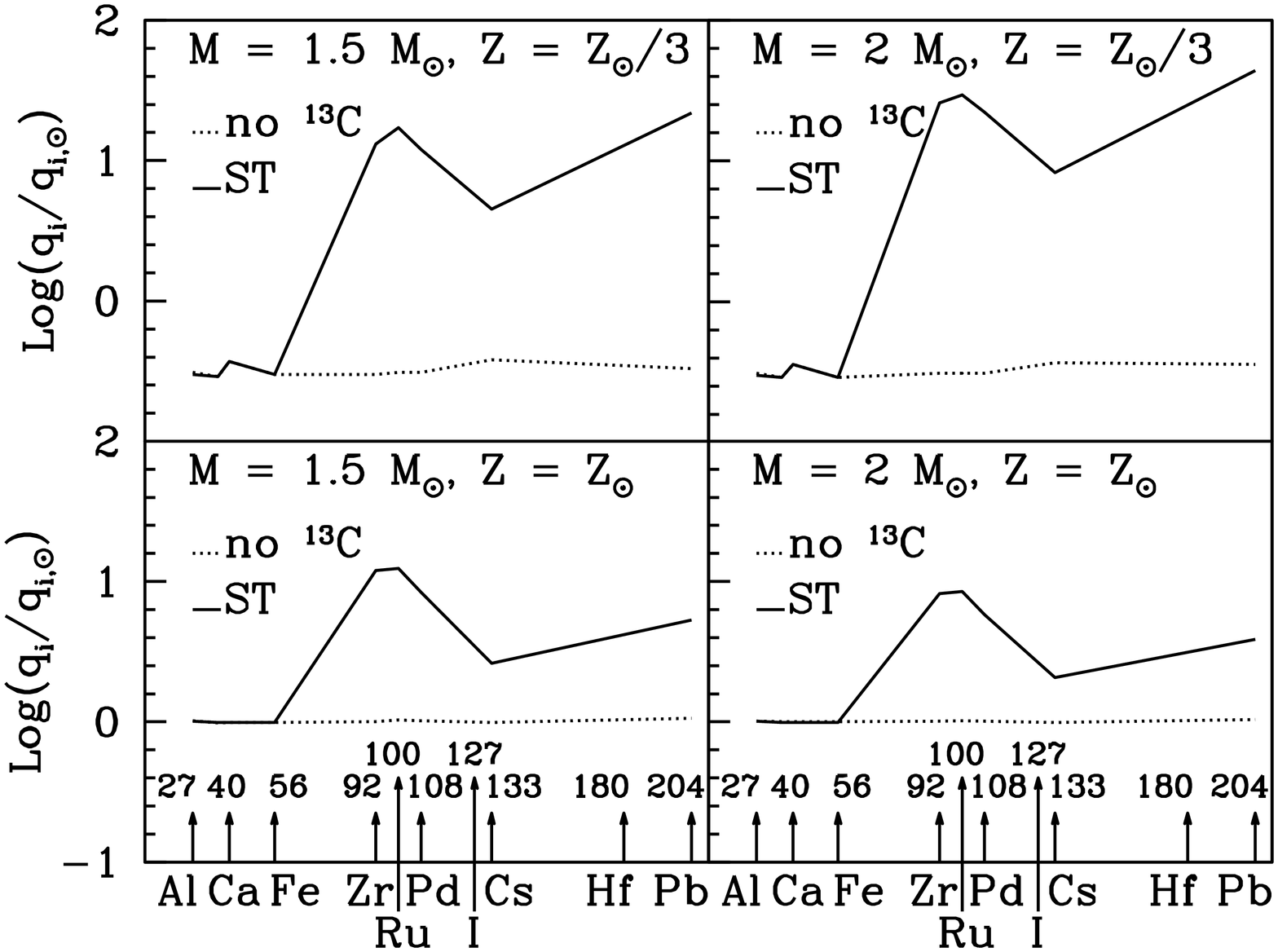,angle=270,width=5in}}
 
\centerline{\epsfig{file=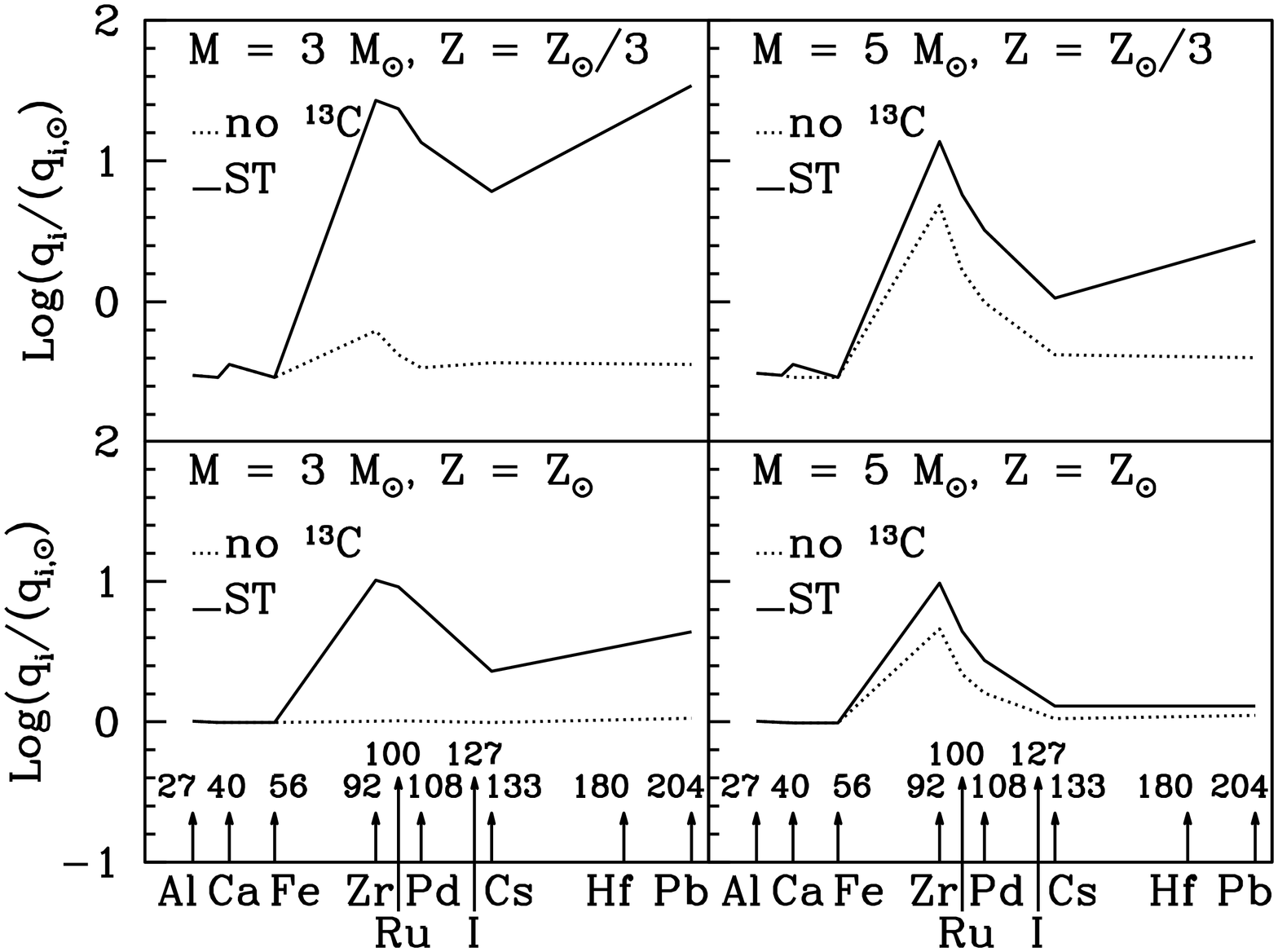,angle=270,width=5in}}

\end{figure}

\begin{figure}
\begin{center}
\end{center}
\caption{The dilution factor $f_0$ necessary to give
($^{107}$Pd/$^{108}$Pd)$_{ESS}$ with the various AGB models
discussed in the text. The abscissa spans a range of initial
$^{107}$Pd/$^{108}$Pd ratios corresponding to $\Delta_1 +
\Delta_2$ between 0 and 7 Myr. Shaded areas are guides showing the
typical $f_0$ ranges valid for LMS and IMS stars. Note the narrow
range in $f_0$ for LMS stars.} \label{dpd}

\centerline{\epsfig{file=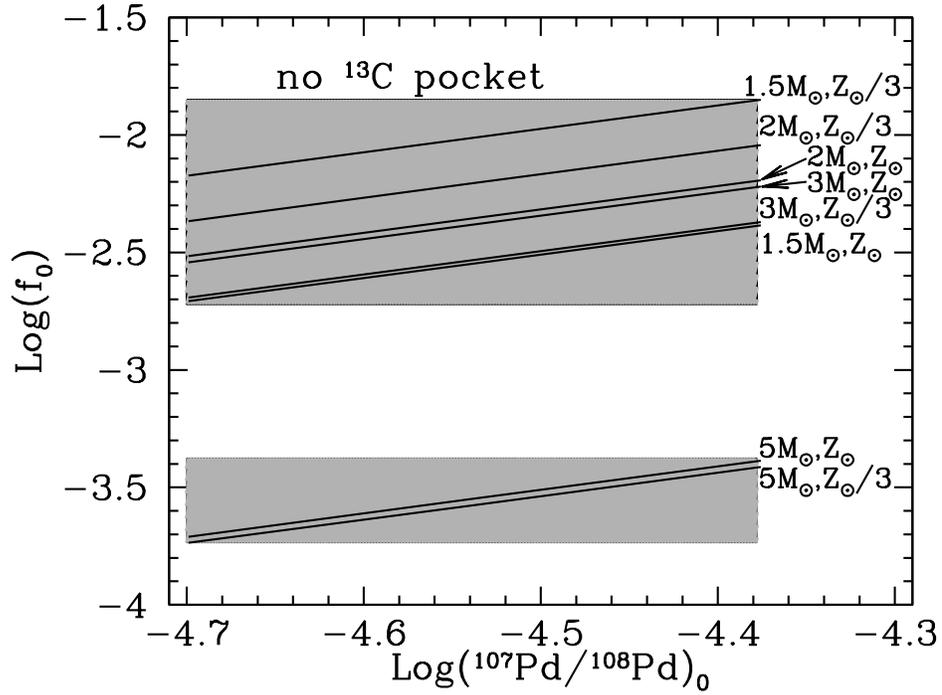,angle=270,width=5in}}

\end{figure}

While \pd~ can be produced in both $r$- and $s$-processes, the solar
inventory can not be attributed to a ``standard'' $r$-process source
due to the long $\Delta_1$ implied by $^{129}$I. We therefore focus on
an AGB source. Production of \pd~ requires a single neutron capture on
stable $^{106}$Pd and thus occurs through both the
$^{22}$Ne($\alpha$,n)$^{25}$Mg and the \ct$(\alpha$,n)$^{16}$O neutron
sources. The laboratory mean life of \pd~ is not significantly changed
in stellar interiors \cite{ty87}. \pd~ in turn captures a neutron to
produce $^{108}$Pd with production ratios roughly governed by the
ratio of the $<\sigma^i >$ values. Our results for \pd~ (and its
stable reference $^{108}$Pd) for an AGB source are shown in Tables
\ref{qenv} and \ref{nrenv}. The results for LMS with a $^{13}$C pocket
are far above those for no $^{13}$C. The choice of the magnitude of
the $^{13}$C pocket is very restricted in LMS by the fact that
increased neutron capture for heavier nuclei greatly increases the Pd
yield without changing the values for \fe, \ca, or \al~ (see
Fig. \ref{qagb}). As a result, a standard \ct~ pocket (sufficient to
explain main solar $s$ abundances with LMS models) does not permit a
simultaneous solution for any lower mass nuclide with those of higher
masses (see \cite{wbgr94} \cite{bgw99}). (The situation is different
for an IMS, as the effects of the \ct-pocket are less important,
cf. Fig. \ref{qagb}). A graph of the dilution factor $f_0$ versus
(\pd/$^{108}$Pd)$_0$ (the subscript refers to time since injection),
for the case of no pocket, is shown in Fig. \ref{dpd}, allowing for a
range of values of the time delay $\Delta_2$. For no $^{13}$C pocket
and $Z = Z_{\odot}/3$ to $Z_{\odot}$, models predict a restricted
range of \pd/$^{108}$Pd in the envelope for low mass stars up to 3
$M_{\odot}$, while a much larger yield is predicted for a 5
$M_{\odot}$ model. For the case of $f_0 \simeq$ 10$^{-2.3}$ and $Z =
Z_{\odot}$ we find (\pd/$^{108}$Pd)$_0 \simeq 3 \times 10^{-5}$. This
is more than sufficient to account for the \pd~ found in the Early
Solar System. It can be seen that for low mass stars with no $^{13}$C
pocket the possible range of $f_0$ values is $2 \times 10^{-3} \leq
f_0 \leq 1.5 \times 10^{-2}$. For a 5 $M_{\odot}$ model the range is
$2 \times 10^{-4} < f_0 < 4 \times 10^{-4}$. These ranges of $f_0$
will now be used to estimate the contributions for all other nuclei
produced by AGB models.

\subsection{Calculated Abundances from an AGB source}

\begin{figure}
\begin{center}
\end{center}
\caption{The dilution factor $f_0$ required to produce different
values of $^{60}$Fe/$^{56}$Fe by addition of AGB materials to
the protosolar cloud for the models discussed in the text. The
abscissa spans values of initial $^{60}$Fe/$^{56}$Fe in the range
(0.1 - 1)x10$^{-6}$. Shaded areas are guides showing the regions
covered by the dilution factor obtained for $^{107}$Pd in figure
\ref{dpd}.} \label{dfe}

\centerline{\epsfig{file=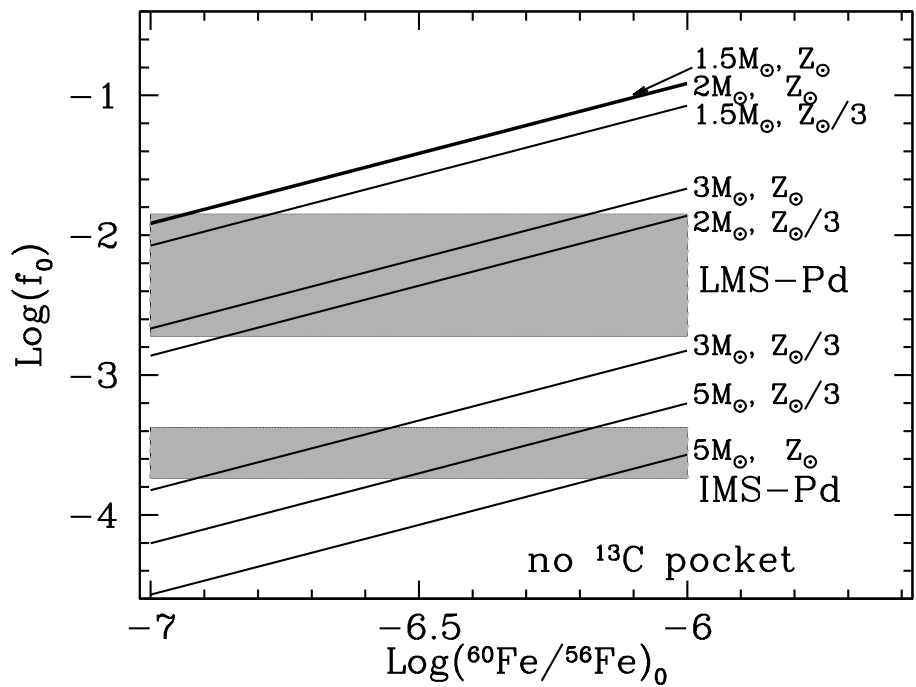,angle=270,width=5in}}

\end{figure}

Using the composition of the winds for different AGB stellar models
and the dilution factor obtained from $^{107}$Pd, we may now calculate
the abundances of short-lived nuclei to be expected at a time
$\Delta_1$ in a mix of AGB ejecta and ISM material.  Examples of these
results are shown in Tables \ref{ejecta} and \ref{cloud}. Table
\ref{ejecta} refers to the solar metallicity models using the range of
dilution factors shown in Fig.  \ref{dpd}.  Similarly, the range of
$f_0$ found from \pd\ is superimposed on the $^{60}$Fe/$^{56}$Fe
versus f$_0$ graph in Fig.  \ref{dfe} to show the range of
$^{60}$Fe/$^{56}$Fe corresponding to these bounds.  A value of $f_0$
from its permitted range can then be combined with
(\ca/$^{40}$Ca)$_{ESS}$ to determine $\Delta_1$ and match the ESS
inventory for this very short-lived nuclide. The corresponding values
for all the other isotopes are then presented at that $f_0$ and
$\Delta_1$, along with the $\Delta_2$ for an exact match to \pd, in
Tables \ref{ejecta} and \ref{cloud}.  These can be compared with the
ESS values reported in Table \ref{mlt}.

From inspection of Table \ref{ejecta}, we find the following
general rules:

i) LMS models (1.5 or 2 \ms) can be a suitable source for \ca~ and
\pd~ and can also predict  a substantial  inventory of all other
species listed;

ii) It can again be seen that the typical values of
(\al/$^{27}$Al)$_{ENV} \sim (3-5) \times 10^{-3}$, which derive
from pure H-shell burning without CBP, are insufficient to match
the ESS values by a factor of $\sim$ 3;

iii) \fe~ can be accounted for by a late addition from a low mass
AGB only at the lowest of the recently suggested ESS ratios;

iv) IMS models can easily produce abundant  \fe, sufficient to
account for almost the highest (\fe/$^{56}$Fe)$_{ESS}$ that has
been proposed; they also copiously produce other isotopes.
However, as shown for a 5 $M_{\odot}$ star, the value of
$\Delta_1$ required by \ca~ is very short.

Both standard (i.e. without CBP) LMS and IMS models end up with
the requirements that other mechanisms or sources must supplement
AGB production for $^{26}$Al and one or another isotope of the
short list presented in Tables \ref{qenv} to  \ref{ejecta}. As
indicated earlier \cite{nbw03}, CBP produces large amounts of \al.
For \al~ we have calculated the (\al/$^{27}$Al)$_{ENV}$ required
in order to obtain (\al/$^{27}$Al)$_{ESS} = 5 \times 10^{-5}$. The
exact envelope value required to provide a solution is indicated
in parentheses for each case in Table \ref{ejecta}. These levels
are well within the reach of CBP models without stretching to high
$T_P$ values.

We thus consider the $^{26}$Al observations to be explained by an
AGB star with some CBP. With regard to possible matches with ESS
values, we note that a compromise exists between the two cases
(1.5-2)$M_\odot$ and 5 \ms. The model of a 3 \ms\ star provides
the best balance and accounts for the largest number of ESS
radioactivities, including \ca~ and \fe, with reasonable values of
the time delays. As an example of this, we show in Table
\ref{cloud} the result of applying the above approach to a 3 \ms,
$Z = Z_{\odot}/3$ case. As can be seen, a late contamination by
such a star could explain the ESS abundances of \al, \ca, \pd, and
\fe, (even for quite high ESS values). It would also give
substantial contributions of $^{135}$Cs. The matter of $^{36}$Cl
is discussed in Section \ref{36cl} devoted to that isotope.

All the computations of Tables \ref{ejecta} and \ref{cloud} predict
  upper bounds for $^{205}$Pb. This last datum cannot be better
  determined, due to the high uncertainty of $^{205}$Pb survival after
  a thermal pulse (see Mowlavi et al. \cite{mow98} for a
  discussion). As in our original calculations \cite{wbgr94}, we note
  that a large fraction of this $^{205}$Pb decays before being
  ejected, at least in LMS cases. Due to the complex behavior of the
  couple $^{205}$Pb-$^{205}$Tl in He-shell conditions, the higher
  temperature of IMS models (including that of Table \ref{cloud})
  would favor the survival of $^{205}$Pb in the critical interval
  after a pulse (a few hundred years) before it is transferred to the
  envelope by dredge up. Hence IMS models are probably the most
  promising site for $^{205}$Pb formation. Thorough and detailed
  calculations are required to obtain precise estimates of the rates
  for the weak interactions involved.

\section{Comments on some critical isotopes}

The overall results presented above give a reliable and broad
assessment of what different sources could provide. There are
important, critical issues involving the abundances of certain
nuclei inferred from measurements in meteorites. In some cases
there are uncertainties in the network calculations. In this
section we present some considerations of the reliability or
uncertainty of the results used above.

\subsection{Estimating the \fe~ inventory}

The isotope \fe~ plays a key role in determining $\Delta_1$ for \mn~
and \fe. Both are produced in abundance by SNe II and are present in
high levels in the ISM for the UP model. As \fe~ (but not \mn) is also
produced by AGB stars, they would add to the inventory from SNe II.
Because of its importance, we present a short review of the history of
the searches for \fe.

The first hint of \fe~ in the solar system was found as excesses of
$^{60}$Ni in CAIs \cite{blug88}. These data can be interpreted to
indicate that the ratio at the time of some CAI formation was
\fe/$^{56}$Fe $= (1.6 \pm 0.5) \times 10^{-4}$. However, the measured
$^{60}$Ni excesses might reflect isotopically anomalous Ni from
pre-existing stellar condensates rather than decay of
\fe~\cite{sluga}. Further, it was found by V\"{o}lkening \&
Papanastassiou \cite{vp89} that some CAIs contained anomalous Fe with
conspicuous excesses of $^{58}$Fe. Thus the straightforward
interpretation of the CAI data remains unclear. The first clear
evidence indicating \fe~ was found in eucrites \cite{sluga}
\cite{slugb}, which are basaltic rocks. This study was well-designed
as eucrites are greatly depleted in Ni from the formation of an Fe/Ni
core in their parent planetesimal.  However, the data showed no good
correlation for $^{60}$Ni$^\ast$ versus $^{56}$Fe. The inferred
\fe/$^{56}$Fe for eucrites was far lower than that inferred for CAIs
and differs considerably from one eucrite to another: $3.9 \times
10^{-9}$ to $4.3 \times 10^{-10}$.  In contrast, the abundances of
\mn~ $(\bar\tau = 5.3$ Myr) inferred for these same eucrites were
typically uniform (\mn/$^{55}$Mn $\approx 1-4 \times 10^{-6}$)
\cite{ls98}. Even assuming a value of \fe/$^{56}$Fe for the eucrites,
there remained the question of $\Delta_2$ for formation/metamorphism
of these basaltic rocks. As $\bar\tau_{^{60}\mathrm{Fe}} = 2.2 \times
10^6$ yr, if $\Delta_2 = 5 \times 10^6$ yr, the initial value would
then be ten times greater than for the eucrite.

Several searches for evidence of \fe\ were made in chondrites, which
did not totally melt and are far more primitive than eucrites
\cite{slugb} \cite{chw99} \cite{knt00}. Troilite (FeS) from a
relatively metamorphosed chondrite, Ste. Marguerite, gave an upper
limit of $2.4 \times 10^{-8}$ for \fe/$^{56}$Fe \cite{slugb}. Ion
microprobe studies of unmetamorphosed olivine phenocrysts in
chondrules (a major constituent of chondrites) gave no resolvable
$^{60}$Ni$^\ast$ in unequilibrated ordinary chondrites \cite{knt00}
\cite{cwh99}. The upper limit for \fe/$^{56}$Fe in a Semarkona
chondrule was $3.4 \times 10^{-7}$ \cite{knt00}, while that for
Bishunpur and Semarkona chondrules was $1.6 \times 10^{-6}$
\cite{cwh99}. Measurements of a sulfide-rich opaque inclusion and
spinels within a CAI \cite{cwh99} gave the upper limit on
\fe/$^{56}$Fe of $1.7 \times 10^{-6}$. Very recently, there have been
important observations in the search for \fe\ that show clear
correlations of $^{60}$Ni/$^{61}$Ni versus $^{56}$Fe/$^{61}$Ni with
very high values of Fe/Ni $(\sim 10^5 - 10^6)$ \cite{th03}
\cite{htac04} \cite{mlh04}. The results of \cite{th03} and
\cite{tac05} gave \fe/$^{56}$Fe $\simeq 1 \times 10^{-7}$ for several
troilite samples. The results by Mostefaoui et al. \cite{mlh04} on
troilites gave $(^{60}\mathrm{Fe}/^{56}\mathrm{Fe})_0 \simeq 1 \times
10^{-6}$. Most recently, measurements by Tachibana et al. \cite{tac05}
on silicates in chondrules of very low metamorphic grade chondrites
gave \fe/$^{56}$Fe $\simeq 2 \times 10^{-7}.$ If $\Delta_2 = 2 \times
10^6$ yr, this implies (\fe/$^{56}$Fe)$_0 = 5 \times 10^{-7}$. This
matter is under intense study by several groups at present. As noted
in \cite{bgw03}, the very high \fe~ abundances now indicated should
readily permit high precision isochrons to be established using TIMS
or ICPMS techniques with only small Fe-Ni fractionation between
phases.

Fig. \ref{dfe} shows a plot of the dilution factor $f_0$ versus the
ratio (\fe/$^{56}$Fe)$_0$ in a cloud in which the ejecta from an AGB
star (see previous section) were instantaneously mixed.  Assuming a
dilution factor of $f_0 = 10^{-2}$ we find that the values of
(\fe/$^{56}$Fe)$_0$ range from $10^{-7}$ to $4 \times 10^{-7}$ for $M
< 3 M_\odot$ and $Z = Z_{\odot}$.  The acceptable ranges of $f_0$ as
determined by \pd~ are shown as shaded areas on the graph of $f_0$
versus \fe/$^{56}$Fe.  For a model of 5 \msb and $ Z_{\odot}$,
(\fe/$^{56}$Fe)$_0$ values in excess of 10$^{-6}$ would be permitted
with $f_0$ in the range allowed by \pd. There is thus a wide range of
(\fe/$^{56}$Fe)$_0$ possible (more than an order of magnitude) from
AGB stars with solar abundances at low dilutions. The repeated
statements in the literature that \fe~ is a clear mark of SNe is
incorrect as shown long ago.  It is evident that if
(\fe/$^{56}$Fe)$_{ESS}$ lies between $2 \times 10^{-6}$ and $2 \times
10^{-7}$, then a 1.5 \msb\ case is excluded. The possible AGB sources
are then restricted to 2-3 \ms.

Note that in earlier efforts to determine possible AGB
contributions, Wasserburg et al. \cite{wbgr94}   focused on a 1.5
\msb source. It was found that the local ratio in the production
zone was $^{60}$Fe/$^{56}$Fe = 10$^{-4}$ to 10$^{-3}$
\cite{wbgr94}. This gave fully adequate \fe~ to provide the
inventory then current of (\fe/$^{56}$Fe)$_{\Delta_2} \sim
10^{-8}$ with reasonable choices of $\Delta_1 \lesssim 10^7$ yr.
Those calculations are still applicable for a 1.5 \msb source with
only minor changes for the prediction of other short-lived nuclei.

From the existing data sets, we now must consider a rather wide range
of much higher values of \fe~ abundance in our models.  For the
purposes considered here, we will consider the range $2 \times
10^{-7}$ to $2 \times 10^{-6}$, as shown in Table 1.

With the improved models discussed in Section \ref{revagb} and for no
\ct~ pocket at solar metallicities, there is a very restricted range
of \fe/$^{56}$Fe in the envelope for $M \leq 3 M_\odot$ (see Tables
\ref{qenv} to \ref{ejecta}).  It follows that for $M \lesssim 3
M_\odot$ and $Z_\odot$, the (\fe/$^{56}$Fe)$_0$ is then essentially
fixed by the dilution factor (see equation \ref{ENV}).
On the other hand, there is a large increase in $^{60}$Fe production
at 5 \msb due to the greatly increased temperatures of the He shell.
For this case note that $q_{ENV}$ also drops with decreasing $Z$ due
to consumption of the Fe seed.

We now explore how robust these results are in terms of the nuclear
physics.  A major uncertainty in the \fe~ production lies in the
neutron capture cross sections of both $^{59}$Fe and \fe, which are
based on theoretical estimates only. The Maxwellian-averaged cross
section of \fe~ at 30 keV is rather low: $< \sigma >$(30 keV)(\fe) =
3.65 mbarn, according to Woosley et al. \cite{woos78}. Consequently,
only a small fraction of the \fe~ produced can undergo further neutron
captures. On the other hand, \fe~ synthesis directly depends on the
cross section of $^{59}$Fe, whose value at 30 keV, as estimated by
Woosley et al.  \cite{woos78} is 12.3 mbarn. More recent calculations,
based on a different theoretical approach, have been performed by
Rauscher \& Thielemann \cite{rth00}; they obtain $< \sigma >$(30
keV)($^{59}$Fe) = 22.5 mbarn. We adopt here these recent estimates,
but we verified that, if we had instead used the old cross sections
for both $^{59}$Fe and \fe, our production of \fe~ (as given in Tables
\ref{qenv}, \ref{nrenv}) would have been reduced by a factor $\approx
1.8$ for all the $M = 1.5$ and 3 \msb models, and by a factor of $\sim
1.2$ for the $M = 5 M_\odot$ models. We conclude that variations of
the neutron capture cross sections cannot change our predictions by
more than a factor of 2.

\subsection{\mn}\label{mn}

The nuclide \mn~ is particularly important since it can only be
produced in SNe or by proton bombardment. It was first shown by Birck
\& All\`{e}gre \cite{bir85} \cite{bir88} that excess $^{53}$Cr was
present in CAIs. While there appears to be a correlation of
$^{53}$Cr$^\ast$ with Mn in CAIs, there is in addition a widespread
occurrence of variations in $^{54}$Cr which can only be ascribed to
endemic, that is widespread, nuclear effects in meteorites due to
incomplete mixing from different sources \cite{rba92}. As recognized
by Birck \& All\`{e}gre \cite{bir85}, it is not obvious that the
$^{53}$Cr$^\ast$ can be unambiguously attributed to in situ decay of
\mn. A recent report by Papanastassiou et al. \cite{dap05} on a CAI
known to contain \al\ showed with precise data a correlation of
$^{53}$Cr$^\ast$ with Mn/Cr giving \mn/$^{55}$Mn $ = 1.5 \times
10^{-4}$, in agreement with Birck \& All\`{e}gre \cite{bir85}.  This
sample also showed large excesses of $^{54}$Cr/$^{52}$Cr. We consider
this value as the upper estimate of \mn/$^{55}$Mn in the ESS. In
contrast, from the well-defined study of Birck \& All\`{e}gre
\cite{bir88} on a pallasite (stony-iron), it is clear that $^{53}$Mn
was present in differentiated planets at a level of \mn/$^{55}$Mn =
2.3$\times10^{-6}$. Extensive studies of the \mn-$^{53}$Cr system in
eucrites (basaltic achondrites), which are planetary differentiates,
show very well-defined correlations of $^{53}$Cr/$^{52}$Cr versus
$^{55}$Mn/$^{52}$Cr.  This work demonstrates the presence of \mn~ in
this stage of planetary evolution and yields
(\mn/$^{55}$Mn)$_{\Delta_2} = (4.7 \pm 0.5) \times 10^{-6}$ for some
meteorites. In some cases it was possible to determine internal
isochrons with slopes corresponding to about the same value (see the
extensive report by Lugmair \& Shukolyukov \cite{ls98}). Studies of
whole-rock chondrites with very limited fractionation of Mn from Cr
indicate \mn/$^{55}$Mn $\sim 9 \times 10^{-6}$ \cite{nyq01}
\cite{ls98}. We take this value as a reasonably sound estimate of the
ESS value, although somewhat higher values have been found in
late-stage alteration products in a CM2 chondrite \cite{brearley02}.
The issue of the times $\Delta_2$ at which the eucrites formed and the
problem of thermal metamorphism persist.  This is well recognized by
these authors and is evident in the eucrite samples where \mn~ is
essentially dead (cf. the meteorite Caldera, \cite{wlug96}). The
reader is directed to the extensive reviews by McKeegan \& Davis
\cite{md04}, and Lugmair \& Shukolyukov \cite{ls98}. In our discussion
we will consider two values of (\mn/$^{55}$Mn)$_{\Delta_1}$,
specifically $10^{-5}$ and $10^{-4}$ (roughly covering the range of
data shown in Table \ref{mlt}). The higher value corresponds to the
high value in CAIs and is uncertain due to endemic nuclear effects in
CAIs. The lower value is somewhat above the highest value found in
planetary differentiates.

\subsection{\al}

\begin{figure}
\begin{center}
\end{center}
\caption{ a) Evolution diagram for an extinct nuclide with an initial
abundance ratio $(N^P/N^S)^0$. P and S are isotopes of the same
element. P decays to the daughter D of the same chemical element as
I. If a system were originally isotopically homogeneous but with
phases A, B, ... E having different ratios of the species S and I
(representing different elements) then the existence of P with an
abundance ($N^P/N^S$)$^0$ would require the data on A, B,... E to lie
on a straight line. b) The results of a CAI from Allende, showing the
excellent correlation of $^{26}$Mg/$^{24}$Mg with
$(^{26}\mathrm{Al}/^{27}\mathrm{Al})_0 = 5\times 10^{-5}$. Note small
but real deviations in the inset. After \cite{lpw77}.}\label{iso26}

\centerline{\epsfig{file=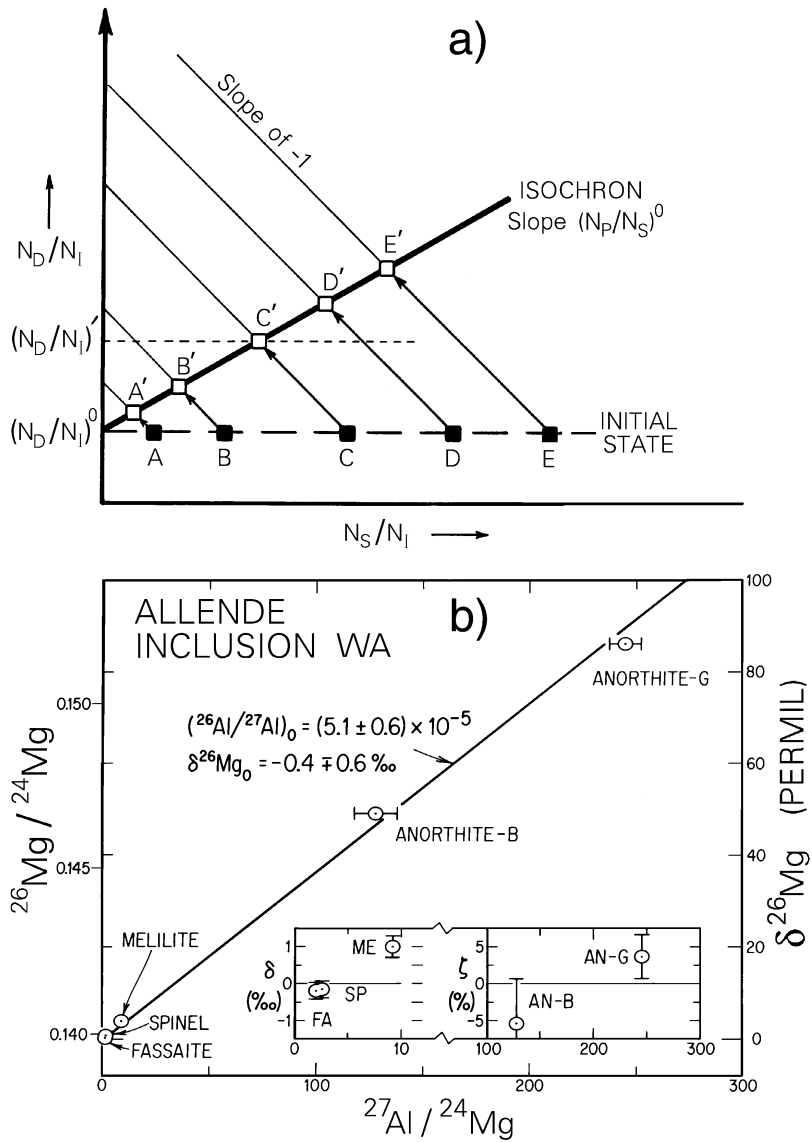,height=6in}}

\end{figure}

The decay of \al~ ($\bar\tau = 1.03$ Myr) to $^{26}$Mg has a special
role. The discovery of $^{26}$Mg$^\ast$ correlated with $^{27}$Al in
the CAIs from the Allende CV3 meteorite showed that a nuclide with the
chemistry of Al was present in the early solar system with an
abundance of $^{26}\mathrm{Al}/^{27}\mathrm{Al} = 5 \times 10^{-5}$ (see
Fig. \ref{iso26}). This is analogous to the $^{129}$I case discussed
earlier, but because of the short mean life of this nuclide, a rather
immediate synthesis of \al~ and rapid formation of the solar system is
required. The discovery of $^{26}$Al changed the time scales from
$\sim 10^2$ Myr to $\sim 1 $Myr. These results were followed by
studies that found evidence of \al~ in other CV meteorites
\cite{stegg} with abundant CAIs. A considerable number of analyses on
Allende and other CV meteorites showed that the evidence for \al~ in
CAIs was widespread, with high values of
$^{26}\mathrm{Al}/^{27}\mathrm{Al}\sim 5 \times 10^{-5}$ and many
lower values. The results compiled by MacPherson et al. \cite{mdz95}
showed that these observations are a general rule. The search for CAIs
in different chondrite classes showed that these lithic types were
almost universally present albeit in very small amounts. There was
also clear evidence of some live \al~\cite{bischoff} in these CAIs in
typical chondrites. It has now been shown that CAIs in chondrites
ranging from unequilibrated to equilibrated (i.e., varying degrees of
thermal metamorphism to permit element diffusion and some
recrystallization), all show varying degrees of \al/$^{27}$Al
(cf. \cite{huss01} \cite{guan00}). It follows that CAIs containing
\al~ with a maximum value of \al/$^{27}$Al$ \sim 5 \times 10^{-5}$ are
present in all chondritic meteorite classes.

The broader issue is how widespread the evidence is for \al~ in the
bulk material in the chondritic meteorites that essentially represent
the relative solar abundances of condensible elements. The most
abundant lithic components of chondrites are the
spherical-subspherical chondrules which are formed by crystallization
of liquid droplets of silicates.

The first definitive evidence for \al~ in a chondrule was discovered
by Hutcheon \& Hutchison \cite{hh89}. These workers found a clear
excess of $^{26}$Mg well correlated with Al/Mg and yielding
$^{26}\mathrm{Al}/^{27}\mathrm{Al} \approx 8 \times 10^{-6}$. This ratio
is a factor of 6 below the typical high values found in
CAIs. Intensive studies by other workers established that some
chondrules have evidence of \al~ but almost always in the range $5
\times 10^{-6}$ to $10^{-6}$. In most chondrules $^{26}$Mg$^\ast$ was
not detectable at the precision available (cf.
\cite{shw91}\cite{russ96} \cite{knt00}, see also discussion and Fig. 3
in \cite{md04}). The general rule was that chondrules were formed with
much less \al~ than typical CAIs. If the \al/$^{27}$Al ratios are
interpreted as due to a time difference, this would require that most
chondrules were formed or metamorphosed 1-5 Myr later than CAIs. We
note that the low \al~ found by Russell et al. \cite{russ96} is in a
sodium-rich glass in a chondrule.  This chondrule clearly did not form
under ``nebular'' conditions and, as Na is very volatile, required a
very high Na partial pressure in the gas phase to condense silicate
droplets with such a high Na concentration.  It is also evident that
there are CAIs with low or no \al. The issue is whether the lower
values represent radioactive decay with the passage of time or if they
represent the range of \al~ available from a state that was highly
heterogeneous with some materials that were formed without \al.  The
mechanisms for producing CAIs are then an issue. If one assumes a
homogeneous initial state, then it follows that CAIs can be formed at
later times. This is, of course, possible. If the systems were highly
heterogeneous (i.e., no well-defined initial state) then there is the
problem of explaining the rather sharp cut-off and peak at
\al/$^{27}$Al$ \approx 5 \times 10^{-5}$.

With significant improvements in the precision of Mg isotopic
measurements using MC-ICPMS techniques, it has become possible to
measure small Mg isotopic fractionation effects and to measure
$^{26}$Mg$^\ast$ in bulk chondrules with low $^{27}$Al/$^{24}$Mg
\cite{galy00}. These workers found evidence for \al~ in several
chondrules. A recent report by Bizzarro et al. \cite{bbh04} found that
\al~ was present in chondrules (in addition to CAIs) with
\al/$^{27}$Al ranging from $5.7 \times 10^{-5}$ to $1.4 \times
10^{-5}$. These are all important results and require considerable
care in verification; they have not been substantiated and may be
subject to serious revision.  Nonetheless, there is now abundant
evidence for the presence of \al~ in broadly distributed chondritic
material. In planetary differentiates, evidence for $^{26}$Al was
first found by Srinivasan et al. \cite{sgb99}. This report was
confirmed \cite{spw00}. However, the \al/$^{27}$Al in this planetary
differentiate is uncertain because $^{26}$Mg$^\ast$ was enriched
compared to $^{24}$Mg in the plagioclases, but was essentially uniform
(i.e. independent of the Al/Mg ratio) while the pyroxene shows normal
$^{26}$Mg/$^{24}$Mg. This system appears to be isotopically
homogenized within the plagioclase and indicates low temperature
metamorphism resulting in partial isotopic equilibration. Presence of
$^{26}$Al would not be detectable in the Mg-rich phases.

A most important study of a eucrite by Nyquist et al. \cite{nrw03}
reports an isochron with \mn/$^{55}$Mn = ($4.6\pm1.7$)$\times10^{-6}$
and also an $^{26}$Al isochron with \al/$^{27}$Al =
($1.18\pm0.14$)$\times10^{-6}$. A more recent report by Wadhwa et
al. \cite{wa05} using multicollector ICPMS techniques has confirmed
these results with much higher precision.  It also provided a
$^{207}$Pb/$^{206}$Pb age of (4565.03$\pm$0.85) Myr, improved in
precision due to using the computer printout.  These results
unambiguously demonstrate the presence of two short-lived nuclei in an
ancient basaltic meteorite in self-consistent amounts. Thus, there is
the general rule that both $^{26}$Al and $^{53}$Mn were present at
appropriate abundances in early planets. There must also have been
$^{60}$Fe present. These are the heat sources required for melting of
protoplanets \cite{ur55} and must have been widely and generally
distributed through the solar system.

In general, \al~ appears to have been present in CAIs, chondrules, and
planetary differentiates. Thus, it was widespread throughout condensed
matter in the solar system at different stages. The ranges in
\al/$^{27}$Al may reflect differences in time or an initial
heterogeneous distribution. If the source of \al~ is local (i.e.,
within the solar system), then it must provide this nuclide to the
bulk of materials forming ``condensed'' planetary matter. If the
source is an external one (e.g., an AGB star), then it was rather well
mixed with the material that formed solids.

The peak in the observed distribution of
$^{26}\mathrm{Al}/^{27}\mathrm{Al} = 5 \times 10^{-5}$ is called the
``canonical value''. The distribution below this peak is in general
easily attributable to redistribution of Mg as a result of
metamorphism, but some unmetamorphosed CAIs having almost no \al~
(e.g. HAL) do exist.  The latter observation requires that some CAIs
formed quite late, or that the distribution of \al~ was very
heterogeneous, with a sharp high peak. Also this well-defined and
highly populated peak shows a range: this could be due to metamorphism
from an initial state that had a uniform and fixed value of
\al/$^{27}$Al.  However, a study by El Goresy et al. \cite{eaw85}
showed that one unaltered CAI was made up of two petrographically
distinct CAI units that were later entrapped in a third liquid of CAI
composition. Isotopic studies of these lithic subunits showed that
they had clearly distinguishable \al/$^{27}$Al ratios (3.3 to
6)$\times$10$^{-5}$ that correlated with the sequence defined
petrographically \cite{hwh00}. If they started with the same
\al/$^{27}$Al initial value, then this requires that these three units
(in the same droplet) formed over a time interval of $\approx 4\times
10^{5}$ yr. Alternatively, they could very possibly represent multiple
droplet formation in a single event, but from a starting material that
was heterogeneous.  Recent results by Galy, Hutcheon \& Grossman
\cite{galy04}, by Liu, Iizuka \& McKeegan \cite{liu05} and by Young et
al. \cite{y05} using ``bulk'' CAI samples in a new generation of
high-precision measurements show that the spread in \al/$^{27}$Al is
rather general, with some samples having a ``supra-canonical'' value
of \al/$^{27}$Al $\approx 6 \times 10^{-5}$.  These more recent
studies point to a higher value of $^{26}\mathrm{Al}/^{27}\mathrm{Al}$
in the ESS.  However, there are no data indicating \al/$^{27}$Al
ratios that are a factor of two or more higher than the canonical
value.

With regard to measurements on ``bulk'' inclusions, these results are
like ``total rock'' ages and require considerable attention.  Shifts
in the initial $^{26}\mathrm{Mg}/^{24}\mathrm{Mg}$ of 0.02\% would
cause 10--20\% shifts in the estimated
$^{26}\mathrm{Al}/^{27}\mathrm{Al}$.  The argument that the ``total''
systems are closed is not readily defensible.  The reported effects
appear to be real and suggest that variations in the initial
$^{26}\mathrm{Mg}/^{24}\mathrm{Mg}$ should be found in careful studies
of internal isochrons.  There is certainly evidence of ``subnormal''
$^{26}\mathrm{Mg}/^{24}\mathrm{Mg}$ ($\sim 1$ per mil) in pyroxenes in
some Type B CAIs as found using TIMS techniques.  These effects are an
order of magnitude greater than from the $^{26}$Al inventory (see Fig.
10 in \cite{wp82}).  A re-examination of CAI phases appears to be in
order using improved techniques and considering the presence of
nuclear anomalies.

The question remains as to whether this range in \al/$^{27}$Al is due
to somewhat incomplete mixing of presolar stellar debris (a small
range in the dilution factor), or if the \al~ is produced by an
extremely intense irradiation, with some small local differences in
the shielding and thorough mixing, but not producing ``hot spots.''

In our discussion we treated the \al~ problem as if the bulk solar
system initially had (\al/$^{27}$Al)$_{\odot} \cong 5 \times
10^{-5}$. A small local variation in the dilution factor (e.g.
incomplete mixing) could readily account for small deviations.  The
range of possible shifts to higher values mentioned above will not
greatly affect the arguments given here. The fundamental question is:
do the samples represent the bulk solar value or, do they represent
only solids and materials in parts of the accretion disc?

\subsection{\ca}
\begin{figure}
\begin{center}
\end{center}
\caption{Demonstration of the presence of $^{41}$Ca in the ESS and its
correlation with \al. Panel a) shows Al-Mg data for the CAI samples
analyzed from different meteorites which exhibit the presence of
$^{26}$Al, with $^{26}$Al/$^{27}$Al = 5$\times$10$^{-5}$ (filled
symbols). Note that some samples (open symbols) show no evidence of
$^{26}$Al.  Panel b) shows $^{41}$K/$^{39}$K measurements on the same
samples plotted versus $^{40}$Ca/$^{39}$K$\times$10$^6$. The line
showing $^{41}$Ca/$^{40}$Ca = 1.4$\times$10$^{-8}$ is from the
original reports \cite{sug94} \cite{sri96}. The other data are from
Sahijpal et al. \cite{sgd98}. These workers showed that the samples
with $^{26}$Al had $^{41}$Ca and that those without $^{26}$Al had no
$^{41}$Ca. Note that there is limited data. The line corresponds to
radioactive decay from an initial state with \ca/$^{40}$Ca$ = 1.4
\times 10^{-8}$. These data show that \al~ and \ca~ are well
correlated and must be either co-produced or well-mixed after their
production.  Figure reproduced from Ref. \cite{sgd98}, copyright
Nature Publishing Group.}\label{iso41}

\vskip 1in
\centerline{\epsfig{file=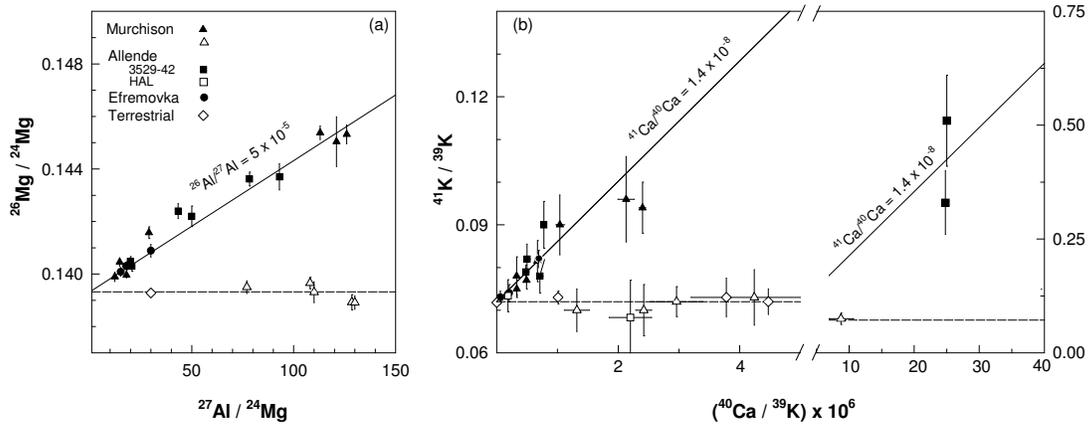,angle=270,width=6in}}

\end{figure}

The first hint of \ca~ was found by Hutcheon et al. \cite{haw84}, who
could only establish an upper bound of \ca/$^{40}$Ca $\le$ (8$\pm$3)
$\times 10^{-9}$ in CAIs. Clear evidence for the presence and
abundance of \ca~ in the ESS was found in CAIs by Srinivasan et
al. \cite{sug94} \cite{sri96}.  Further evidence was found by Sahijpal
et al. \cite{sgd98} and a good correlation line was found for hibonite
crystals in CAIs from Murchison, Allende and Efremovka. These results
showed \ca/$^{40}$Ca $= 1.4 \times 10^{-8}$ for CAIs that had
\al/$^{27}$Al $\cong 5 \times 10^{-5}$. The hibonite samples that
showed essentially no \al~ also showed no $^{41}$K$^\ast$ (see
Fig. \ref{iso41}). These important experiments showed that \ca~ and
\al~ were present at the same time, and also absent at the same
time. Generalizing from these observations, it is necessary that both
\al~ and \ca~ had to be present at fixed abundance levels at some
early time. Therefore, the mechanisms responsible for these two nuclei
appear to be coupled.

The very short lifetime of \ca~ ($\bar\tau_{^{41}\mathrm{Ca}} = 0.15$
Myr) makes it particularly important. It can be produced both in AGB
stars and by proton irradiation. Thus it might be diagnostic of an
irradiation model if the $\Delta_1$ inferred from an AGB source were
so short as to be dynamically unreasonable.  However, \ca~ is
abundantly produced in AGB stars, with \ca/$^{40}$Ca $\sim 2 \times
10^{-2}$ in the production zone, as can be seen in \cite{bgw99}.  Even
for a dilution of $5 \times 10^{-3}$, $\Delta_1$ seems not to be
forced to implausibly low values by the abundance level of $^{41}$Ca
in the ESS, as long as the deduced $\Delta_1$ = 0.5 to 0.8 Myr
suffices for injection and collapse.

\subsection{$^{36}$Cl}\label{36cl}

The nuclide $^{36}$Cl ($\bar\tau_{^{36}\mathrm{Cl}} = 0.43$ Myr) has a
branched decay $(\beta^- $ to $^{36}$Ar and $\beta^+$ to $^{36}$S)
with the $e.c./\beta^+$ branch having only a 1.9\% yield. Efforts to
find evidence for this nuclide were made by several groups. A study by
G\"obel, Begemann and Ott \cite{gbo82} of $^{36}$Ar abundances in
Allende samples (including CAIs) was done along with Cl
measurements. These workers found large shifts in $^{36}$Ar/$^{38}$Ar
(up to 90 at intermediate-temperature releases).  These results
correlated with Cl and $^{60}$Co. If interpreted as due to $^{36}$Cl
decay, their results would give \cl/$^{35}$Cl $\sim 2\times 10^{-8}$
(U. Ott, personal communication).  A report by Villa et
al. \cite{villa81} found a sodalite-rich fine-grained inclusion called
``the Pink Angel'' that gave $^{36}\mathrm{Cl}/^{35}\mathrm{Cl} =
(0.2-1.2)\times 10^{-8}$.  This is the same material that gave
$^{129}\mathrm{I}/^{127}\mathrm{I} \approx 1.0\times 10^{-4}$.  A more recent
effort \cite{mur97} also found effects in $^{36}$Ar that were
attributed to \cl. However, no strong case could be made for \cl~ from
this report. Interpretation of these data is unclear \cite{rai03}.
Recent reports by Lin et al. \cite{lgl04}\cite{lgl05} demonstrated an
impressive correlation of $^{36}$S/$^{34}$S with $^{35}$Cl/$^{34}$S in
late-formed halogen-rich phases in CAIs and showed it to be decoupled
from $^{26}$Al.  The results appear to demonstrate the presence of
$^{36}$Cl in the ESS with the very high values $^{36}$Cl/$^{35}$Cl
$\cong 5 \times 10^{-6}$. If $\Delta_1 + \Delta_2$ = (0.5 -- 1) Myr,
then the required amount is outside the range possible for an AGB
source or an SN II source.  We note that the problem of isobaric
interferences of HCl with $^{36}$Cl requires attention. This is a well
known and persistent interference and might occur during
sputtering. This would of course give a very good correlation of
H$^{35}$Cl with $^{35}$Cl.  The mass resolution used by Lin et
al. \cite{lgl05} should resolve this isobaric interference.  However,
because of the importance of this result, such experimental matters
should be specifically addressed.  The obvious conflict between the
earlier reports and that of \cite{lgl05} is not easily reconcilable.
The results by \cite{lgl04} and \cite{lgl05} are, if correct, of
considerable importance and should be verified by further experiments
on halogen-rich alteration phases in other CAIs and in appropriate
phases in chondrites. Measurements of halite as done by Whitby et
al. \cite{whit0} for $^{129}$I should be carried out. There is
difficulty in assigning a ``time'' $\Delta_2$ to these alteration
phases. It may be possible to make determinations on phosphates or
other minerals rich in halogens or other minerals where independent
``age'' assignments can be made from the available data.  Very low or
no \al~ has been previously detected in sodalites, as was also found
in \cite{lgl05}, who showed that major phases with no Cl show the
canonical value of \al\ in the same CAI.  From the evidence available,
it appears that \cl, like \be, is decoupled from the production of
\al.  There is clear evidence of $^{129}$I/$^{127}$I $ = 1.0 \times
10^{-4}$ in CAI sodalite, but the $^{129}$I life time is far too long
to be useful. Correlation of $^{36}$Cl with other short-lived species
is much needed.

Inspection of Tables \ref{qenv} and \ref{nrenv} shows that in AGB
stars ($^{36}$Cl/$^{35}$Cl)$_{ENV}$ is not dependent on the presence
or absence of a \ct~ pocket or on $Z$ for all stellar masses and $Z$
ranges studied here. Inspection of Tables \ref{ejecta} and \ref{cloud}
shows that ($^{36}$Cl/$^{35}$Cl)$_{\Delta_1 = 0} = (1.7 - 5.4) \times
10^{-6}$ for all cases. However, due to the short lifetime, it drops
to $(4.7 - 9.4) \times 10^{-7}$ for all LMS cases if $\Delta_1 \sim
0.7$ Myr. Thus it is not evident that an AGB source could provide the
inventory that appears to be required by the data reported by Lin et
al. \cite{lgl04} \cite{lgl05}. If $\Delta_1 + \Delta_2 = (1-2)$ Myr,
then the values obtained here and in our earlier calculations
\cite{wbgr94} \cite{bgw99} would be low by factors from 10 to 100.
Attribution of the source of $^{36}$Cl to SNe is not appropriate,
considering the arguments given in Section \ref{SNeII} -- the
$^{36}$Cl case is much like that of $^{26}$Al for a SN source. 

The basic issue is where the volatile halogen-rich alteration, common
in CAIs, took place. It is not at all evident that this is the result
of nebular processes. It might be the result of metamorphism and
transport of volatile-rich materials from a heated interior
penetrating the near surface layers of a protoplanet heated by
$^{26}$Al and $^{60}$Fe.

There is a further complexity related to the production of Cl isotopes
from the abundant $^{32}$S seeds. This involves branching points at
$^{33}$S, $^{35}$S, $^{35}$Cl, $^{36}$Cl, and $^{37}$Cl.  These
involve (n,$\gamma$) (n,p), and (n,$\alpha$) captures competing with
$\beta^\pm$ decays and electron captures. This is a complex issue as
it depends on the decomposition of $s$-processes into main and weak
components, and the general modeling of the total $s$ inventory for
these nuclei. The calculations used here were based on the cross
section measurements for $^{34}$S by Reifarth et al. \cite{rsk00} (see
full discussion in that work) and the revised solar abundances
\cite{lod03}. For more details on the nuclear parameters and
abundances of the isotopes involved in this region of the $s$-path,
see Mauersberger et al. \cite{mau04}. It remains to be seen whether
the production of $^{36}$Cl in AGB stars could conceivably be greatly
increased by possible revisions to the model.

The level of $^{36}$Cl production by irradiation of dust has been
presented by Marhas \& Goswami \cite{mg04} and Leya et al.
\cite{lhw03}. An extensive analysis of an irradiation model that does
not consider $^{36}$Cl is given in \cite{mgs01}. If the Lin et
al. \cite{lgl04} \cite{lgl05} results are confirmed, then the
theoretical treatment of an irradiation model must be reconsidered as
the level of $^{36}$Cl production may have to be higher in comparison
with other short-lived nuclei. In one report \cite{lhw03} the fluence
required to provide the $^{10}$Be appears to be commensurate with that
needed to produce the $^{36}$Cl abundances of Lin et al. \cite{lgl04}
\cite{lgl05}. This gives $^{36}$Cl/$^{35}$Cl $= 1.3 \times 10^{-4}$
for the case without saturation, but there is also a considerable
shortfall in \al/$^{27}$Al, which is in accord with the absence of
\al\ in the sodalite with abundant \cl. In short, there is a strong
need for more complete and definitive measurements of the $^{36}$Cl
abundance to guide future work and more efforts at modelling the
production of these short-lived nuclei by particle bombardment.
Calculations directed to $^{36}$Cl, $^{10}$Be, and $^{26}$Al
production at the level $^{36}\mathrm{Cl}/^{35}\mathrm{Cl}\sim
10^{-5}$ would be useful.  Careful consideration as to the conditions
under which volatile-rich alteration processes might take place is
urgently needed.

{\it Note added in proof:} The presence of $^{36}$Cl in the Pink Angel (see
Section 11.5) at the level of $^{36}$Cl/$^{35}$Cl=$(4\pm 1)\times
10^{-6}$ has been established by W. Hsu et al. (personal
communication).  This sample has $^{26}$Al/$^{27}$Al $<2\times
10^{-6}$.  The $^{36}$Cl is not correlated with the $^{26}$Al.  This
confirms the results by Lin et al. [252].  These results show the
necessity of energetic particle bombardment in the ESS as shown by the
$^{10}$Be results [50].  Further, no $^{36}$Ar was present in the Pink
Angel, so almost complete loss of $^{36}$Ar is required.

\subsection{\be\ and Irradiation}

\begin{figure}
\begin{center}
\end{center}
\caption{Results by McKeegan, Chaussidon, and Robert \cite{mcr00}
proving the presence of $^{10}$Be in the ESS with an abundance of
$^{10}$Be/$^{9}$Be $\sim$ 10$^{-3}$. The insert shows data at low
$^{9}$Be/$^{11}$B values. The $^{10}$Be is only produced by
particle irradiation, not by stellar nucleosynthesis.  Reprinted
with permission from Ref. \cite{mcr00}.  Copyright (2000) AAAS.}
\label{iso10}

\centerline{\epsfig{file=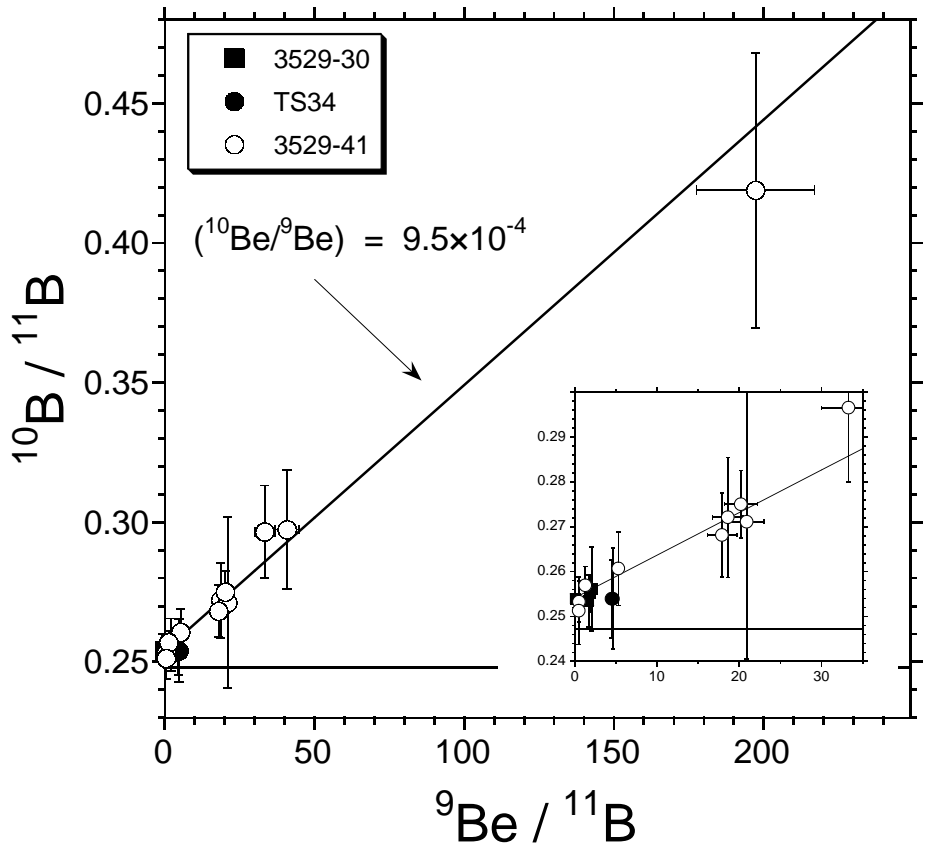,width=6in}}

\end{figure}

The nuclide \be~ $\beta^-$ decays to $^{10}$B
$(\bar\tau_{^{10}\mathrm{Be}} = 2.3 \times 10^6$ yr) and is not the
product of stellar nucleosynthesis. It is produced by energetic H and
He nuclei by spallation reactions on a variety of targets (mostly
O). The discovery of \be~ in ESS materials by McKeegan, Chaussidon \&
Robert \cite{mcr00} in meteorites demonstrates the significance of
early energetic particle bombardment by cosmic rays in providing
short-lived nuclei. These cosmic rays may be from the T-Tauri phase of
the proto-sun, or from outside sources \cite{dcs04}. As shown by
McKeegan et al.  \cite{mcr00}, there is a clear correlation of
$^{10}$B/$^{11}$B with $^9$Be/$^{11}$B in several CAIs. This clearly
establishes the presence of \be~in the ESS with \be/$^9$Be $= 9.5
\times 10^{-4}$ as shown in Fig. \ref{iso10}.  These CAIs were
previously shown to have \al/$^{27}$Al $\approx 5 \times 10^{-5}$. The
fundamental question is whether the irradiation that produced the \be~
could also be responsible for other short-lived nuclei (in particular
\al, $^{36}$Cl, and \ca). We note that the total number of \be~nuclei
is very small in any sample.

\begin{figure}
\begin{center}
\end{center}
\caption{Comparison of $^{10}$Be/$^{9}$Be and $^{26}$Al/$^{27}$Al in
refractory inclusion data from several publications, compiled in
Ref. \cite{md04}. These results show that samples with low or very low
$^{26}$Al/$^{27}$Al contain $^{10}$Be/$^{9}$Be at a level compatible
to that found by McKeegan et al. \cite{mcr00}. This shows that
$^{10}$Be may be present when $^{26}$Al is absent. The production
mechanisms or sites thereby appear unrelated.  Note that there is only
a limited amount of data. The full line is the time trajectory from an
initial state with $^{26}\mathrm{Al}/^{27}\mathrm{Al} = 4.5 \times
10^{-5}$ and $^{10}$Be/$^9$Be$ = 6.7 \times 10^{-4}$.  (Reprinted from
Ref. \cite{md04}, Copyright (2003), with permission from Elsevier.)}
\label{1026}

\centerline{\epsfig{file=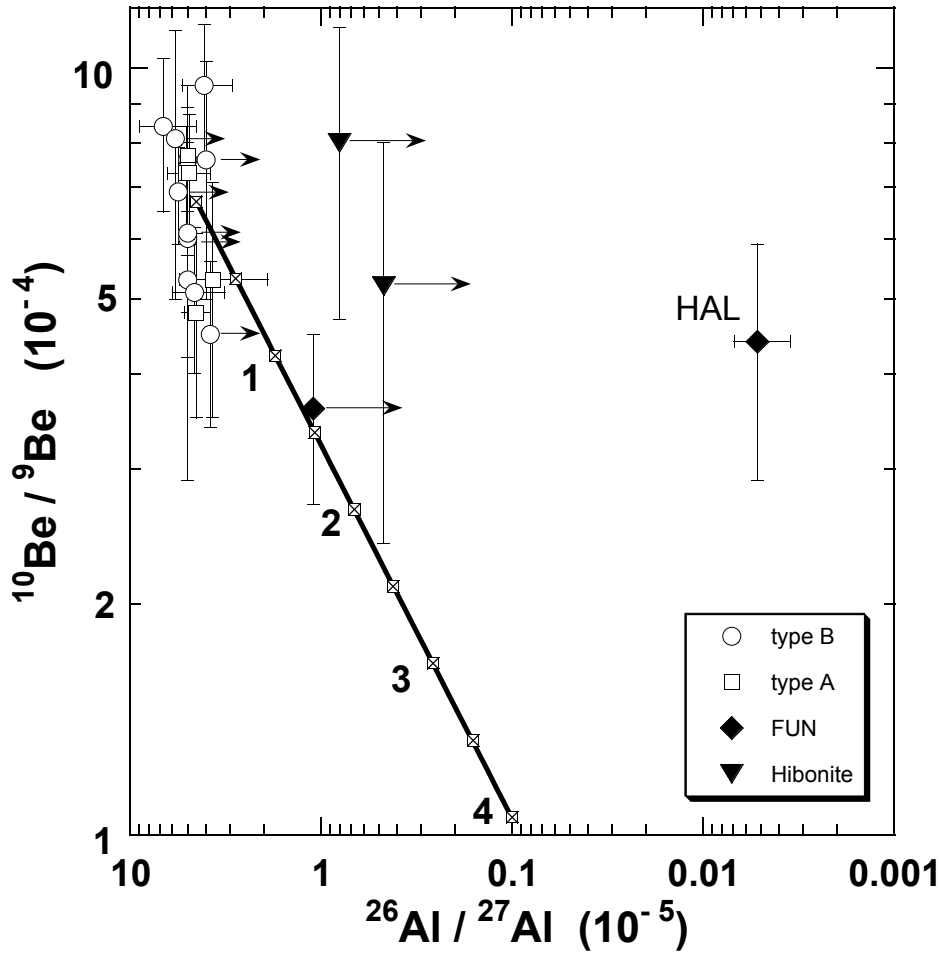,width=6in}}

\end{figure}

To establish whether there is a correlation between \be\ and \al\ or
\ca, studies of ultra-refractory phases in CAIs with varying levels of
\al~ and \ca~ were conducted by Marhas et al.  \cite{mar02}
\cite{mg03} \cite{mgd02}. In particular, Marhas \& Goswami \cite{mg03}
found that the well known FUN inclusion HAL and a few other CAIs had
\be/$^9$Be similar to what was found in other CAIs but with two to
three orders of magnitude lower \al/$^{27}$Al. This demonstrated the
decoupling of \be~ from \al. The available data on \be~vs. \al~are
shown in Fig.  \ref{1026}. The conclusion from that presentation is
that $^{10}$Be and $^{26}$Al are decoupled.  This decoupling from \al\
is also supported by the results for \cl\ \cite{lgl05}.  No single or
uniform solar cosmic ray bombardment can explain these results, so
some much more complicated scenarios must be considered.

There are now data indicating that \al/$^{27}$Al $\sim 5 \times
10^{-5}$ in some chondrules \cite{bbh04}. The search for \be~ should
be extended to such objects. Certainly the \fe~ and \pd~ and possibly
\mn~ require extra-solar-system nucleosynthetic sources. The isotopes
\pd, \mn, and \al~ are found in planetary differentiates and require a
significant inventory. The demonstration of both \al~ and \mn~ in the
same eucrites by Nyquist et al. \cite{nrw03} and the possible
correlation with $^{207}$Pb/$^{206}$Pb ages is most intriguing. It now
appears clear that the material in differentiated planetary bodies had
significant inventories of short-lived nuclei. These nuclei must have
been present in the precursor dust and other material at sufficient
levels so that they were still present during planetesimal melting.
This melting must be attributed to radioactive heating by \al~ and
some \fe. These nuclei must then have been generally present in the
planet-forming material of the ESS. If \al~ is from an irradiation
source, then this heat source must be supplied to the condensed matter
making planets at least at a level of \al/$^{27}$Al $\sim$ 10$^{-5}$.

Models of irradiation of solids in the ESS have been extensively
discussed by several workers \cite{ssl96} \cite{mg04}
\cite{lhw03} \cite{ssg97} \cite{ssgg01}. There has been a major
effort to seek an explanation for a large number of the
short-lived nuclei by proto-solar cosmic rays impinging on dust or
small rocks. There has been a particular emphasis on the X-wind
model of Shu et al. \cite{ssl96}.  Gounelle et al. \cite{gss01}
have explored this matter using different irradiation parameters
in consideration of the \be~ issue and with efforts to match the
observed values of other isotopes.

It has been proposed that $^{10}$Be might have its origin in
galactic cosmic ray bombardment, during the collapse phase
\cite{dcs04}. This idea is very appealing; however, it does not
appear to explain the lack of correlation of $^{10}$Be with \al.
Indeed, if all of the material were well mixed, these isotopes
would still be correlated.

The fundamental issue is whether some small fraction of debris in the
early solar system was irradiated in several different episodes and
not responsible for the average inventory of \al\ and \mn, or if the
irradiation was extensive and accounts for a substantial part of what
we assume is the bulk solar inventory. In any case, the irradiation
episodes had to be in a series of events to give the relationships
found for $^{10}$Be, \al~ and \ca.  The irradiation conditions
proposed should be directed toward explaining and addressing the
observations made on meteorites, including the decoupled production of
inventories of the several nuclei.

\subsection{\hf}

We review the situation concerning $^{182}$Hf, which $\beta^-$ decays
to $^{182}$W.  A greatly improved measurement of
$\bar\tau_{^{182}\mathrm{Hf}}$ by Vockenhuber et al. \cite{vob04} has
now established $\bar\tau_{^{182}\mathrm{Hf}} = 12.8 \pm 0.1$ Myr. The
presence of $^{182}$Hf in the early solar system was first determined
by Harper \& Jacobsen \cite{hj96} and by Lee \& Halliday \cite{lh95}
from the deficiency of $^{182}$W/$^{184}$W in iron meteorites. For
some iron meteorites $^{182}$W/$^{184}$W was found to be $4~
\epsilon$u lower than the values for the earth's crust. The inferred
ESS abundance was $^{182}$Hf/$^{180}$Hf $\approx 2.84 \times
10^{-4}$. More recent studies by Kleine et al. \cite{kle02} and Yin et
al. \cite{yin02} have demonstrated from internal isochrons on two
chondrites and data on bulk samples of ordinary chondrites, a eucrite,
and CAIs, that the \hf~ is well-correlated with Hf-W
fractionation. These workers showed that the abundance of
$^{182}$Hf/$^{180}$Hf $= (1.00 \pm 0.08) \times 10^{-4}$ and that the
initial $(^{182}$W/$^{183}$W)$_{ESS}$ corresponds to $\epsilon_W =
-3.4$ relative to samples of terrestrial rocks. It was further shown
by these workers that the terrestrial samples are enriched in
$^{182}$W relative to the bulk solar values resulting from the
chemical fractionation of Hf and W between the earth 's metallic core
and the silicate mantle while \hf~ was still present (see Jacobsen
\cite{jsb05}). These important results significantly change the ESS
inventory of $^{182}$Hf shifting $^{182}$Hf/$^{180}$Hf to a lower
value by a factor of $\sim 2.8$ compared to the original estimate.

A well-defined study of Hf-W by Srinivasan et al. \cite{sri04} on
zircon crystals (ZrSiO$_4$, rich in Hf) from eucritic meteorites
showed clear large excesses of $^{182}$W that correlated with Hf/W
and gave $^{182}$Hf/$^{180}$Hf$ = 1.4 \times 10^{-4}$. This is in
good general agreement with the higher results cited above. This
study is another example where phases with enormous enrichments of
parent element relative to daughter element in ancient materials
permit large isotopic effects to be observed and avoid dependence
on small differences. The matter of $\Delta_2^{^{182}\mathrm{Hf}}$ then
remains an issue.

Studies of AGB sources of $^{182}$Hf showed that it was not
possible to produce the observed $^{182}$Hf/$^{180}$Hf, by a
factor of $\sim 10^{-2}$ \cite{wbgr94}. The new results cited
above do not alter this conclusion.  However, they are pertinent
to the question of diverse SN sources for $r$-process nuclei.  In
their original report, \cite{wbg96} showed for a model of uniform
production of $r$-process nuclei that
$(^{182}$Hf/$^{180}$Hf)$_{\rm UP} = 4.8 \times 10^{-4}$.  This was
based on an estimate of 57\% for the $r$ contribution to $^{182}$W
obtained by Gallino et al. \cite{gabl98} from updated cross
sections and AGB models.  (The previous best value, estimated by
K\"appeler, Beer \& Wisshak \cite{kbw89}, had been 33 \%.)  The
revised measurement of $(^{182}$Hf/$^{180}$Hf$ )_{ESS} = 1.0
\times 10^{-4}$ discussed above still remains in sharp contrast to
that obtained for the time for free decay of $^{129}$I required by
$(^{129}$I/$^{127}$I$)_{ESS} = 1.0 \times 10^{-4}$.  The time required
 of $\Delta_1^{^{129}{\rm I}} \sim 70 \times 10^6$ yr
would give $(^{182}$Hf/$^{180}$Hf)$_{\rm UP} \times e^{-\Delta_1 /
\bar\tau_{182}} = 2 \times 10^{-6}$.  It follows that the
$r$-process source of $^{182}$Hf and that of $^{129}$I are not the
same, as argued by \cite{wbg96}. Further, the new value of
$(^{182}$Hf/$^{180}$Hf$ )_{ESS}$, combined with a continuous
production model, requires that the time interval between the last
$r$-process contribution to solar \hf~ and the formation of solids
was $18 \times 10^6$ years.  It is now important to re-evaluate
the $r$ and $s$ contributions to $^{182}$W. A significant shift in
the value of $^{182}$W$_r$ used would change the time scale for
$\Delta^{^{182}{\rm Hf}}_1$.

\subsection{$^{92}$Nb}

Evidence for the $p$ process nuclide \nb~ was shown by Harper
\cite{harp96}, who found enhanced $^{92}$Zr/$^{90}$Zr in rutile
(TiO$_2$) with a high value of Nb/Zr in an iron meteorite with
silicate inclusions. Later work by Yin et al. \cite{yin00} found clear
excesses of $^{92}$Zr in a rutile. M\"unker et al.  \cite{mwm00} found
a deficiency of $^{92}$Zr in some CAIs enriched in Zr relative to Nb
and an excess of $^{92}$Zr in phases with high Nb relative to
Zr. There is a serious conflict regarding the initial \nb/$^{93}$Nb
ratio. The value of \nb/$^{93}$Nb determined by \cite{yin00} and
\cite{mwm00} was two orders of magnitude higher than reported by
Harper \cite{harp96}. It is difficult to attribute this value to
differences in $\Delta_2$, as $\bar\tau_{^{92}{\rm Nb}} \approx 52$
Myr, unless the sample analyzed by Harper \cite{harp96} was formed
very late ($\Delta_2 > 100$ Myr). A further conflict has appeared with
the report of Sch\"{o}nb\"achler et al. \cite{sch02}, who determined
internal isochrons on three meteorites giving results in clear support
of \cite{harp96}. This matter requires further study, particularly of
mineral phases showing substantial Nb/Zr fractionation. It is not
evident whether the conflict is due to analytical difficulties or to
element redistribution. The possible cosmo-chronologic implications of
\nb~ were early considered by Minster \& All\`{e}gre
\cite{ma82}. However, as the yield of \nb~ is quite unknown in terms
of any nuclear systematics, and there is no other Nb $p$-only isotope,
it is only possible to come to broad chronologic conclusions.

We note that $^{93}$Nb is dominantly produced by $s$-processing in
AGB stars, due to the decay of $^{93}$Zr in the ISM after the AGB
phase has ceased. The ratio for UP is:
\begin{equation}\label{nb}
\left ( {^{92}\mathrm{Nb} \over ^{93}\mathrm{Nb}} \right )_{UP} =
{{\bar \tau_{^{92}\mathrm{Nb}}} \over T} \left ({p^{^{92}\mathrm{Nb}}
\over p^{^{93}\mathrm{Nb}}}\right )_p f^{^{93}\mathrm{Nb}}_p = 5\times
10^{-3} \left ({p^{^{92}\mathrm{Nb}} \over p^{^{93}\mathrm{Nb}}}\right
)_p f^{^{93}\mathrm{Nb}}_p
\end{equation}

Here $({p^{^{92}{\rm Nb}}/p^{^{93}{\rm Nb}}})_p$ is the ratio of the
$p$-process production rates and $f^{^{93}{\rm Nb}}_p$ is the fraction
of $^{93}$Nb from the $p$ process. From consideration of the $s$
process it is reasonable to assume $f^{^{93}{\rm Nb}}_p \sim 0.15$
\cite{ar99}. A much larger value would cause conflicts with the Zr
isotope systematics. Taking $({p^{^{92}{\rm Nb}}/p^{^{93}{\rm Nb}}})_p
\sim 1$, the expected ratio is $\sim 10^{-3}$, roughly matching some
of the values claimed in the literature.  A much larger value is not
easily understood.  This matter requires clarification by well-defined
efforts to find internal isochrons.

\section{The Actinides}

There are some serious problems relating to models of actinide
production in ``the $r$-process'' and the observed relative abundances
of the nuclides $^{232}$Th, $^{238}$U, $^{235}$U, \pu, and \cm~ in
meteorites. The production rates of the actinides are not fully
understood in terms of models of $r$-process nucleosynthesis (see
e.g. Goriely and Arnould \cite{ga01}). For an extensive review see
Cowan et al. \cite{cpk99} and Thielemann et al. \cite{thi02}.  For
more general considerations see \cite{qian02} regarding the problems
of $r$-process nucleosynthesis, including the role of neutrinos after
freezeout, and fission recycling.  Observations on low-metallicity
stars by Hill et al. \cite{hill01} show clear evidence of non-solar
$r$-process yields at high U-Th abundances. From considerations of
theoretical studies we note that the yields of the actinides from
their neutron rich precursors depend on models of nuclei that are far
off the trajectory of beta stability, and the masses and nuclear
structure of these precursor nuclei are not known.  The usual approach
to evaluate the abundances along the neutron flow path has been to
assume that the abundance pattern of the precursors in the actinide
region is smooth and slowly varying. The approach first used in
Burbidge, Burbidge, Fowler, \& Hoyle \cite{b2fh} for the actinides
assumes that the precursors, very neutron-rich nuclei, are essentially
in equal abundance along the flow path and that the number of source
nuclei may be obtained by summing the number of progenitors that do
not fission and may decay into the longer-lived actinides listed
above.

In all calculations presented here, we have used relative yields for
the actinides and trans-actinides from Seeger et al.  \cite{sfc65},
including the estimates of odd-even effects. A recent calculation of
the relative abundances of the actinides and trans-actinides by
Lingenfelter et al. \cite{ling03} reports essentially the same values
as used here by counting progenitors following \cite{b2fh} and
\cite{sfc65}. The yields calculated by \cite{ling03} were obtained in
a parametric $r$-process study, using a detailed model of neutron flow
and beta decay and assuming that this process takes place within a
core collapse SN, in a high-entropy region.

 We recognize that these quantities are not well-defined in terms
either of experimental observations or of pure theory. Nonetheless,
the implications of these values with respect to ESS values and
implications for nucleosynthesis are very large. Thus we will pursue
this approach assuming that all the estimates are correct. Note that
\pu~ is in the chain feeding $^{232}$Th and $^{247}$Cm is in the chain
feeding $^{235}$U.

Observations on phases in meteorites give (\pu/$^{238}$U)$_{ESS} \sim
6 \times 10^{-3}$ but with some range. A very important result by
Turner et al. \cite{thh04} on ancient terrestrial zircon samples
unequivocally demonstrates the presence of \pu~ in the early earth and
gives a value (\pu/$^{238}$U)$_{ESS} = 0.0066 \pm 0.0010$ at 4.56 Gyr
ago. This is important in two ways: first, it confirms the values
obtained from meteorites; second, it requires that \pu~ be taken into
account in all models of terrestrial evolution, in particular with
consideration of the isotopic abundances of $^4$He and Xe in the
earth's interior \cite{pw95}. The actual value of
(\pu/$^{232}$Th)$_{ESS}$ or (\pu/$^{238}$U)$_{ESS}$ depends on the
fractionation factor for Pu relative to Th and U in the samples
studied. A comparative study of \pu~ fission products, U, Th, and
light rare earth elements (LREE) was done to better assess
fractionation effects and to obtain a reliable estimate of Pu and
Pu/Nd (cf \cite{hag90}).  This general fractionation issue has been
discussed extensively by Burnett et al. \cite{bsj82}. Direct
experimental work on fractionation between phases by Jones \& Burnett
\cite{jb87} showed that Pu behaves like a LREE and Pu/Sm can be
enhanced by a factor $\sim 2$.  Again, the measured value of
\pu/$^{238}$U may be shifted from the bulk solar system value by
chemical fractionation, and this remains as an uncertainty in all
estimates. As \pu~ was present but is now extinct in the meteorite
samples, all correlations of the chemistry must use a surrogate
element that is now present and behaves like Pu (see
\cite{hag90}). This is not an unambiguous assignment. Respecting this
uncertainty, the measured values are all about a factor of two below
the UP values (see also Table \ref{mlt}). This could be due to: 1)
errors in the assumed relative yield pattern at \pu; 2) a time delay
of $\sim 10^8$ yr prior to solar system formation and the last
production of actinides; or 3) an error in estimating the chemical
fractionation of Pu and Th in the meteoritic mineral samples
analyzed. The results {\it per se} suggest a very long $\Delta_1 \sim
10^8$ yr for \pu.

With regard to \cm, if there are samples of mineral phases or bulk
meteorites in which there is substantial fractionation of Cm relative
to U by a factor $F_{\mathrm{Cm/U}}$ over the bulk solar system value,
then the fractional shift in $^{235}$U/$^{238}$U,
$\delta(^{235}\mathrm{U}/^{238}\mathrm{U})/(^{235}\mathrm{U}/^{238}\mathrm{U})
_{ESS}$, observed today is
\begin{equation}
\frac{\delta(^{235}\mathrm{U}/^{238}\mathrm{U})}{(^{235}\mathrm{U}/^{238}
\mathrm{U})_{\rm
ESS}} = (^{247}\mathrm{Cm}/^{235}\mathrm{U})_{ESS} F_{\mathrm{Cm/U}}
\end{equation}
(cf. \cite{bs73}). Estimates of $F$ depend on using a surrogate
element for Cm. From the upper bounds given for the variations of
$^{235}$U/$^{238}$U of $4 \times 10^{-3}$ \cite{cw81a} \cite{cw81b},
and using a reasonable estimate of $F$, (\cm/$^{235}$U)$_{ESS} << 4
\times 10^{-3}$, or more strictly $< 2 \times 10^{-3}$, was obtained.
The samples included meteoritic phosphates with initially abundant
\pu~ present and ages of 4.56 Gyr with large fractionation of Pu, U,
Th and REE. This value may be compared with (\cm/$^{235}$U)$_{\rm UP}
= 8.9 \times 10^{-3}$ (see Table \ref{mlt}) which would imply
$\Delta_1 > 2 \times 10^7$ years. The very recent results by Stirling
et al. \cite{stir05} on bulk chondritic meteorites using Nd as the
surrogate with a range of U/Nd of a factor of 3.5 indicate
(\cm/$^{235}$U)$_{ESS} \leq 10^{-4}$. This result would require
$\Delta_1 > 5\times 10^7$ yr and is in accordance with the \pu~
results. In no way could the gross deficiency of \cm~ in the ESS found
by these workers be attributed to structure in the yield pattern of
the actinides, even considering the range estimated by \cite{ga01}.
This result then forces a longer $\Delta_1$ for all the actinides.

More high-precision analyses of appropriate meteoritic mineral phases
showing large Nd, U, Th fractionation and containing \pu~ must be done
to better establish the bound. The extent to which Nd/U or Th/U
fractionation may be used as a precise estimate of Cm/U fractionation
needs some laboratory studies.  This can be done directly by
determining Pu, Cm, U, Th, Gd, and Nd fractionation in phases
crystallizing in synthetic silicate systems using ion probe techniques
and should be undertaken. Such a laboratory study can be done at
relatively low activity levels.

The use of $^{232}$Th as the long-lived cosmo-chronometer was
pioneered and explored by \cite{b2fh} and \cite{fh60}. This
application is still an issue of importance. For $^{238}$U/$^{232}$Th
we must first consider the solar system value. There have been
numerous measurements of this ratio in various chondritic meteorites
and there is a clear dispersion in the observations. The question is:
Which chondrites determine the solar ratio? The most extensive data
set is by \cite{cw92} \cite{cwp93}. It was found that for all
chondrites, $^{232}$Th/$^{238}$U $= 3.9 \pm 0.8$ (today). For C1 and
CM2 carbonaceous chondrites, the value is $3.7 \pm 0.1$. The values of
$^{232}$Th/$^{238}$U estimated for the bulk earth and the moon using
the abundances of $^{208}$Pb and $^{206}$Pb indicate values close to
3.8 - 4.0. Because the terrestrial and lunar materials studied are
magmatic differentiates representing stages of planetary evolution, it
is not possible to fix more precise values. Insofar as the
carbonaceous chondrites represent the ``holy grail,'' we use this
ratio in our discussion.

From Table  \ref{mlt} we see that $(^{238}{\rm U}/^{232}{\rm
Th})_{ESS} = 0.438$ while for UP it is 0.388. To explain the 13\%
difference would require either changing $T$ to times
significantly shorter than $10^{10}$ yr, thereby changing the other
ratios of actinides, or requiring a value of
($^{232}$Th/$^{238}$U$)_\odot = 4.18$. Alternatively, the
production ratios for the actinides would have to be changed. This
could most simply be accomplished by decreasing
$p^{^{232}\mathrm{Th}}$ but respecting the contributions within
each decay chain. Theoretical estimates of the production of
actinides will not resolve this issue.

\section{ Conclusions and Major Problems - Potential and
Real}\label{majprob}

In Table \ref{sites} we summarize the inferred sources of short-lived
nuclei not associated with the $r$-process. We have taken into
consideration the ranges of abundances that exist in estimating the
ESS inventories. We will continue to assume that the ESS values used
for most nuclei represent the bulk solar inventory. Exceptions will be
noted. The list gives possible sources (stellar types or particle
irradiation) that can produce the nuclide or that are excluded from
being the source from arguments given earlier in the text.

The exclusive cases are $^{10}$Be, which requires irradiation, and
$^{60}$Fe and $^{107}$Pd, which each require stellar sources. The case
of $^{60}$Fe is particularly important in this regard. If the source
is an AGB star (with $f_0$ $\sim$ 5$\times$10$^{-3}$) then it also
requires that the $^{53}$Mn be from the ISM inventory that is the
product of longer-term nucleosynthesis by SNe. There is the further
requirement that the last time this ISM was replenished with \mn~ by a
SN was less than 10$^7$ yr prior to formation of the ESS. The total
$^{60}$Fe would then be a sum of what the AGB injected and what was
left after \fe~ decay in the ISM. This model would give a quantitative
or semi-quantitative account for all of the other nuclei listed
($^{60}$Fe, $^{26}$Al, $^{41}$Ca, $^{107}$Pd) that came from stellar
sources and require that they be correlated in occurrence.

Certainly the high abundance of AGB debris found in meteorites is
evidence of AGB components, but does not prove they were the
carriers of live nuclei. A demonstration of the presence of
$^{205}$Pb would strongly support the idea that this stellar
source was an AGB star.

The $^{10}$Be requires that some of the material in the solar
system was subject to irradiation, possibly producing $^{36}$Cl,
but not responsible for the major inventory of $^{26}$Al, at least
according to current irradiation models.

If a SN source provided all the $^{60}$Fe and \mn, the \al~ would then
have to come from irradiation or from a low mass AGB star. In the case
of an AGB producing \al~ and \fe, these nuclei would be the heat
sources required by Urey \cite{ur55}.  For total heat production from
$^{26}$Al we have 6.0$\times$10$^{12}$ erg per gram of Al, if
$(^{26}$Al/$^{27}$Al)$_{ESS} = 5 \times 10^{-5}$.  The corresponding
initial heat production rate for Al is 5.9$\times$10$^6$ erg/yr per
gram of Al. For the total heat production from $^{60}$Fe we have
4.7$\times$10$^{10}$ erg per gram of $^{56}$Fe, if
$^{60}$Fe/$^{56}$Fe = 10$^{-6}$. The initial heating rate is
2.1$\times$10$^{10}\times$($^{60}$Fe/$^{56}$Fe) erg/yr per gram of
$^{56}$Fe.  The net heat from these two nuclides can be calculated for
the bulk composition of choice.  The typical ratio by mass is Al/Fe =
0.047 in chondritic meteorites. For this Al/Fe, the ratio of the
energy produced by \al~ to that by \fe~ is 6. It follows that both are
potent heat sources, but in these proportions, $^{26}$Al is
predominant. The critical matter is what the inventory of \al~ and
\fe~ is in the material that goes to make up the planets.

Another possible scenario is that the formation of the ESS was
immediately triggered by a SN, which provided all the $^{60}$Fe and
$^{53}$Mn (but not \al). These two nuclei are natural products of SN
models, though the quantitative yield of $^{53}$Mn strongly depends on
the unknown ``mass cut'' above which material is expelled. In any
case, the total matter from the SN in the protosolar cloud must now be
quite small ($f_0$ $\sim$ 10$^{-6}$ to 10$^{-7}$). This would then
require that all of the other short-lived nuclei come from another
source. The $^{107}$Pd is certainly present in planetary
differentiates and has not been identified as a product of
irradiation, so it rquires a stellar source. The requirement of no
\ct~ pocket that was placed on possible AGB sources in Section
\ref{agb} was used to seek coverage of several short-lived nuclei
including \al~ and \ca.  However, in this scenario, if \al~ and \ca~
are attributed to irradiation of dust, then the condition of no \ct~
pocket can be removed.  The \pd~ inventory can then be accounted for
by a star with a ``normal'' \ct~ pocket. From equation \ref{ENV} and
the quantities in Tables \ref{qenv} and \ref{nrenv}, we find that a
very small contribution $(f^{^{107}\mathrm{Pd}} = 4 \times 10^{-5})$
from a 1.5 \ms\ star will provide the \pd~ inventory with $\Delta_1 +
\Delta_2 = 8.75 \times 10^6$ yr.  It will also provide $^{205}$Pb with
essentially the same upper bound as given in Table \ref{nrenv}. A
similar result is found for the 3 \ms~ case. As $\bar\tau_{^{107}{\rm
Pd}} = 9.4$ Myr, there is then no strict requirement as to when such
AGB debris needs to be added to the proto-solar cloud. This scenario
is then one of a SN trigger feeding \fe~ and \mn~ to the cloud that
had been salted with a very small amount of ``normal'' AGB debris
within some extended interval, possibly within a 15-20 Myr window. The
\al, $^{10}$Be, $^{36}$Cl, and other related nuclei would then all be
attributed to energetic particle irradiation of the disk. This
scenario would appear to be a self-consistent story, and its
implications are clear. There can only be a fraction of the \mn~
inventory from the irradiation model. The requirement of an adequate
and physically plausible irradiation scenario must provide $^{10}$Be
at some time but, more significantly, provide \al~ to the disc at an
appropriate level for a heat source. The rather sharply defined
canonical value of \al/$^{27}$Al must then be explained.

There are in any case overwhelmingly strong astrophysical observations
and theoretical considerations that such particle bombardments occur
around an active young star. In particular, Feigelson and his
colleagues have consistently presented both observational and
theoretical evidence that they must occur (see recent report by
\cite{fei02}). There are extensive recent studies of irradiation
models referred to earlier in the text. The issue is what is a
plausible and self-consistent astrophysical scenario (certainly one
that does not try to explain everything, including the
petrochemistry). With regard to irradiation models, we feel that some
effort should be made to determine what fraction of matter in the disk
was irradiated and whether this represents the total material that
went into making the protoplanets and planets or is a small fraction
of it. Certainly a single irradiation scenario is not acceptable
considering the observations on meteorites.

In the case of a SN source for both $^{60}$Fe and $^{53}$Mn, these
nuclei should represent the abundances of the total solar system
including the sun. For an AGB source, and with $^{53}$Mn inherited
from the ISM, the abundances should also represent global values.
In the case of irradiation models, the material exposed to the
particle bombardment does not represent the total solar system.
There are then issues concerning the extensiveness of the
irradiation. The $^{26}$Al and $^{53}$Mn considered here as ESS
values are then only for the material that went to make up
protoplanetary bodies and planets; the effective average
irradiation must then be sufficient to achieve these values for
the amount of ``dust'' required. We note that a universal value of
(\al/$^{27}$Al)$_{ESS} \sim 5 \times 10^{-5}$ has serious
implications for the formation and history of cometary bodies
because of its intense heat source. It is also most reasonable to
expect that irradiation of material occurred only for local
parcels of matter or after most protoplanetary objects had formed.
In the latter case, the irradiation effects do not apply to the
contents in the planets and $^{26}$Al would not be a significant
heat source. The $^{60}$Fe coming from a stellar source would have
to suffice.

The relative levels of abundances of the different nuclei and their
correlation or non-correlation need better definition in irradiation
models. The amounts in all materials (i.e. bulk effects seen in
planets) need to be more explicitly evaluated. Scenarios with
sharply-defined predictions would greatly aid in future work.  Some
significant efforts have been made in this direction since the work of
Shu et al. \cite{ssg97} \cite{ssgg01}.  However the physical
plausibility of these models is not evident. The consideration of the
size distribution of chondrules is almost certainly a consequence of
surface tension and not of X-wind dynamics and ejection. There is now
no doubt (cf. $^{10}$Be) that irradiation processes are important, but
precisely what observations can be simply explained and what is
predicted is not yet clear. The role of reprocessing of materials in
and on planetesimals cannot be ignored.

The $^{60}$Fe requires a stellar source; if it is from SNe, then
$^{53}$Mn will certainly be co-produced and the $^{53}$Mn produced
by irradiation is then supplemental. From the observation of
chondrules \cite{tac05} there is good reason to consider that
$^{60}$Fe was omnipresent in the ESS. Its presence in planets as
first investigated by Shukolyukov \& Lugmair \cite{sluga}
\cite{slugb} needs to be addressed again in order to firmly
establish the bulk ESS $^{60}$Fe inventory. The general presence of
$^{60}$Fe in chondrites is being actively pursued.

If we consider the results on \pu/$^{232}$Th and \cm/$^{235}$U at face
value then several major issues appear. If in fact the value of
$(^{244}\mathrm{Pu}/^{232}\mathrm{Th})_{ESS} = $ \\
$1/2(^{244}\mathrm{Pu}/^{232}\mathrm{Th})_{\rm UP}$, then this
requires a $\Delta_{^{244}\mathrm{Pu}} \approx 8 \times 10^7$ yr
(quite close to $\Delta_{^{129}\mathrm{I}}$). For this case, at the
time of solar system formation
$(^{247}\mathrm{Cm}/^{235}\mathrm{U})_{ESS} =$
$(^{247}\mathrm{Cm}/^{235}\mathrm{U})_{\rm
UP}e^{-\Delta_1/\bar\tau_{^{247}\mathrm{Cm}}} =$ $ 2.5 \times
10^{-4}$. This would be in reasonable accord with the recent bound on
\cm/$^{235}$U required by Stirling et al. \cite{stir05} from their
data base.  It appears evident that the last $r$-process episode
producing the actinides and trans-actinides took place $\sim$ 10$^8$
yr before the sun formed.  The critical piece of data here is the
bound on $(^{247}\mathrm{Cm}/^{235}\mathrm{U})_{ESS}$.  This number
requires a long delay after the last actinide-producing event, and is
independent of uncertainties in the details of production.

Assuming the above to be valid, there is a further conflict with the
\hf~ results, which require $\Delta_1^{^{182}\mathrm{Hf}} \lesssim 2
\times 10^7$ yr. This suggests that the actinide group is made in an
$r$-process distinct from the intermediate heavy mass $r$-nuclei.
There would then be three regimes of $r$-process production - light
$r$-nuclei (Ba and below), intermediate (Ba to intermediate heavy
nuclei), and heavy (intermediate heavy nuclei -- Pt group? -- to
actinides).  Scenarios involving a blend of various $r$-process
components have already been advanced \cite{qian02}.  As established
from observations on low metallicity halo stars with high heavy
$r$-process enrichment, there is no evidence of any gross variations
in the abundances of all the heavy $r$-nuclei relative to the solar
value. This includes U and Th. However, it is now recognized that some
real variability of yields exists and is to be expected across this
wide mass region from Ba to U. This affects the validity of Th/Eu as a
chronometer (see
\cite{qw02}\cite{qw03}\cite{hill02}\cite{cay01}\cite{qianrev}\cite{qian02}). These
works have demonstrated that Th/Eu is rather variable and the notion
of an exact robust heavy $r$-pattern (Ba -- Th) is not
valid. Consideration of the ESS inventory of $r$-nuclei is not as
simple as it was long supposed to be. It is clear that no significant
very heavy $r$-process event occurred close to the time that the
proto-solar system formed. Some intermediate mass $r$-process must
have contributed $^{182}$Hf and other nuclides.  The apparent
coincidence of the time scale inferred for $^{129}$I and that for the
trans-actinides suggests that this may be due to fission
\cite{qian02}.  The last heavy $r$ contribution would then be from an
$r$-process that produced very heavy nuclei without fission recycling
so that the yields at Ba and below (including I) were governed by
fission.  This matter requires further investigation.

\begin{figure}
\begin{center}
\end{center}
\caption{Cartoon showing the complex astrophysical scenario
required. There are more ancient $r$-process sites producing the
actinide group and $^{129}$I ($\sim$10$^8$ yr before formation of the
solar system); a molecular cloud that was replenished by a SN at or
after $\sim 10^7$ yr before the solar system formed; input from an AGB
star within $\sim 10^6$ yr before the solar system formed; energetic
$p$- and $\alpha$-capture irradiation (or several irradiations) in the
early solar system. The various stellar agents that may be or may not
be contributing are discussed in the text. The $r$ sources are not
known, but may include low mass SNeII and accretion-induced collapse
(AIC) of a binary companion to produce a neutron star.} \label{cart}

\centerline{\epsfig{file=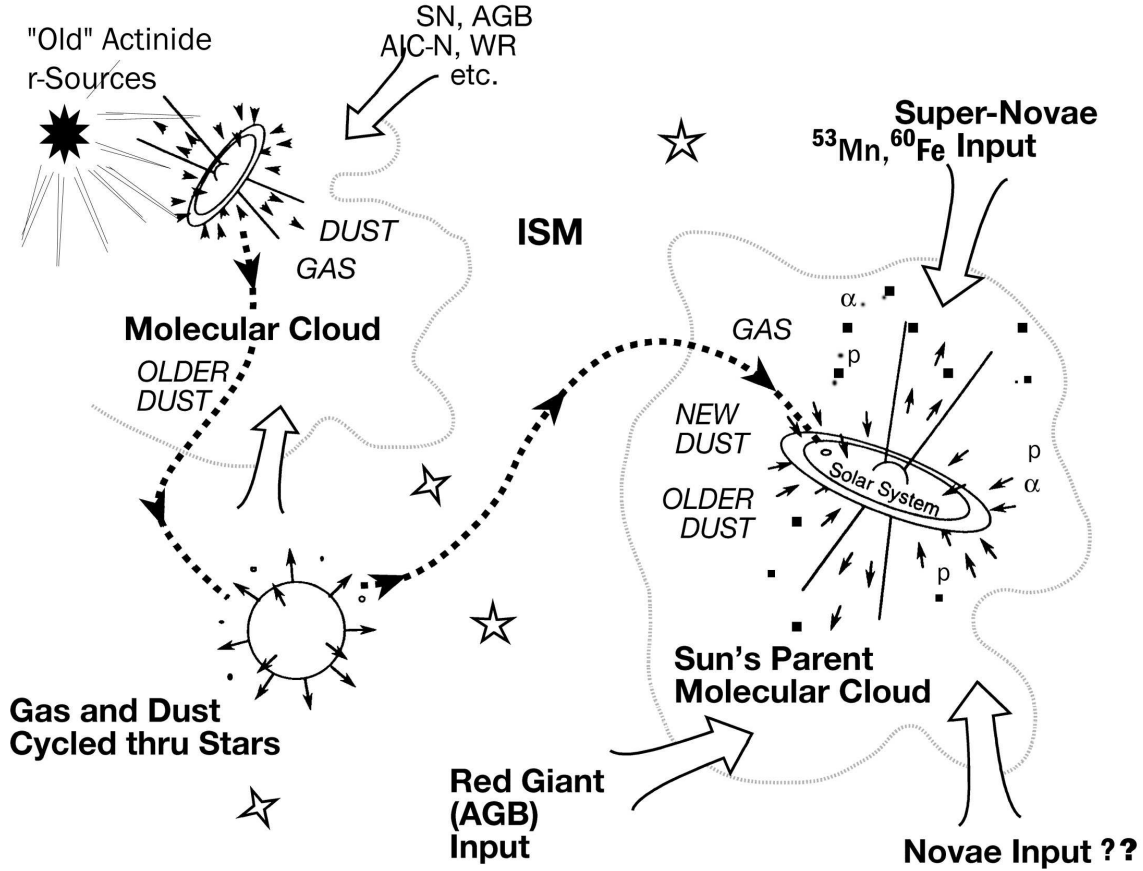,width=6in}}

\end{figure}

The most important issue now appears to be to divine a coherent and
plausible astrophysical scenario that could explain the times of
different additions, in particular the late (or very late) additions
of the shorter-lived nuclei from an AGB source, and the availability
of matter in the ISM that was replenished by SNe II debris within
$\sim 10^7$ years before the hypothesized AGB injection. These
processes must involve neighboring molecular clouds that interacted
after one was dispersed from a SNe II event, and the subsequent onset
of further astration in its neighbor (see Fig. \ref{cart}). It is
evident from the sampling of circumstellar dust grains that material
from very different molecular clouds with different metallicities must
have occurred.  As the lifetimes of molecular clouds are far shorter
than the lifetimes of low- and intermediate-mass stars, it follows
that any injection from these sources must come from stars that formed
long before the molecular cloud in which the solar nebula formed.

In summary, the various radioactive nuclei present in the early solar
system appear to be dominated by several distinct sources.  These are:
1) The production of actinides and trans-actinides up to $\approx
10^8$ yr before the solar nebula formed, which then provided the ISM
with the overall inventory of these nuclides.  Essentially none of
this group of nuclei were added later; 2) the addition of $^{53}$Mn
and $^{60}$Fe from a more recent stellar event occurring less than
10$^7$ yr prior to formation of the sun; 3) The addition or production
of a host of other short-lived nuclei, some of which appear to require
a stellar source (\pd) and some of which could be produced locally in
the disk by intense irradiation of dust; 4) Some irradiation of
material that occurred after protoplanet formation.

There exists a compact commercial device called a ``multipurpose
tool'' that has combinations of many individual tools all folded into
a single compact pocket. It is not commonly used for repairs.  Whether
such an all-function device is available in the astrophysical tool box
to use for the formation of the solar system remains to be seen. It
always seems that one also needs some special independent tool when
doing a job.

The clarification of possible and realizable scenarios now
requires more active study.

{\bf Acknowledgments} \noindent The authors thank the editors for
their indulgence in handling this tardy submission of a very long
report. The efforts of Uli Ott in providing a detailed, thorough,
incisive and scholarly review have greatly improved this work.
Extremely useful comments and criticisms were provided by Ian Hutcheon
as well as a possible but thankfully not realized SNe shock.  We are
indebted to useful comments by F. K\"appeler, J. Truran, A.  McKeegan,
A. Davis and M. Thiemens, as well as several other
colleagues. M.E. Johnson did her usual superior work in cryptography
and a special effort in re-encoding on the many versions of this
manuscript. M.B. and R.G. acknowledge financial support in Italy by
MIUR, under contract PRIN2004-025729. G.J.W.  acknowledges the support
of DOE-DE-FG03-88ER13851.  K.M.N. was supported by the U.S. DOE,
Nuclear Physics Division, under contract No. W-31-109-ENG-38. Caltech
Division Contribution 9014(1119).

\newpage

\clearpage


\begin{landscape}
\scriptsize
\begin{longtable}{|c|c|c|c|c|c|c|c|c|}

\caption{Mean life times and abundances of short-lived nuclides,
uniform production (UP) and early solar system
inventory}\label{mlt}\\\hline
Radioactive  & Reference  & Process & Mean Life  &
$(N^R/N^I)_{ESS}$ & \multicolumn{4}{c|}{$(N^R/N^I)_{UP}$}\\
\cline{6-9} Isotope (R) & Isotope (I) & & $\bar\tau_R$ (Myr) & &
$\Delta_1 = 0$ & $\Delta_1 = 5$ Myr & $\Delta_1 = 10$ Myr & $\Delta_1 =
70$ Myr\\ \hline $^{238}$U & $^{232}$Th & r;r &
$6.45\times10^3;2.03\times10^4$ & 0.438 & 0.388 & 0.388 & 0.388 &
0.388\\ \hline $^{235}$U & $^{238}$U & r;r &
$1.02\times10^3;6.45\times10^3$ & 0.312 &
0.289 & 0.289 & 0.289 & 0.270\\
\hline
  $^{244}$Pu  & $^{232}$Th & $r;r$ & 115; $2.03 \times 10^4$ & $3 \times
10^{-3}$ &
  $5.6 \times 10^{-3}$ & $5.4 \times 10^{-3}$ & $5.1\times10^{-3}$ & $3.1
\times 10^{-3}$\\
  \cline{2-9}
& $^{238}$U & $r;r$ & 115; $6.45 \times 10^3$ &  $6 \times 10^{-3}$
& $1.4 \times 10^{-2}$ & $1.3 \times 10^{-2}$ &  $1.3 \times
10^{-2}$ & $7.6 \times 10^{-3}$\\ \hline $^{247}$Cm & $^{235}$U &
$r; r$ & 22.5; $1.02 \times 10^3$ & $(<2 \times 10^{-3};
<10^{-4})$ & $8.9 \times 10^{-3}$ & $7.2 \times 10^{-3}$ & $5.7
\times 10^{-3}$ & $4 \times 10^{-4}$\\ \hline $^{182}$Hf &
$^{180}$Hf & $r; r,s$ & 13; stable & $2.0 \times 10^{-4}$ & $4.5
\times 10^{-4}$ & $3.8 \times{10^{-4}}$ & $2.1 \times10^{-4}$ & $2
\times 10^{-6}$\\
\hline $^{146}$Sm & $^{144}$Sm & $p; p$ & 148; stable & $1.0
\times 10^{-2}$ & $1.5 \times 10^{-2}$ & $1.5 \times 10^{-2}$ &
$1.4 \times 10^{-2}$ & $9.4 \times 10^{-3}$\\
\hline $^{92}$Nb & $^{93}$Nb & $p; s$ & 52; stable &  ? & $1.0
\times 10^{-4}$ & $9.0 \times 10^{-5}$ &
$8.2 \times 10^{-5}$ & $2.6 \times 10^{-5}$\\
\hline $^{135}$Cs & $^{133}$Cs & $r,s; r,s$ & 2.9; stable & $1.6
\times 10^{-4}$ ? & $2.1 \times 10^{-4}$ & $3.7 \times 10^{-5}$ &
$7 \times10^{-6}$ & 0 \\ \hline
$^{205}$Pb & $^{204}$Pb & s; s & 22; stable & ? & -- & -- & -- & -- \\
\hline $^{129}$I & $^{127}$I & $r; r,s$ & 23; stable & $1.0 \times
10^{-4}$ & $(2-5) \times 10^{-3}$ & $(1.6-4.0) \times 10^{-3}$ &
$(1.4-3.5) \times 10^{-3}$ & $(1-2) \times 10^{-4}$\\ \hline
$^{107}$Pd & $^{108}$Pd & $s,r; r,s$ & 9.4; stable & $2.0 \times
10^{-5}$ & $6.2 \times 10^{-4}$ & $3.7 \times 10^{-4}$ & $2.2
\times 10^{-4}$ & $4 \times 10^{-7}$ \\ \hline
  $^{60}$Fe & $^{56}$Fe & $eq,exp,s$ & 2.2; stable
& $(2\times10^{-7}; 2\times10^{-6})$ & $5 \times 10^{-7}$ & $5.2
\times 10^{-8}$ & $ 5.3 \times 10^{-9}$ & 0 \\ \hline $^{53}$Mn &
$^{55}$Mn & $p,exp; exp$ & 5.3; stable & $(\sim 6\times10^{-5};
5\times10^{-6})$ & $\sim 1 \times 10^{-4}$ & $4 \times 10^{-5}$ &
$1.6 \times 10^{-5}$ & 0 \\ \hline $^{41}$Ca & $^{40}$Ca & $s,exp;
exp$ & 0.15; stable & $1.5 \times 10^{-8}$ & $2 \times 10^{-8}$ & 0
& 0 & 0 \\ \hline
  $^{36}$Cl & $^{35}$Cl & $s; exp$ & 0.43; stable & $5 \times
  10^{-6}$ & $3.8 \times 10^{-7}$ & 0 & 0 & 0 \\ \hline $^{26}$Al &
$^{27}$Al & $p; exp$ & 1.03; stable & $5 \times 10^{-5}$ & $\sim
10^{-7}$ & 0 & 0 & 0 \\ \hline $^{10}$Be & $^{9}$Be & spallation
& 2.3; stable & $1 \times 10^{-3}$ & 0 & 0 & 0 & 0\\  \hline
\end{longtable}
\normalsize
\end{landscape}

  \clearpage
\begin{table}
\caption{Short-lived nuclei from a 15 \msb solar-metallicity SN source$^\ast$}
\begin{tabular}{cccccc}\hline
Rad. & Ref. & $q^I_{ENV}/q^I_0$ & $(N^R/N^I)_{ENV}$ & A & B\\
&  &   &   &  $(N^R/N^I)_{\Delta_1}$ & $(N^R/N^I)_{\Delta_1}$\\
\hline $^{26}$Al & $^{27}$Al & 80.7 & $5.7 \times 10^{-3}$ & $5.0
\times 10^{-5}$ & $1.2 \times 10^{-6}$\\
$^{41}$Ca & $^{40}$Ca & 4.7 & $1.5 \times 10^{-2}$ & $1.5 \times
10^{-8}$ & $3.5 \times 10^{-10}$\\
$^{53}$Mn & $^{55}$Mn & 95.6 & 0.15 & $3.5 \times 10^{-3}$ & $8
\times 10^{-5}$\\
$^{107}$Pd & $^{108}$Pd & 1.2 & $3.1 \times 10^{-2}$ & $1.1 \times
10^{-5}$ & $2.6 \times 10^{-7}$ \\
$^{60}$Fe & $^{56}$Fe & 107.8 & $2.4 \times 10^{-3}$ & $4.7 \times
10^{-5}$ & $1.1 \times 10^{-6}$\\
$^{36}$Cl & $^{35}$Cl & 4.45 & $9.9\times 10^{-3}$  & $1.1\times 10^{-6}$ & 
$2.5\times 10^{-8}$\\

\hline
\end{tabular}\label{sln}
\end{table}

$^\ast$Calculated from the model of Rauscher et al. \cite{rau02};
$ENV$ represents the ratio in the ejected stellar envelope.

A. Calculated to match $(^{26}$Al/$^{27}$Al)$_{\Delta_1} = 5 \times
10^{-5}$ with a dilution factor $f_0 = 3.05 \times 10^{-4}, \Delta_1
= 1.09$ Myr.

B. Calculated to match $(^{53}\mathrm{Mn}/^{55}\mathrm{Mn})_{\Delta_1}
= 1 \times 10^{-4}$ with $f_0 = 7 \times 10^{-6}, \Delta_1 = 1.09$
Myr.

\clearpage

\begin{longtable}{|c|cc|cc|}
\caption{ $(q^I_{ENV}/q^I_0)$ values for a TP-AGB
star}\label{qenv}\\\hline

\multicolumn{5}{|c|}{Initial mass 1.5 \msb}\\
\hline {$Z/Z_\odot$} & \multicolumn{2}{|c}{1} &
\multicolumn{2}{c|}{1/3}\\ \hline Isotope & ST & no \ct~ pocket & ST
& no \ct~ pocket\\ \hline
$^{27}$Al & 1.01 & 1.02 & 0.30 & 0.31\\
$^{35}$Cl & 0.99 & 0.98 & 0.29 & 0.29\\
$^{40}$Ca & 0.99 & 0.99  & 0.37 & 0.37\\
$^{56}$Fe & 0.99 & 0.99 & 0.30 & 0.30\\
$^{82}$Kr & 11.60 & 1.13 & 2.70 & 0.34\\
$^{92}$Zr & 11.98 & 1.00 & 13.1 & 0.30\\
$^{100}$Ru & 12.36 & 1.03 & 17.2 & 0.31\\
$^{108}$Pd & 8.37 & 1.02 & 12.02 & 0.31\\
$^{133}$Cs & 2.62 & 0.99 & 4.50 & 0.38\\
$^{204}$Pb & 5.30 & 1.06 & 21.90 & 0.33\\ \hline
\multicolumn{5}{|c|}{Initial mass 2 \msb}\\ \hline
$^{27}$Al & 1.01 & 1.01 & 0.30 & 0.31\\
$^{35}$Cl & 0.99 & 1.00 & 0.29 & 0.29\\
$^{40}$Ca & 0.99 & 1.00 & 0.36 & 0.36\\
$^{56}$Fe & 0.99 & 1.00 & 0.29 & 0.29\\
$^{82}$Kr & 7.89 &1.09 & 4.60 & 0.42\\
$^{92}$Zr & 8.19 & 1.01 & 25.91 & 0.31\\
$^{100}$Ru & 8.48 & 1.02 & 29.48 & 0.31\\
$^{108}$Pd & 5.83 & 1.01 & 22.16 & 0.31\\
$^{133}$Cs & 2.07 & 0.99 & 8.26 & 0.37\\
$^{204}$Pb & 3.88 & 1.04 & 43.78 & 0.36\\ \hline \pagebreak\hline
  \multicolumn{5}{|c|}{Initial mass 3 \msb}\\
\hline
  $^{27}$Al & 1.01 & 1.01 & 0.30 & 0.30\\
$^{35}$Cl & 0.99 & 0.99 & 0.29 & 0.29\\
$^{40}$Ca & 0.99 & 0.99 & 0.36 & 0.36 \\
$^{56}$Fe & 0.99 & 0.99 & 0.29 & 0.29 \\
$^{82}$Kr & 8.49  & 1.12 & 3.72 & 1.63\\
$^{92}$Zr & 10.25 & 1.01 & 26.89 & 0.62\\
$^{100}$Ru & 9.18 & 1.02 & 23.46 & 0.42 \\
$^{108}$Pd & 6.57 & 1.01 & 13.54 & 0.34\\
$^{133}$Cs & 2.29 & 0.99 & 6.06 & 0.37\\
$^{204}$Pb & 4.37 & 1.06 & 34.32 & 0.36\\ \hline
\multicolumn{5}{|c|}{Initial mass 5 \msb}\\ \hline $^{27}$Al & 1.01
& 1.01 & 0.31 & 0.31\\
$^{35}$Cl & 0.99 & 0.99 & 0.30 & 0.30\\
$^{40}$Ca & 0.98 & 0.98 & 0.36 & 0.29\\
$^{56}$Fe & 0.98 & 0.98 & 0.29 & 0.29\\
$^{82}$Kr & 10.58 & 7.50 & 4.40 & 4.18\\
$^{92}$Zr & 9.74 & 4.60 & 13.77 & 4.83\\
$^{100}$Ru & 4.41 & 2.19 & 5.74 & 1.66\\
$^{108}$Pd & 2.74 & 1.60 & 3.22 & 0.99\\
$^{133}$Cs & 1.29 & 1.05 & 1.06 & 0.42\\
$^{204}$Pb & 1.29 & 1.11 & 2.70 & 0.40\\ \hline
\end{longtable}

\clearpage

\begin{longtable}{|c|cc|cc|}
\caption{ $(N^R/N^I)_{ENV}$ values for a TP-AGB star}\label{nrenv}
\\\hline
\multicolumn{5}{|c|}{Initial mass 1.5 \msb}\\ \hline {$Z/Z_\odot$} &
\multicolumn{2}{c|}{1} & \multicolumn{2}{c|}{1/3}\\ \hline Isotope
Pair & ST & no \ct~ pocket & ST & no \ct~ pocket\\ \hline
\al/$^{27}$Al & $5.8 \times 10^{-3}$  & $5.6 \times 10^{-3}$
  & $4.4 \times 10^{-3}$ & $4.3 \times 10^{-3}$\\
$^{36}$Cl/$^{35}$Cl & $1.4 \times 10^{-3}$ & $1.1 \times 10^{-3}$ &
$7.2 \times 10^{-4}$ & $7.8 \times 10^{-4}$ \\
\ca/$^{40}$Ca & $4.2 \times 10^{-4}$ & $4.5 \times 10^{-4}$ & $4.5
\times 10^{-4}$ & $4.7 \times 10^{-4}$\\
\fe/$^{56}$Fe & $7.1 \times 10^{-6}$ & $1.2 \times 10^{-5}$ & $1.0
\times 10^{-5}$ & $2.7 \times 10^{-5}$\\
$^{81}$Kr/$^{82}$Kr & $1.1 \times 10^{-2}$ & $3.0 \times 10^{-3}$ &
$4.2 \times 10^{-3}$ & $1.3 \times 10^{-3}$\\
$^{93}$Zr/$^{92}$Zr & $2.6 \times 10^{-1}$ & $1.1 \times 10^{-2}$ &
$2.6 \times 10^{-1}$ & $1.1 \times 10^{-2}$\\
$^{99}$Tc/$^{100}$Ru & $1.1 \times 10^{-1}$ & $7.0 \times 10^{-3}$ &
$6.1 \times 10^{-2}$ & $3.9 \times 10^{-3}$\\
\pd/$^{108}$Pd & $1.4 \times 10^{-1}$ & $1.0 \times 10^{-2}$ & $1.5
\times 10^{-1}$ & $9.6 \times 10^{-3}$\\
$^{135}$Cs/$^{133}$Cs & $3.3 \times 10^{-1}$ & $1.8 \times 10^{-2}$
& $5.1 \times 10^{-1}$ & $1.6 \times 10^{-2}$\\
$^{205}$Pb/$^{204}$Pb & $1.0 \times 10^0$ & $1.0 \times 10^{-1}$ &
$1.2 \times 10^0$ & $1.1 \times 10^{-1}$\\ \hline
\multicolumn{5}{|c|}{Initial mass 2 \msb}\\ \hline
\al/$^{27}$Al &
$3.6 \times 10^{-3}$ & $3.4 \times 10^{-3}$ & $5.0 \times 10^{-3}$
& $5.1 \times 10^{-3}$\\
$^{36}$Cl/$^{35}$Cl & $6.5 \times 10^{-4}$ & $ 7.3 \times 10^{-4}$ &
$1.5 \times 10^{-3}$ & $1.2 \times 10^{-3}$\\
\ca/$^{40}$Ca & $2.6 \times 10^{-4}$ & $2.8 \times 10^{-4}$ & $7.6
\times 10^{-4}$ & $2.5 \times 10^{-4}$\\
  \fe/$^{56}$Fe & $4.6 \times 10^{-6}$ & $8.3 \times 10^{-6}$ & $1.3 \times
  10^{-4}$ & $2.5 \times 10^{-4}$ \\
$^{81}$Kr/$^{82}$Kr & $7.9 \times 10^{-3}$ & $1.5 \times 10^{-3}$ &
$2.7 \times 10^{-3}$ & $2.4 \times 10^{-3}$\\
$^{93}$Zr/$^{92}$Zr & $1.3 \times 10^{-1}$ & $6.5 \times 10^{-3}$ &
$2.6 \times 10^{-1}$ & $2.3 \times 10^{-2}$\\
$^{99}$Tc/$^{100}$Ru & $7.6 \times 10^{-2}$ & $3.1 \times 10^{-3}$ &
$5.0 \times 10^{-2}$ & $5.4 \times 10^{-3}$\\
\pd/$^{108}$Pd & $1.3 \times 10^{-1}$ & $6.5 \times 10^{-3}$ &
$1.5 \times 10^{-1}$ & $1.5 \times 10^{-2}$ \\
$^{135}$Cs/$^{133}$Cs & $2.7 \times 10^{-1}$ & $1.0 \times 10^{-2}$
  & $6.6 \times 10^{-1}$ & $3.1 \times 10^{-2}$\\
  $^{205}$Pb/$^{204}$Pb & $9.4 \times 10^{-1}$ & $6.7 \times 10^{-2}$
  & $1.3 \times 10^0$ & $2.3 \times 10^{-1}$\\ \hline
  \pagebreak\hline
\multicolumn{5}{|c|}{Initial mass 3 \msb}\\ \hline
\al/$^{27}$Al &
$2.9 \times 10^{-3}$ & $2.8 \times 10^{-3}$ & $3.5 \times 10^{-3}$ &
$3.5 \times 10^{-3}$\\
$^{36}$Cl/$^{35}$Cl  & $6.8 \times 10^{-4}$ & $7.3 \times 10^{-4}$ &
$1.5 \times 10^{-3}$ & $1.5 \times 10^{-3}$\\
\ca/$^{40}$Ca & $3.0 \times 10^{-4}$ & $3.2 \times 10^{-4}$ & $4.0
\times 10^{-4}$ & $4.1 \times 10^{-4}$\\
\fe/$^{56}$Fe & $2.8 \times 10^{-5}$ & $4.7 \times 10^{-5}$ & $1.6
\times 10^{-3}$ & $2.3 \times 10^{-3}$\\
$^{81}$Kr/$^{82}$Kr &$4.8 \times 10^{-3}$ & $1.2 \times 10^{-3}$ &
$5.7 \times 10^{-3}$ & $1.0 \times 10^{-2}$\\
$^{93}$Zr/$^{92}$Zr & $2.4 \times 10^{-1}$ & $9.1 \times 10^{-3}$ &
$2.6 \times 10^{-1}$ & $1.3 \times 10^{-1}$\\
$^{99}$Tc/$^{100}$Ru & $7.6 \times 10^{-2}$ & $3.7 \times 10^{-3}$ &
$1.2 \times 10^{-1}$ & $6.3 \times 10^{-2}$\\
\pd/$^{108}$Pd & $1.3 \times 10^{-1}$ & $6.9 \times 10^{-3}$ & $1.5
\times 10^{-1}$ & $2.9 \times 10^{-2}$\\
$^{135}$Cs/$^{133}$Cs & $4.4 \times 10^{-1}$ & $1.8 \times 10^{-2}$
& $8.7 \times 10^{-1}$ & $3.0 \times 10^{-2}$\\
$^{205}$Pb/$^{204}$Pb & $1.0 \times 10^0$ & $9.1 \times 10^{-2}$ &
$1.1 \times 10^0$ & $2.4 \times 10^{-1}$\\ \hline
\multicolumn{5}{|c|}{Initial mass 5 \msb}\\ \hline \al/$^{27}$Al &
$5.3 \times 10^{-4}$ & $5.3 \times 10^{-4}$ & $6.6 \times 10^{-4}$ &
$6.3 \times 10^{-4}$\\
$^{36}$Cl/$^{35}$Cl & $1.1 \times 10^{-3}$ & $1.1 \times 10^{-3}$ &
$1.9 \times 10^{-3}$ & $1.4 \times 10^{-3}$\\
\ca/$^{40}$Ca & $1.2 \times 10^{-4}$ & $1.2 \times 10^{-4}$ & $1.0
\times 10^{-4}$ & $7.9 \times 10^{-5}$\\
\fe/$^{56}$Fe & $3.6 \times 10^{-3}$ & $3.8 \times 10^{-3}$ & $4.8
\times 10^{-3}$ & $5.5 \times 10^{-3}$\\
$^{81}$Kr/$^{82}$Kr & $7.1 \times 10^{-3}$ & $7.5 \times 10^{-3}$ &
$6.8 \times 10^{-3}$ & $9.4 \times 10^{-3}$\\
$^{93}$Zr/$^{92}$Zr & $2.3 \times 10^{-1}$ & $1.9 \times 10^{-1}$ &
$2.6 \times 10^{-1}$ & $2.5 \times 10^{-1}$\\
$^{99}$Tc/$^{100}$Ru & $1.3 \times 10^{-1}$ & $9.7 \times 10^{-2}$ &
$1.6 \times 10^{-1}$ & $1.5 \times 10^{-1}$\\
\pd/$^{108}$Pd & $1.1 \times 10^{-1}$ & $6.4 \times 10^{-2}$ & $1.5
\times 10^{-1}$ & $1.1 \times 10^{-1}$\\
$^{135}$Cs/$^{133}$Cs & $2.9 \times 10^{-1}$ & $8.3 \times 10^{-2}$
& $8.0 \times 10^{-2}$ & $3.0 \times 10^{-1}$\\
$^{205}$Pb/$^{204}$Pb & $3.0 \times 10^{-1}$ & $1.6 \times 10^{-1}$
& $9.1 \times 10^{-1}$ & $3.1 \times 10^{-1}$\\ \hline
\end{longtable}

\clearpage \renewcommand{\thefootnote}{\fnsymbol{footnote}}
\begin{longtable}{|c|c|c|c|}
\caption{ Abundances of short-lived nuclei in a cloud salted with
ejecta from an AGB star (no $^{13}$C)}\label{ejecta}
\\\hline
\multicolumn{4}{|c|} {$f_0 = 5 \times 10^{-3}; (1.5 M_\odot, Z_\odot)$}\\
\hline $(N^R/N^I)_{\Delta_1}$ & $\Delta_1 = 0$ & $\Delta_1 = 0.75$
Myr & $\Delta_1 + \Delta_2 = 8.75$ Myr \\ \hline\hline
  $^{26}$Al/$^{27}$Al ($2 \times
10^{-2})$\footnotemark[1] & $(1.0 \times 10^{-4})$ & $(5 \times
10^{-5})$ & $(2.5 \times 10^{-8})$\\ \hline $^{36}$Cl/$^{35}$Cl &
$5.4 \times 10^{-6}$ & $ 9.4 \times 10^{-7}$ &
---\\ \hline
$^{41}$Ca/$^{40}$Ca & $2.2 \times 10^{-6}$ & $(1.5 \times
10^{-8})$\footnotemark[2] & ---\\
  \hline \fe/$^{56}$Fe & $5.9 \times
10^{-8}$ & $4.2 \times 10^{-8}$ & $1.1 \times 10^{-9}$\\ \hline
$^{81}$Kr/$^{82}$Kr & $1.7 \times 10^{-5}$ & $1.4 \times 10^{-6}$ &
---\\ \hline $^{93}$Zr/$^{92}$Zr & $5.5
\times 10^{-5}$ & $3.9 \times 10^{-5}$ & $1.0 \times 10^{-6}$\\
\hline $^{99}$Tc/$^{100}$Ru & $3.6 \times 10^{-5}$ & $2.7 \times
10^{-6}$ & ---\\ \hline \pd/$^{108}$Pd & $5.1 \times 10^{-5}$ & $4.7
\times 10^{-5}$ & $(2.0 \times 10^{-5})$\footnotemark[7]\\ \hline
$^{135}$Cs/$^{133}$Cs &
$8.9 \times 10^{-5}$ & $6.9 \times 10^{-5}$ & $4.4 \times 10^{-6}$\\
\hline
$^{205}$Pb/$^{204}$Pb & $<5.3 \times 10^{-4}$ & $<5.1 \times
10^{-4}$ &
$<3.5 \times 10^{-4}$\\
\hline\hline
\multicolumn{4}{|c|} {$f_0 = 5 \times 10^{-3}; (2.0 M_\odot, Z_\odot)$}\\
\hline $(N^R/N^I)_{\Delta_1}$ & $\Delta_1 = 0$ & $\Delta_1 = 0.68$
Myr & $\Delta_1
+ \Delta_2 = 4.78$ Myr\\
\hline\hline \al/$^{27}$Al $(1.9 \times 10^{-2})$\footnotemark[1] &
$(9.6 \times 10^{-5})$ & $(5.0 \times 10^{-5})$ & $(1.0 \times 10^{-6})$\\
\hline
$^{36}$Cl/$^{35}$Cl & $3.6 \times 10^{-6}$ & $7.4 \times
10^{-7}$ & ---\\ \hline
\ca/$^{40}$Ca & $1.4 \times 10^{-6}$ & $(1.5 \times
10^{-8})$\footnotemark[2] & ---\\
\hline \fe/$^{56}$Fe  & $4.2 \times 10^{-8}$ & $3.1 \times 10^{-8}$
& $4.8 \times 10^{-9}$\\ \hline $^{81}$Kr/$^{82}$Kr & $8.2 \times
10^{-6}$ & $8.6 \times 10^{-7}$ & ---\\ \hline
$^{93}$Zr/$^{92}$Zr & $3.3 \times 10^{-5}$ & $2.4 \times 10^{-5}$ &
$3.7 \times 10^{-6}$\\ \hline $^{99}$Tc/$^{100}$Ru & $1.6 \times
10^{-5}$ & $1.5 \times 10^{-6}$ & ---\\ \hline \pd/$^{108}$Pd & $3.3
\times 10^{-5}$ & $3.1 \times 10^{-5}$ & $(2.0 \times
10^{-5})\footnotemark[7]$\\
\hline $^{135}$Cs/$^{133}$Cs & $4.9 \times 10^{-5}$ & $3.9 \times
10^{-5}$ & $9.5 \times 10^{-6}$\\ \hline
$^{205}$Pb/$^{204}$Pb & $<3.5 \times 10^{-4}$ & $<3.4 \times 10^{-4}$
& $<2.8
\times 10^{-4}$\\
\hline
\pagebreak\hline
\multicolumn{4}{|c|} {$f_0 = 5 \times 10^{-3}; (3.0 M_\odot, Z_\odot)$}\\
\hline $(N^R/N^I)_{\Delta_1}$ & $\Delta_1 = 0$ & $\Delta_1 = 0.70$
Myr & $\Delta_1
+ \Delta_2 = 5.1$ Myr\\
\hline\hline \al/$^{27}$Al $(1.9 \times 10^{-2})$ & $(9.7 \times
10^{-5})$ & $(5.0 \times 10^{-5})$ & $(7.6 \times 10^{-7})$\\ \hline
$^{36}$Cl/$^{35}$Cl & $3.6
\times 10^{-6}$ & $7.1 \times 10^{-7}$ & ---\\
\hline \ca/$^{40}$Ca & $1.6 \times 10^{-6}$ & $(1.5\times
10^{-8})$\footnotemark[2]
&
---\\ \hline \fe/$^{56}$Fe  & $2.3 \times 10^{-7}$ & $1.7 \times
10^{-7}$ & $2.3 \times 10^{-8}$\\ \hline $^{81}$Kr/$^{82}$Kr & $6.7
\times 10^{-6}$ & $6.6 \times 10^{-7}$ & ---\\
\hline $^{93}$Zr/$^{92}$Zr & $4.6 \times 10^{-5}$ & $3.3 \times
10^{-5}$ & $4.5 \times 10^{-6}$\\ \hline $^{99}$Tc/$^{100}$Ru & $1.9
\times 10^{-5}$ & $1.7 \times 10^{-6}$ & ---\\ \hline \pd/$^{108}$Pd
& $3.5 \times 10^{-5}$ & $3.2 \times 10^{-5}$ & $(2.0 \times
10^{-5})\footnotemark[7]$\\ \hline $^{135}$Cs/$^{133}$Cs & $8.9
\times 10^{-5}$ &
$7.0 \times 10^{-5}$ & $1.5 \times 10^{-5}$\\ \hline
$^{205}$Pb/$^{204}$Pb & $<4.8 \times 10^{-4}$ & $<4.6 \times 10^{-4}$
& $<3.8
\times 10^{-4}$\\
\hline\hline \multicolumn{4}{|c|} {$f_0 = 3.16 \times 10^{-4};
(5.0 M_\odot, Z_\odot)$}\\ \hline $(N^R/N^I)_{\Delta_1}$ &
$\Delta_1 = 0$ & $\Delta_1 = 0.14$ Myr & $\Delta_1
+ \Delta_2 = 4.54$ Myr\\
\hline\hline \al/$^{27}$Al (0.18)\footnotemark[1] & $(5.7 \times
10^{-5})$ & $(5.0\times 10^{-5})$ & $(7.6 \times 10^{-7})$\\
\hline $^{36}$Cl/$^{35}$Cl & $3.3 \times 10^{-7}$ & $2.4 \times
10^{-7}$ &
---\\ \hline \ca/$^{40}$Ca & $3.7 \times 10^{-8}$ & $(1.5 \times
10^{-8})$\footnotemark[2] & --- \\ \hline \fe/$^{56}$Fe & $1.2
\times 10^{-6}$ &
$1.1 \times 10^{-6}$ & $1.5 \times 10^{-7}$\\ \hline
$^{81}$Kr/$^{82}$Kr & $1.8 \times 10^{-5}$ & $1.1 \times 10^{-5}$ &
---\\ \hline $^{93}$Zr/$^{92}$Zr & $2.6 \times
10^{-4}$ & $2.4 \times 10^{-4}$ & $3.2 \times 10^{-5}$\\ \hline
$^{99}$Tc/$^{100}$Ru & $6.4 \times 10^{-5}$ & $4.1 \times 10^{-5}$ &
--- \\ \hline \pd/$^{108}$Pd &
$3.2 \times 10^{-5}$ & $3.2 \times 10^{-5}$ & $(2.0 \times
10^{-5})$\footnotemark[7]\\
\hline $^{135}$Cs/$^{133}$Cs & $2.8 \times 10^{-5}$ & $2.6 \times
10^{-5}$ & $5.8 \times 10^{-6}$\\ \hline
$^{205}$Pb/$^{204}$Pb & $<5.6 \times 10^{-5}$ & $<5.6 \times
10^{-5}$ &
$<4.6 \times 10^{-5}$\\
\hline
\end{longtable}
\footnotetext[1]{For all of Table 5 for each model this is the
value required in the envelope to give ($^{26}$Al/$^{27}$Al) $= 5
\times 10^{-5}$ at the chosen $\Delta_1$.} \footnotetext[2]{This
is the value assumed to determine $\Delta_1$.}
\footnotetext[7]{Value used to determine $f_0$ for a selected
$\Delta_1$.  All values that are {\it assumed} are shown in
parenthesis.}

  \clearpage
  \renewcommand{\thefootnote}{\alph{footnote}}
\begin{table}\caption{ Abundances of short-lived nuclei in a cloud
salted with ejecta from a 3 \msb AGB star with $Z/Z_\odot = 1/3$
(no $^{13}$C)}\label{cloud}
\begin{tabular}{|c|c|c|c|c|}\hline
\multicolumn{5}{|c|}{$f_0 = 4 \times 10^{-3}$; (3 \ms,$Z_\odot =
1/3 Z_\odot$)\footnotemark[1]}\\ \hline &
$(N^R/N^I)_{ENV}(q^I_{ENV}/q^I_0)$\footnotemark[2] &
\multicolumn{3}{c|}{$(N^R/N^I)_\Delta$}\\ \cline{3-5}
&  & $\Delta_1 = 0$ Myr & $\Delta_1 = 0.55$ Myr & $\Delta_1 = 6.7$ Myr \\
\hline \al/$^{27}$Al & $(2.0 \times 10^{-2})$ & $(8.0 \times
10^{-5})$ & $(5.0\times 10^{-5})$ & $(8.5 \times 10^{-8})$\\
\hline $^{36}$Cl/$^{35}$Cl & $4.4 \times 10^{-4}$ & $1.7 \times
10^{-6}$ & $4.7 \times 10^{-7}$ & --- \\ \hline \ca/$^{40}$Ca &
$1.5 \times 10^{-4}$ & $5.9 \times 10^{-7}$ & $(1.5 \times
10^{-8})$ & ---\\ \hline \fe/$^{56}$Fe & $6.7 \times 10^{-4}$ &
$2.7 \times 10^{-6}$ & $2.1 \times 10^{-6}$ & $1.0 \times
10^{-7}$\\ \hline $^{81}$Kr/$^{82}$Kr & $1.6 \times 10^{-2}$ &
$6.4 \times 10^{-5}$ & $1.0 \times 10^{-5}$ & ---\\ 
\hline $^{93}$Zr/$^{92}$Zr & $8.1 \times 10^{-2}$ &
$3.2 \times 10^{-4}$ & $2.5 \times 10^{-4}$ & $1.2 \times
10^{-5}$\\ \hline $^{99}$Tc/$^{100}$Ru & $2.6 \times 10^{-1}$ &
$1.1 \times 10^{-3}$ & $1.9 \times 10^{-5}$ & ---\\ \hline
\pd/$^{108}$Pd & $9.9 \times 10^{-3}$ & $4.1 \times 10^{-5}$ &
$3.8 \times 10^{-5}$ & $(2.0 \times 10^{-5})$\\ \hline
$^{135}$Cs/$^{133}$Cs & $1.1 \times 10^{-2}$ & $4.4 \times
10^{-5}$ & $3.6 \times 10^{-5}$ & $3.5 \times 10^{-6}$\\ \hline
$^{205}$Pb/$^{204}$Pb & $<8.6\times 10^{-2}$ & $<3.5 \times
10^{-4}$ & $<3.4 \times 10^{-4}$ & $<2.5 \times 10^{-4}$\\ \hline
\end{tabular}
\end{table}
\footnotetext[1]{Calculated to match (\al/$^{27}$Al)$_0$,
(\ca/$^{40}$Ca)$_{0.55 \mathrm{Myr}}$,
(\pd/$^{108}$Pd)$_{6.7\mathrm{Myr}}$ .} \footnotetext[2]{Values in
the envelope calculated for $Z = 1/3 Z_\odot$ with the factor of
$q^I_{ENV}$ for this $Z$ divided by $(q^I_0)_\odot$ for the
unsalted protosolar cloud.}
  \clearpage

  \clearpage

\centering{
\begin{table}
\caption{Possible Sources of Short-Lived Nuclei}

{Requiring Late Addition ($\le$ 10 Myr)}\label{sites}

\begin{tabular}{|c|c|c|}\hline
NUCLIDE&POSSIBLE&EXCLUDED \\ \hline
$^{10}$Be & IRRAD & AGB, SNe \\
  & & \\
  $^{26}$Al & AGB, IRRAD. & SNe \\
  & & \\
  $^{36}$Cl & AGB?, IRRAD. & SNe \\
  & & \\
$^{41}$Ca & AGB, IRRAD. & SNe \\
  & & \\
$^{53}$Mn & SNe, IRRAD. & AGB \\
  & & \\
$^{60}$Fe & AGB, SNe & IRRAD. \\
  & & \\
$^{107}$Pd & AGB & SNe, IRRAD. \\
  & & \\
$^{135}$Cs? & AGB & SNe, IRRAD.\\
  & & \\
  $^{205}$Pb? & AGB & SNe, IRRAD. \\
\hline
\end{tabular}
\end{table}
}

\end{document}